\documentclass[a4paper,11pt]{article}
\include{definitions}
\pdfoutput=1 

\usepackage{jheppub} 

\usepackage[T1]{fontenc} 
\usepackage{amssymb,amsmath,amsfonts}
\usepackage{mathtools}
\usepackage{mathrsfs}
\usepackage{bbm}
\usepackage{slashed}
\usepackage{nicefrac}
\usepackage{graphicx}
\usepackage{caption}
\usepackage{subcaption}
\usepackage{esint}
\usepackage{verbatim}

\usepackage{graphicx}
\usepackage[dvipsnames]{xcolor}
\usepackage{array}

\title{Precise description of medium-induced emissions}

\author[a]{Johannes Hamre Isaksen}
\author[a]{and Konrad Tywoniuk}
\affiliation[a]{Department of Physics and Technology, University of Bergen, 5007 Bergen, Norway}

\emailAdd{johannes.isaksen@uib.no}
\emailAdd{konrad.tywoniuk@uib.no}

\abstract{
We study jet fragmentation via final-state parton splittings in the medium. These processes are usually calculated theoretically by invoking the large-$N_c$ limit. In this paper we perform the first computation of a $1\to2$ parton splitting in a thermal medium at finite numbers of colors $N_c$, for arbitrary momentum-sharing fraction $z$ and with full transverse dynamics. We show how the problem can be transformed into a system of coupled Schrödinger equations, that we solve numerically. The novel numerical results are used to estimate the accuracy of several widely used approximations. We check the error introduced while going from finite $N_c$ (i.e. $N_c =3$) to the large-$N_c$ limit, which we find to be small. For unbalanced splittings, e.g. when $z\to 0$, only one of the partons is affected by transverse momentum exchanges with the medium. The emission process then separates into a term responsible for the $1 \to 2$ splitting and the subsequent independent broadening of the daughter partons. This is also referred to as the factorizable term. For finite $z$, further contributions arise that are responsible for the coherent color dynamics of the two-parton system, and these are referred to as non-factorizable terms. These were argued to be small for soft (unbalanced) splittings and for large media. In this work we therefore determine the accuracy of keeping only the factorizable term of the large-$N_c$ solution. We find that the error is insignificant at a small splitting fraction $z \sim 0$, but can be sizable in a more balanced splitting with $z\sim 0.5$. Finally, we also examine the eikonal approximation, which amounts to approximating the partons' paths through the medium as straight lines. We find that it is associated with a substantial error for the parameter values we explored in this work.
}

\begin{document}

\maketitle

\section{Introduction}

High energy heavy-ion collisions provide a glimpse of a new state of nuclear matter that emerge only when extreme energy densities are achieved. Under these conditions color degrees of freedom, carried by the fundamental quark and gluon excitations, are released and influence the material properties of the system, hence the name quark-gluon plasma (QGP) \cite{Bjorken:1982qr}. One way of measuring the properties of the QGP is to rely on internal probes that were created concurrently with the plasma. So-called ``hard'' probes refer to a class of observables that, due to the large energy or mass scale involved, are created at very short time scales such that their production mechanism is independent of the complicated processes leading to the formation of a QGP and, therefore, are effectively happening in the vacuum. However, in subsequent stages, when these probes traverse a spatially extended and evolving QGP to reach the detector, modifications of their properties can occur. By comparing the same ``hard'' observable in proton-proton ({\it p-p}) collisions and in heavy-ion collisions ({\it A-A}), one can extract information that can be compared to first-principle calculations of the QGP dynamics.

One of the most useful internal probes in the pursuit of studying QCD and the QGP are so-called jets. These are collimated sprays of energetic hadrons originating from the fragmentation of a highly virtual initial parton \cite{Ellis:318585,Dokshitzer:1991wu}. Jets are produced both in {\it p-p} collisions and in heavy ion collisions. In {\it p-p} collisions the fragmentation process brings the initial high virtuality of the parton, which is of the order of its transverse momentum $p_T$, down to the hadronization scale where non-perturbative effects dominate \cite{Larkoski:2017jix,Marzani:2019hun,Dasgupta:2020fwr}. In heavy-ion collisions, however, the medium provides a scale related to the achieved energy density that interferes with the vacuum-like fragmentation. The induced interactions modify the jet properties compared to the vacuum baseline, leading to a set of phenomena that are usually referred to as {\it jet quenching} \cite{Gyulassy:1990ye,Gyulassy:2003mc}, for reviews see \cite{dEnterria:2009xfs,Peigne:2008wu,Mehtar-Tani:2013pia,Blaizot:2015lma,Ghiglieri:2015zma,Apolinario:2022vzg}. 
Given that jets probe a wide range of scales of heavy-ion collisions, they can serve as probes of the initial pre-equilibrium dynamics \cite{Carrington:2021dvw,Avramescu:2023qvv} and anisotropy of the medium \cite{Barata:2022krd,Barata:2022utc}, as well as of the late hydrodynamic evolution \cite{Andres:2022bql}.
There is a substantial experimental effort at RHIC and the LHC in quantifying the effects of jet quenching, focusing on a range of different observables \cite{Armesto:2015ioy,Connors:2017ptx,Andrews:2018jcm,Kogler:2018hem,Cunqueiro:2021wls,Apolinario:2022vzg}.

The dominant driver of jet-medium modifications is induced radiative processes. In dense media, where multiple scatterings are important, the in-medium splitting functions can be understood from the underlying scales that separate the limiting cases~\cite{Baier:1996kr,Arnold:2008iy,Arnold:2009mr,Blaizot:2012fh,Kurkela:2014tla,Dominguez:2019ges,Andres:2020kfg}. This is true whether the spectra are differential in both the longitudinal momentum splitting variable $z$ and the relative transverse momentum $\rmd I/[\rmd z\, \rmd^2\prel]$, or simply $\rmd I/\rmd z$. The full problem can be tackled by numerical  \cite{Zakharov:2004vm,CaronHuot:2010bp,Feal:2018sml,Ke:2018jem,Andres:2020vxs,Schlichting:2020lef} or analytical techniques~\cite{Mehtar-Tani:2019tvy,Mehtar-Tani:2019ygg,Barata:2020sav,Barata:2020rdn,Barata:2021wuf,Isaksen:2022pkj}.

Medium-induced radiation is mainly responsible for the diffusion of jet energy to large angles, leading to energy loss of jets which induces a bias on the observed jet samples. The total jet energy loss is also sensitive to the medium parameters through the number of resolved substructures acting as sources for medium-induced emissions \cite{Mehtar-Tani:2017ypq,Mehtar-Tani:2017web}. When emitted at small angles, medium-induced emissions can also modify jet substructure \cite{Mehtar-Tani:2016aco,Caucal:2019uvr,Caucal:2021cfb,Andres:2022ovj}. In both cases described above, it is desirable to have a precise description of medium-induced emissions to exploit to a maximal extent jet observables as probes of the QGP.

It is well understood that radiative processes in the medium are non-local, meaning that they extend over finite longitudinal distances in the medium \cite{Baier:1996kr,Zakharov:1996fv,Zakharov:1997uu,Wiedemann:2000za,Arnold:2002ja}. The characteristic {\it branching}, or \textit{formation}, time arises from the transverse momentum accumulated during the splitting. In dense media, dominated by Gaussian diffusion, the dispersion in transverse momentum grows linearly with time, i.e. $\langle k_\perp^2 \rangle = \hat q t$, where $\hat q$ is the so-called jet quenching parameter. The resulting branching time, scaling as $t_{\rm br} \sim \omega /\langle k_\perp^2 \rangle \sim \sqrt{\omega/\hat q}$, can therefore extend up to the length of the medium, $t_{\rm br}\sim L$, for sufficiently energetic splittings, with $\omega \sim \hat q L^2$. On the contrary, sufficiently soft splitting splittings should occur quasi-instantaneously in the medium, i.e. with $t_{\rm br} \ll L$.

However, the non-locality of medium-induced emissions also has a more profound consequence. It turns out \cite{Blaizot:2012fh} that the splitting products can remain  {\it color} correlated to each other over an extended period of time. During this time, the pair interacts coherently with the medium as it is allowed to explore different color representations. After the pair de-correlates, both daughters continue to broaden independently until they exit the medium. Using analytical calculations in the large-$N_c$ limit for gluon-to-gluon branching, the decoherence time was estimated to be of the same order as the branching time $t_{\rm br}$ \cite{Blaizot:2012fh}, see also \cite{Apolinario:2014csa} for a similar discussion. The large-$N_c$ approximation consists of taking the number of colors, $N_c$, to infinity which greatly simplifies the color structure of the problem, making it easier to solve.  Hence, for sufficiently short branching (and decoherence) times, non-factorizable contributions are expected to scale as $t_{\rm br}/L$ and quasi-instantaneous $1\to 2$ splittings factorize from long-distance transverse momentum broadening. This factorization, which we will critically examine below, lends support to a probabilistic, Markovian picture of multiple emissions \cite{Blaizot:2013vha}, see also \cite{Schlichting:2020lef} for similar rate equations in the context of thermal effects.

Contributions that violate this probabilistic picture become important for long formation, or branching, times. Multiple emissions with overlapping formation times lead to interesting factorization-breaking effects \cite{Arnold:2015qya,Arnold:2016kek,Arnold:2016mth,Arnold:2016jnq,Arnold:2018yjd,Arnold:2020uzm}. In particular, in the regime of strongly ordered formation times, one can reinterpret part of these effects, namely the ones as contribute with logarithms of the medium length, i.e. $\log^2 L$ and $\log L$, as radiative corrections to the jet quenching parameter $\hat q$ \cite{Liou:2013qya,Blaizot:2014bha,Wu:2014nca,Iancu:2014kga,Blaizot:2019muz}. It is still an open question how to extend the description of multiple in-medium splitting beyond the probabilistic picture, see e.g. \cite{Barata:2021byj,Isaksen:2022pkj,Arnold:2023qwi}.

On the other side of the spectrum we can consider very hard emissions in the medium. These occur on short timescales and are expected to be vacuum-like, giving rise to new color sources that can be affected by interactions over long distances \cite{Mehtar-Tani:2010ebp}. Kinematically, short formation times correspond to balanced splittings occurring at large relative angles. In order to overcome some of these problems and to focus on hard $1\to 2$ emissions in the medium, i.e. when both daughter partons carry a large longitudinal momentum ($z \approx 1/2$ and $E \to \infty$), one introduced the eikonal approximation \cite{Dominguez:2019ges}. This consists of fixing the straight-line trajectories of the partons through the medium, thus neglecting completely their transverse momentum broadening. In this approximation, the problem reduces most clearly to that of decoherence of the intermediate partons that move apart from each other in transverse space as $\sim \theta t$, where $\theta$ is the relative angle of the pair. Going beyond the large-$N_c$ approximation was then achieved in \cite{Isaksen:2020npj}. For a further discussion of color dynamics in the QGP, see also \cite{Nikolaev:2005zj,Zakharov:2018hfz,Arnold:2019qqc}. Nevertheless, the eikonal approximation becomes unreliable exactly where the medium scales become important, where transverse momentum broadening starts playing an important role.

\begin{figure}[t!]
\centering
\includegraphics[width=0.85\textwidth]{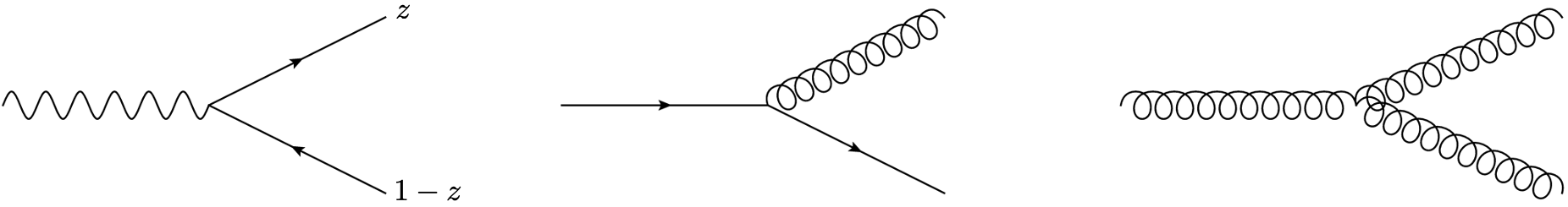} 
\caption{Three different splitting processes where a parton with LC longitudinal momentum $p^+\equiv E$ splits into two partons with momenta $zE$ and $(1-z)E$. The processes shown are $\gamma \to q\bar q$, $q \to g q$, and $g \to gg$.
}
\label{fig:three_splittings}
\end{figure}
In this work, we address the full calculation of the medium-induced splitting function in relative angle $\theta$ (or transverse momentum $\prel$) and longitudinal momentum-sharing fraction $z$ of one single parton into two, for some relevant examples see Fig.~\ref{fig:three_splittings}, for dilute as well as dense media, surpassing all previous approximations.\footnote{In-medium splitting functions (or exclusive two-particle cross sections) were also computed in the so-called {\it opacity} expansion \cite{Sievert:2018imd,Sievert:2019cwq}, appropriate for dilute media \cite{Isaksen:2022pkj}.} This includes effects from a finite number of colors $N_c$ and comprises all non-factorizable contributions. We arrive at a simple formula that generalizes well-known results obtained earlier only in the limit of soft emissions, i.e. $z \ll 1$ \cite{Wiedemann:2000za,Andres:2020vxs,Barata:2021wuf}, and resembles the heuristic picture of a two-step process of emission and subsequent broadening argued for above. However, it involves a novel building block, the quadrupole correlation function, that describes the full dynamics of the daughter parton pair after the splitting. The quadrupole function is governed by a hierarchy of coupled Schrödinger equations, which describes all possible intermediate color representations of the daughter parton pair. Our novel approach allows us to critically examine the conventional approximations and shed light on the dynamics of large-angle and balanced emissions.

We also analyze the deviations from the 
{\it factorization} of the two-parton emission spectrum, and examine the size of the corrections stemming from non-factorizable processes, whether at large- or finite-$N_c$, in a wide kinematic range.

The paper is structured in the following way: in Sec. \ref{sec:notation} we will introduce the basic elements and notation that we will use throughout the paper. In Sec. \ref{sec:splitting} we will derive the spectrum for medium-induced emissions on general grounds, and show how the different approximations simplify the calculations. Lastly, in Sec. \ref{sec:concrete calculation} we do the calculation for a specific splitting process, and in Sec. \ref{sec:numerics} we show the results of our numerical calculation.

\section{Basic elements and notation}\label{sec:notation}

To describe the medium interaction we consider how hard QCD partons behave in an external classical colored field. We work in light-cone gauge $A^{+}=0$.\footnote{We use the conventions $x^\mu = (x^+,x^-,\x)$, where $x^+ = (x^0 + x^3)/2$, $x^- = x^0 - x^3$ and $\x = (x^1,x^2)$, and similarly for other variables.} A parton with large light-cone (LC) ``energy'' $E \equiv p^+$ (or, more precisely, longitudinal momentum) couples mainly to the reciprocal background field component $\Ac(t,\r) \equiv A^{-}(x^+,x^-\simeq 0,\r)$, where we have introduced the light-cone ``time'' $t \equiv x^+$ and we have neglected the extent of the background field in the $x^-$ direction. This immediately guarantees that no longitudinal momentum is exchanged between the parton and the background field, restricting the dynamics to the two-dimensional transverse space at each instant of light-cone time.

Then, the in-medium propagation is described by the propagator,
\begin{equation}
\label{eq:prop-G}
    (\x|\cG_R(t,t_0)|\x_0)=
    \int^{\x}_{\x_0} \cD\r \exp\left[i\frac{E}{2}\int_{t_0}^t \rmd s\,  \dot \r^2(s) \right]\, V_R(t,t_0;[\r])\,,
\end{equation}
describing the transition of a parton from transverse position $\x_0$ at time $t_0$ to the transverse position $\x$ at time $t$ (it is implicitly assumed that $t>t_0$).
The parton is constantly ``kicked'' in the transverse plane by the medium interaction, and it can therefore take an infinite number of possible paths through the medium. These are summed up in the path integral.\footnote{Switching between transverse coordinate basis and transverse momentum basis is straightforward by applying $| \x ) = \int_\p \, \rme^{- i \p \cdot \x} | \p)$ and$(\x| = \int_\p \, \rme^{ i \p\cdot \x} (\p|$ for the initial and final state bases, respectively. Throughout, we use the notation $\int_\q \equiv \int \frac{\rmd^2 \q}{(2\pi)^2}$.} 

The interaction with the medium also involves an exchange of color. The color rotation is encapsulated in a Wilson line along the parton trajectory, given by
\begin{equation}\label{Wilson-definition}
    V_R\left(t, t_{0} ;[\r]\right)=\mathcal{P} \exp \left[i g \int_{t_{0}}^{t} \mathrm{d} s \,\Ac^a(s, \r(s))T^{a}_R\right] \,.
\end{equation}
Here, the label $R$ refers to the fundamental representation of SU(3) for quarks, i.e. $R=F$ with $(T^a)_{ij} \equiv t^a_{ij}$, and the adjoint representation for gluons, i.e. $R=A$ with $(T^b)^{ac} \equiv i f^{abc}$. 

Finally, for the anti-quark, the propagator is given by
\begin{equation}
    (\x_0|\bar \cG_F(t_0,t)|\x)=
    \int^{\x}_{\x_0} \cD\r \exp\left[i\frac{E}{2}\int_{t_0}^t \rmd s\,  \dot \r^2(s) \right]\, V_F^\dagger(t,t_0;[\r])\,.
\end{equation}
Here, we have used the fact that the background field is real, i.e. ${\Ac^a}^\ast(s,\r) = \Ac^a(s,\r)$.

In vacuum, i.e. setting $\Ac = 0$, the propagator simply reduces to its vacuum counterpart
\begin{equation}
\label{eq:G0-pos}
    (\x | \Gc_0(t,t_0) |\x_0) = 
    \frac{E}{2\pi i(t-t_0)} \rme^{i \frac{E}{2} \frac{(\x-\x_0)^2}{(t-t_0)}} \,.
\end{equation}

This can be written in momentum space as 
\begin{equation} \label{eq:G0-mom}
    (\p|\Gc_0(t,t_0)|\p_0) = (2 \pi)^{2} \delta(\boldsymbol{p}-\boldsymbol{p}_{0}) \mathrm{e}^{-i \frac{\p^{2}}{2 E}\left(t-t_{0}\right)}\,.
\end{equation}
Although we have skimmed over some details, the adiabatic turn-off prescription, which is crucial when the time arguments tend to $\pm \infty$, can easily be reinstated.

At high energy, i.e. $E \to \infty$, the parton becomes increasingly constrained to the classical trajectory between the end-points, and the path integral in Eq. \eqref{eq:prop-G} becomes trivial, reducing to a product of a free propagator and a Wilson line,
\begin{equation}\label{eq:eikonal}
    \Gc_R(t,t_0)\simeq \Gc_0(t,t_0)\, V_R(t,t_0;[\x_{\rm cl}])  \,,
\end{equation}
where $\x_{\rm cl}(s) = \frac{t-s}{t-t_0} \x_0 + \frac{s-t_0}{t-t_0} \x$.
This is called the eikonal approximation and can be used to simplify calculations when considering highly energetic partons. Correspondingly, deviations from the straight-line trajectory are called non-eikonal corrections, see also \cite{Altinoluk:2014oxa,Altinoluk:2015gia} for a systematic expansion in non-eikonal corrections. However, in the general case for jet quenching observables in dense heavy-ion collisions the full propagators, as given in Eq.~\eqref{eq:prop-G}, are used.

When computing observables, combining the parton evolution in the amplitude and the complex-conjugate amplitude, one has to account for the fluctuations of the background field. Assuming that the amplitude of the background field is Gaussian, the medium average over the classical field is given by 
\begin{equation}\label{eq:med-avg}
    \langle \Ac^{a}(t, \r) {\Ac^{b}}^\ast(t', \r')\rangle = \delta^{a b} n(t) \delta (t-t') \,\gamma(\r-\boldsymbol{r}^{\prime}) \,,
\end{equation}
where $n(t)$ is the density of scattering centers and the function $\gamma(\r)$ is given by

\begin{equation}\label{eq:gamma-potential}
    \gamma(\r) = \int_\q \, \rme^{ i \q \cdot \r} \frac{\rmd^2 \sigma_{\rm el}}{\rmd^2 \q} \,,
\end{equation} 
and $\rmd^2 \sigma_{\rm el}/\rmd^2 \q$ is the elastic scattering cross-section in the medium.

\section{Describing parton splitting in the medium}
\label{sec:splitting-main}

\subsection{Derivation of the emissions spectrum}\label{sec:splitting}

\begin{figure}[t!]
\centering
\includegraphics[width=0.7\textwidth]{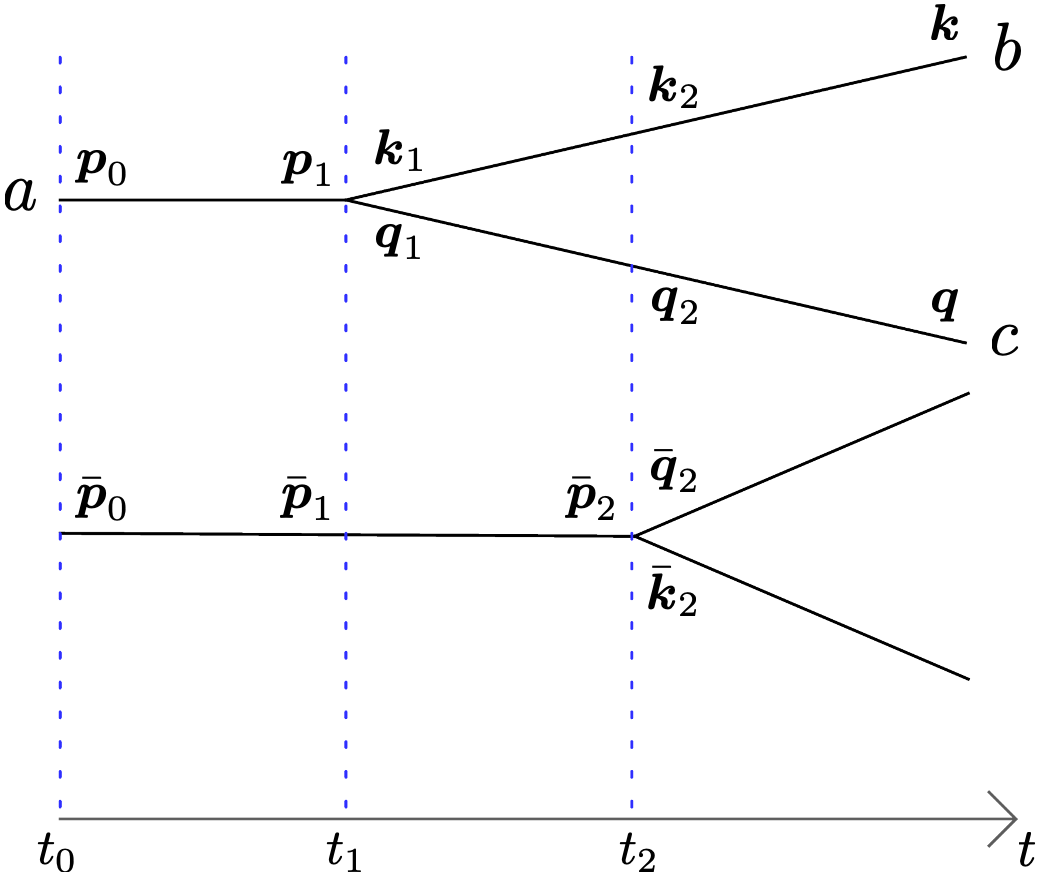} 
\caption{The process of a parton splitting into two, where the time runs from left to right. The amplitude is depicted on the top and the complex conjugate amplitude is on the bottom. All lines refer to propagators that include arbitrarily many medium interactions. Due to the lack of exchange of longitudinal momentum with the medium, the energy stays constant along the lines but the transverse momentum (or transverse position) is continuously updated according to Eq.~\eqref{eq:prop-G}. The parton is created at the initial time $t_0=0$, and the splitting occurs at time $t_1$ in the amplitude and at a later time $t_2$ in the complex conjugate amplitude. Due to the transverse momentum conservation in the vertices, $\q_1 = \p_1- \k_1$ and $\bar \q_2 = \bar \p_2 - \bar \k_2$. 
}
\label{fig:general_splitting}
\end{figure}

In this section, we discuss the $1 \to 2$ splitting process on the partonic level in a deconfined medium. We consider the splitting of a parton of type $a$ with initial transverse momentum $\p_0$ and energy $E$ into two partons of types $b$ and $c$, with transverse momenta and energies $(k^+=zE,\k)$ and $(q^+=(1-z)E,\q)$, respectively, where $z$ is the energy sharing fraction. The conservation of energy is implicitly accounted for in all the vertices.
This is depicted in Fig.~\ref{fig:general_splitting}, where we have placed the process occurring in the amplitude upstairs, and the corresponding process in the complex-conjugate amplitude downstairs. It could be one of many possible splitting processes involving photons, quarks and gluons, but here we choose to keep the notation general and postpone the discussion of a concrete process, namely $\gamma \to q \bar q$, to Sec. \ref{sec:pair_prod_calculation}. For further details about the calculation, we refer to App.~\ref{app:calc-of-processes}, where we explicitly have derived the double-inclusive cross sections for $\gamma \to q \bar q$ and $g \to gg$ splittings.

After performing the medium averages and simplifying the color structure, we can write the inclusive splitting cross section as
\begin{align}
\label{eq:emission-spectrum-not-simplified}
    \frac{\rmd \sigma}{\rmd \Omega_\k \rmd \Omega_\q} &= \frac{g^2 C_R}{(2E)^2}  \rmR \int_0^\infty \rmd t_2 \int_0^{t_2} \rmd t_1
    \int_{\p_0\p_1\k_1\k_2 \bar p_0 \bar\p_2\bar\k_2}  
    {\bs\Gamma}^{i}(\k_1-z\p_1)\cdot{\bs\Gamma}^{\bar i}(\bar \k_2-z\bar \p_2) \nn
    &\times (\k,\q;\k,\q|S^{(4)}(t_\infty,t_2)|\k_2,\q_2;\bar \k_2,\bar \p_2-\bar \k_2) \nn
    &\times (\k_2,\q_2;\bar \p_2|S^{(3)}(t_2,t_1)|\k_1,\p_1- \k_1;\bar p_1)\nn
    &\times (\p_1;\bar \p_1|S^{(2)}(t_1,0)|\p_0;\bar \p_0)\, \langle\Mc_0^{i}(E,\p_0)\Mc_0^{\bar i,\ast}(E,\bar\p_0) \rangle\,,
\end{align}
where the invariant phase space element is $\rmd \Omega_p = \rmd E \rmd^2\p/(2E(2\pi)^3)$. Here, $C_R$ is the squared Casimir operator of the initial parton, or in other words the color charge of the emitter.\footnote{Concretely, $C_R = N_c$ for $g \to gg$, $C_R = C_F$ for $q \to qg$ and $C_R = T_R=\frac12$ for $g \to q\bar q$ and $\gamma \to q \bar q$.} The index ``i'' represents the spin/polarization state of the initial parton before the splitting, which is averaged over, and the product of vertices, i.e. ${\bs \Gamma}^i\cdot {\bs \Gamma}^{\bar i}$, involves a further summation over the final spin/polarization states.
Since we will be interested in the cross section averaged over azimuthal angles, we can simplify the vertex structure as 
\begin{equation}
    C_R {\bs \Gamma}^i(\k_1 - z \p_1) \cdot {\bs \Gamma}^{\bar i}(\bar \k_2 - z \bar \p_2) = \frac{4 P_{ba}(z)}{z(1-z)} \, (\k_1 - z \p_1)\cdot(\bar \k_2 - z \bar \p_2) \delta^{i \bar i} \,,
\end{equation}
for further details see App.~\ref{app:calc-of-processes}. In effect, the product of fundamental vertices becomes directly proportional to the Altarelli-Parisi splitting function $P_{ba}(z)$ \cite{Gribov:1972ri,Altarelli:1977zs,Dokshitzer:1977sg}. The sum over initial polarizations now becomes trivial, allowing to isolate the cross section for the Born process, as $\rmd \sigma_0/\rmd \Omega_{p_0} = \langle \left|\Mc_0(E,\p_0)\right|^2 \rangle$. 

The time integrations in Eq.~\eqref{eq:emission-spectrum-not-simplified} both start at the initial time $t=0$. This corresponds to the creation of an initial hard, energetic particle that propagates through the medium \cite{Wiedemann:2000za,Wiedemann:2000tf}. Then, we can finally write
\begin{equation}
\label{eq:cross-section-P}
    \frac{\rmd \sigma}{\rmd \Omega_q \rmd \Omega_k}=2E\, \int \rmd \Omega_{p_0} \,2\pi\delta(E-k^+-q^+)\,P_2(\k,\q;\p_0) \frac{\rmd \sigma_0}{\rmd \Omega_{p_0}}\,.
\end{equation}
This is the final answer for the two-body cross section in terms of the momenta of the two final-state particles.

The propagation of the partons in the amplitude and complex-conjugate amplitude during the various stages of the splitting process, as depicted in Fig.~\ref{fig:general_splitting}, is encoded in the two-point, three-point, and four-point functions. These are in turn given by correlators of the dressed propagators, namely
\begin{align}
\label{eq:S2-definition-1}
    (\p_1;\bar \p_1 |S^{(2)}(t_1,t_0)|\p_0;\bar \p_0) &= d_a^{(2)} \langle(\p_1|\Gc_a|\p_0)(\bar \p_0|\Gc_a^\dagger|\bar\p_1)\rangle \,, \\
\label{eq:S3-definition-1}
    (\k_2,\q_2;\bar \p_2 |S^{(3)}(t_2,t_1)|\k_1, \p_1-\k_1;\bar \p_1) &=d_{abc}^{(3)} \langle(\k_2|\Gc_b|\k_1)(\q_2|\Gc_c|\p_1-\k_1)(\bar\p_1|\Gc_a^{\dagger}|\bar \p_2) \rangle  \,,
\end{align}
and finally,
\begin{align}
\label{eq:S4-definition-1}
    &(\k,\q;\k,\q|S^{(4)}(t_\infty,t_2)|\k_2,\q_2;\bar\k_2,\bar\p_2-\bar\k_2) \nn
    &=d_{bc}^{(4)} \langle(\k|\cG_b|\k_2)(\q|\cG_c|\q_2)(\bar\p_2-\bar\k_2|\cG_c^\dagger|\q)(\bar\k_2|\cG_b^\dagger|\k)\rangle\,,
\end{align}
where we have dropped the time dependence of the propagators on the right-hand side of the equations. The process-dependent color factors $d_a^{(2)}$, $d_{abc}^{(3)}$ and $d_{bc}^{(4)}$ are responsible for color connecting the propagators and normalizing to the total color charge. We have here also neglected all color indices to be as general as possible. The various color structures for several concrete splitting processes can be inspected in App.~\ref{app:color-summary}.

The $n$-point correlators possess a translation symmetry which renders them invariant under the simultaneous transverse shift of all the coordinates. In momentum space, this becomes manifest as the conservation of momentum incoming and outgoing legs, see App. \ref{appendix:n-point} for details. Concretely, the four-point function then becomes
\begin{align}
\label{eq:s4-simplified}
    &(\k,\q;\k,\q|S^{(4)}(t_\infty,t_2)|\k_2,\q_2;\bar\k_2,\bar\p_2-\bar\k_2)  \nn 
    &= (2\pi)^2\delta^2(\q_2+\k_2-\bar\p_2)\nn
    &\times \Sc^{(4)}\big((1-z)\k-z\q,\k_2-z\bar\p_2,\bar\k_2-z\bar\p_2,\bar\p_2-\k-\q|t_\infty,t_2 \big) \,,
\end{align}
see Eq.~\eqref{eq:S4-structure}. Enforcing the delta-function appearing in \eqref{eq:s4-simplified}, i.e. $\bar \p_2 = \q_2 + \k_2$, the three-point function \eqref{eq:S3-definition-1} further becomes
\begin{align}
\label{eq:k3-simplified}    
    &(\k_2,\bar \p_2-\k_2;\bar \p_2|S^{(3)}(t_2,t_1)|\k_1,\p_1-\k_1;\bar\p_1) \nn
    & = (2\pi)^2\delta(\p_1-\bar\p_1) \, \Sc^{(3)}(\k_2-z\bar \p_2,\k_1-z\p_1,\p_1-\bar \p_2|t_2,t_1)\,.
\end{align}
At this point, it becomes apparent that it makes sense to introduce new momentum variables, namely
\begin{equation}
\label{eq:new-mom-variables}
    \l_1 = \k_1-z\p_1 \,, \quad \l_2 = \k_2-z\bar\p_2 \,, \quad \text{and } \quad \bar\l_2 = \bar\k_2-z\bar\p_2 \,.
\end{equation}
Finally, using that $\p_1 = \bar\p_1$ in \eqref{eq:S2-definition-1}, the two-point function becomes
\begin{equation}
    (\p_1;\p_1|S^{(2)}(t_1-t_0)|\p_0;\bar\p_0) = (2\pi)^2\delta(\p_0-\bar\p_0)\Pc(\p_1-\p_0|t_1,t_0)\,.
\end{equation}
where $\Pc(\p|t_1,t_0)$ is the broadening function. In the new momentum variables, the splitting function becomes
\begin{align}
    P_2(\prel,\P;\p_0) &= \frac{g^2 \,P_{ba}(z)}{z(1-z)E^2} \rmR \int_0^\infty\rmd t_1 \int_{t_1}^\infty \rmd t_2  \nn
    &\times \int_{\p_1\l_1\l_2\bar \p_2\bar \l_2} \l_1\cdot\bar\l_2 \, \Sc^{(4)}(\prel,\l_2,\bar \l_2,\bar\p_2-\P|t_\infty,t_2)\nn
    &\times  \Sc^{(3)}(\l_2,\l_1,\p_1-\bar \p_2|t_2,t_1) \Pc(\p_1-\p_0|t_1,t_0) \,,
\end{align}
where we have introduced the total and the relative transverse momentum of the pair,
\begin{equation}
    \P \equiv \k+ \q \,, \qquad \text{ and} \qquad \prel \equiv (1-z) \k - z \q \,,
\end{equation}
respectively.

In what follows, we will be interested in the information about the relative transverse momentum, while the total momentum $\P$ can be integrated out. The Jacobian from the change of variables yields $\rmd \sigma/[\rmd \Omega_k \, \rmd \Omega_q] = (2\pi)^6 4 z(1-z)E \, \rmd \sigma/[\rmd z \rmd E \, \rmd^2 \p\,\rmd^2\P]$.
The reduced differential cross section in \eqref{eq:cross-section-P} then becomes
\begin{equation}
\int_{\P} \,\frac{\rmd \sigma}{\rmd \Omega_k \, \rmd \Omega_q} = \frac{\rmd I}{\rmd z \, \rmd^2 \p} \frac{\rmd \sigma_0}{\rmd E} \,.
\end{equation}

The integrals over $\bar\p_2$ and $\p_1$ can now be dealt with independently, by shifting separately the integration variables $\bar \p_2-\P \to \P$ and $\bar \p_2 \to \p_1 - \bar \p_2$, and we end up with a rather compact formula from the emission spectrum in momentum space
\begin{align}
\label{eq:p2-simplified-momentum-space}
     (2\pi)^2 \frac{\rmd I}{\rmd z \rmd^2 \p}  &= \frac{1}{4\pi z(1-z)}\int_\P P_2(\p,\P;\p_0) \nn 
     &= \frac{\alpha_s\, P_{ba}(z)}{ \omega^2} \rmR \int_0^\infty\rmd t_1 \int_{t_1}^\infty \rmd t_2 
     \int_{\l_1\l_2\bar \l_2}  \l_1\cdot\bar\l_2 \, \Qc(\p,\l_2,\bar \l_2|t_\infty,t_2)\Kc(\l_2,\l_1|t_2,t_1)\,.
\end{align}
Here we have defined the quadrupole
\begin{equation} 
    \Qc(\p,\l_2,\bar \l_2|t_\infty,t_2) = \int_\P  \,\Sc^{(4)}(\p,\l_2,\bar \l_2,\P|t_\infty,t_2) \,,
\end{equation}
the splitting kernel,
\begin{equation}
    \Kc(\l_2,\l_1|t_2,t_1) = \int_{\bar\p_2} \, \Sc^{(3)}(\l_2,\l_1,\bar \p_2|t_2,t_1)\,,
\end{equation}
and used that $\int_{\p_1}\Pc(\p_1-\p_0|t_1,t_0)=1$. 

Switching to a formulation in transverse position, we apply Fourier transforms of both correlators, to arrive at
\begin{align}
\label{eq:p2-simplified-position-space}
    (2\pi)^2 \frac{\rmd I}{\rmd z \rmd^2 \p} 
    &= \frac{\alpha_s\, P_{ba}(z)}{\omega^2} \rmR \int_0^\infty\rmd t_1 \int_{t_1}^\infty \rmd t_2 \int_{\u_2\u\ub}\,
    \rme^{-i (\u-\bar \u)\cdot\p}\nn
    &\times  (\partial_{\u_1}\cdot \partial_{\bar \u_2})\,
     \Qc(\u,\ub;\u_2,\bar \u_2|t_\infty,t_2)\Kc(\u_2,\u_1|t_2,t_1)\big |_{\u_1=\bar\u_2=0}\,,
\end{align}
where we introduced the notation $\omega \equiv z(1-z)E$.
The $(\u,\ub)$ spatial coordinates are given in terms of the original coordinates in \eqref{eq:coordinate-change-S4}. In this work, we will concretely use Eq.~\eqref{eq:p2-simplified-position-space} to find a numerical solution. Both Eqs.~\eqref{eq:p2-simplified-momentum-space} and \eqref{eq:p2-simplified-position-space} are extremely compact formulas that account for the full kinematics of medium-induced splitting in an arbitrarily dense medium. 

The quadrupole $\Qc(t_\infty,t_2)$ and splitting kernel $\Kc(t_2,t_1)$ are given by the path integrals
\begin{align}
\label{eq:4-point-simplified}
     \Qc(\u_{\text{\tiny{f}}},\ub_{\text{\tiny{f}}}; \u_2,\bar \u_2|t_\infty,t_2) &=\int_{\u_2}^{\u_{\text{\tiny{f}}}} \cD\u \int_{\ub_2}^{\ub_{\text{\tiny{f}}}} \cD{\bar \u}
    \,\rme^{i\frac{\omega}{2}\int_{t_2}^{t_\infty} \dd s\, (\dot \u^2-\dot {\bar \u}^2)} \Cc^{(4)}(\u,{\bar \u})\,, \\
\label{eq:3-point-path-int}
    \Kc(\u_2,\u_1|t_2,t_1) &= \int_{\u_1}^{\u_2} \cD\u 
    \,\rme^{i\frac{\omega}{2}\int_{t_1}^{t_2} \dd s\, \dot \u^2} \Cc^{(3)}(\u)\,.
\end{align}
Hence, after exploiting symmetries of the problem, we have arrived at a description of the problem in terms of a one-body quantum-mechanical propagator $\Kc(t_2,t_1)$, that lives in the time-interval between the splitting time in the amplitude and in the complex-conjugate amplitude, and a two-body propagator $\Qc(t_\infty,t_2)$, that describes the system after the latter splitting time until the end of the medium.\footnote{In the vacuum, the quadrupole in momentum space reduce to a product of delta functions implying the absence of any further transverse momentum broadening of either of the legs.}

The correlators $\Cc^{(n)}$ are given by process-dependent configurations of Wilson lines, e.g. 
\begin{align}\label{eq:C-Wilson-correlators}
    \Cc^{(3)} &= d_{abc}^{(3)} \left\langle V_b V_c V_a^\dagger\right\rangle \nn
    \Cc^{(4)} &= d_{bc}^{(4)} \left\langle V_b V_b^\dagger V_c V_c^\dagger  \right\rangle\,.
\end{align}
In these correlators the color indices are traced over, and the specific form of the color factors $d_{abc}^{(3)}$ and $d_{bc}^{(4)}$ depends on the process. 

It turns out that the three-point function can be represented as a single exponential, namely
\begin{equation}
\label{eq:3-point-path-int-potential}
    \Cc^{(3)}(\u|t_2,t_1) = \rme^{-\int_{t_1}^{t_2}\rmd s \, v_{ba}(\u)}\,,
\end{equation}
where $v_{ba}(\u)$ is the process-dependent potential of a splitting $a\to bc$, and is given as
\begin{align}\label{eq:potential-finite-z}
    v_{ba}(\r,t)=
    n(t) \left[\frac{c_{cba}}{2}\sigma(\r)+\frac{c_{acb}}{2}\sigma(z\r)+\frac{c_{bac}}{2}\sigma((1-z)\r)\right]\,,
\end{align}
where the color factors are $c_{cba} \equiv C_c + C_b - C_a$ etc., and $C_a \equiv C_R$ are the individual color charges of the three partons. The broadening potential $\sigma$ is a combination of real and virtual interactions $\sigma(\r) = g^2[\gamma(0) - \gamma(\r)]$, where $\gamma(\r)$ is defined in Eq. \eqref{eq:gamma-potential}.

The four-point function, or quadrupole, is more complicated since it allows for the mixing between color states \cite{Arnold:2019qqc}. It therefore corresponds not only to a two-body but also a {\it multi-level} quantum-mechanical problem. It can be calculated through a coupled system of Schrödinger-like differential equations, which is derived in App. \ref{appendix:schr-eq}. The specific details of this many-level system depend on the color charges of the involved particles. The correlator $\Cc^{(4)}$ is itself a color singlet, and therefore the number of coupled equations corresponds to the number of possible color singlets one can make out of four partons. Analytical solutions exist only in the large-$N_c$ limit, where the system of coupled equations drastically simplifies. Will discuss different approximations of $\Qc(t_\infty,t_2)$ as well as solve it exactly numerically for the $\gamma \to q \bar q$ process in the following sections.

\subsection{Isolating the medium contribution}
To continue it is convenient to divide the process into three regions, depending on whether the splittings in the amplitude and complex conjugate amplitude happen before or after the system exits the medium. These regions are: 1) $t_1<t_2<L$ (in-in region), 2) $t_1<L<t_2$ (in-out region) and 3) $L<t_1<t_2$ (out-out region). The spectrum then separates into
\begin{equation}
    \frac{\rmd I^{\rm full}}{\rmd z \rmd^2 \p} = \frac{\rmd I^{\rm{in-in}}}{\rmd z \rmd^2 \p} +\frac{\rmd I^{\rm{in-out}}}{\rmd z \rmd^2 \p} +\frac{\rmd I^{\rm{out-out}}}{\rmd z \rmd^2 \p}
\end{equation}
In many cases we want to study the medium contribution, which can be gotten by simply subtracting the vacuum spectrum. The out-out contribution describes a splitting happening entirely out of the medium, which means that it is equivalent to the vacuum spectrum. The medium contribution to the spectrum is then given by
\begin{align}
    \frac{\rmd I^{\textrm{med}}}{\rmd z \rmd^2 \p} &= \frac{\rmd I^{\rm full}}{\rmd z \rmd^2 \p} -\frac{\rmd I^{\rm{vac}}}{\rmd z \rmd^2 \p} \nn
    &=\frac{\rmd I^{\rm{in-in}}}{\rmd z \rmd^2 \p}+\frac{\rmd I^{\rm{in-out}}}{\rmd z \rmd^2 \p} 
\end{align}
One way of isolating the medium contribution to the spectrum is to calculate the so-called medium modification factor $F_{\rm med}$, given by
\begin{align}\label{eq:F-med-definintion}
    \frac{\rmd I^{\rm full}}{\rmd z \rmd^2 \p} = \frac{\rmd I^{\rm{vac}}}{\rmd z \rmd^2 \p}(1+F_{\rm med})\,.
\end{align}
We will put some effort into calculating the three contributions individually, and will in the end use that knowledge to calculate $F_{\rm med}$.

We can immediately deal with the out-out contribution, which is equivalent to the vacuum contribution. Outside of the medium the three- and four-point correlators simply reduce to $\Cc^{(3)}=\Cc^{(4)} =1$, meaning that the path integrals are free. The three-point function becomes the free propagator, i.e. $\Kc(\u_2,\u_1|t_2,t_1)_{g \to0} = \Kc_0(\u_2-\u1,t_2-t_1)$, where
\begin{equation}
    \Kc_0(\u, \Delta t )= \frac{\omega}{2 \pi i\, \Delta t}\rme^{i \frac{\omega}{2 \Delta t}\u^2}\,,
\end{equation}
or, in momentum space
\begin{equation}
    \Kc_0(\p-\p_0,\Delta t) = (2\pi)^2 \delta^2(\p-\p_0) \rme^{-i \frac{\p^2}{2 \omega} \Delta t}\,.
\end{equation}
The quadrupole becomes a product of two free propagators, i.e. $\Qc(t_\infty,t_2)_{g \to 0} = \Qc_0(t_\infty,t_2)$, where
\begin{align}
     \Qc_0(\u_{\text{\tiny{f}}},\ub_{\text{\tiny{f}}},\u_2,\bar \u_2|t_\infty,t_2) = \Kc_0(\u_{\text{\tiny{f}}}-\u_2,t_\infty-t_2)\, \Kc_0^*(\ub_{\text{\tiny{f}}}-\ub_2,t_\infty-t_2)\,.
\end{align}
In momentum space this simply becomes
\begin{equation}
    \Qc_0(\p,\l_2,\bar \l_2|t_\infty,t_2) = (2\pi)^4 \delta(\p-\l_2)\delta(\p-\bar\l_2)\,.
\end{equation}
This simply represents the fact that, in the absence of further transverse momentum exchanges, the transverse momentum of the pair remains the same after the splitting has taken place, i.e. after $t_2$.

The out-out, or vacuum, contribution is reached by inserting the free propagators into \eqref{eq:p2-simplified-momentum-space}, which simply becomes
\begin{align}\label{eq:vacuum-spectrum}
   (2\pi)^2\frac{\rmd I^{\rm{out-out}}}{\rmd z \rmd^2 \p} &= \frac{\alpha_s}{ \omega^2}P_{ba}(z) \rmR \int_{L}^\infty\rmd t_1 \int_{t_1}^\infty \rmd t_2 \,
    \p^2 \, \rme^{-i \frac{\p^2}{2\omega}(t_2-t_1)} \nn
    &= \frac{2 \alpha_s}{\p^2}P_{ba}(z)\,.
\end{align}
In a more familiar form, it reads
\begin{equation}
    \frac{\rmd I^{\rm out-out}}{\rmd z\, \rmd p_t^2} = \frac{\alpha_s}{2\pi}\frac{P_{ba}(z)}{p_t^2} \,,
\end{equation}
where $p_t \equiv |\prel|= \omega \theta$ in the small-angle approximation, which is nothing else than the well-known vacuum splitting function. 

Similarly, the in-out contribution is immediately found to be given by
\begin{align}\label{eq:in-out-general}
    (2\pi)^2\frac{\rmd I^{\rm{in-out}}}{\rmd z \rmd^2 \p}
    = \frac{2 \alpha_s}{\omega}\frac{1}{\p^2}P_{ba}(z)\rmR  \int_{0}^L\rmd t_1 \int_{\u}  \,
    \rme^{-i \u\cdot\p}\p\cdot\partial_{\u_1}\, \Kc(\u,\u_1|L,t_1)|_{\u_1=0} \,.
\end{align}
The equation for $\Kc(L,t_1)$ is given in \eqref{eq:3-point-path-int} and describes a splitting process that starts at time $t_1$ and extends all the way to the end of the medium $L$. For general medium potentials in \eqref{eq:3-point-path-int-potential}, the path integral has no analytical solution, and it is more useful to reformulate the problem as an evolution equation on $\Kc(t,t_1)$, that reads
\begin{equation}
\label{eq:3-point-schrodinger}
    \left[i \frac{\partial}{\partial t} + \frac{{\bs\partial}^2_{\u}}{2 \omega} + i v_{ba}(\u)\right] \Kc(\u,\u_1|t,t_1) = i \delta(\u - \u_1) \delta(t-t_1) \,,
\end{equation}
where $v_{va}(\u)$ is a process dependent potential, see Eq.~\eqref{eq:potential-finite-z}. This correlator is a Green's function, whose evolution equation takes the form of a Schrödinger equation in 2+1 dimensions, describing the transition between initial time $t_1$ and the final time $t$. The evolution can be solved using analytical \cite{Mehtar-Tani:2019tvy,Mehtar-Tani:2019ygg,Barata:2020sav,Barata:2020rdn,Barata:2021wuf,Isaksen:2022pkj} or numerical methods \cite{Zakharov:2004vm,CaronHuot:2010bp,Feal:2018sml,Ke:2018jem,Andres:2020vxs,Schlichting:2020lef}, and is the basic building block for computing the energy emission spectrum $\rmd I/\rmd z$, differential only in the momentum-sharing fraction $z$, see Sec.~\ref{sec:energy-spectrum}.
 
Lastly, the in-in contribution is given by
\begin{align}\label{eq:in-in-general}
    (2\pi)^2\frac{\rmd I^{\rm{in-in}}}{\rmd z \rmd^2 \p} &= \frac{\alpha_s}{ \omega^2}P_{ba}(z) \rmR \int_{0}^L\rmd t_1 \int_{t_1}^L \rmd t_2 \nn
    &\times \int_{\u_2\u\ub}\,
    \rme^{-i (\u-\bar \u)\cdot\p} (\partial_{\u_1}\cdot \partial_{\bar \u_2})\,
     \Qc(\u,\ub,\u_2,\bar \u_2|L,t_2)\Kc(\u_2,\u_1|t_2,t_1)\big |_{\u_1=\bar\u_2=0}\,.
\end{align}
It is clear that to calculate the in-in spectrum, we have to be able to calculate the quadrupole $\Qc(\u,\ub,\u_2,\bar \u_2|L,t_2)$. In the next subsection we will show one way of doing this numerically.

Now we   will show how to calculate the four-point function in position space
\begin{align}\label{eq:4-point-pos-space}
     \Qc(\u_{\text{\tiny{L}}},\ub_{\text{\tiny{L}}},\u_2,\bar \u_2|L,t_2)=\int_{\u_2}^{\u_{\text{\tiny{L}}}} \cD\u \int_{\ub_2}^{\ub_{\text{\tiny{L}}}} \cD{\bar \u}
    \,\rme^{i\frac{\omega}{2}\int_{t_2}^L \dd s\, (\dot \u^2-\dot {\bar \u}^2)} \Cc^{(4)}(\u,{\bar \u})\,.
\end{align}
The four-point correlator $\Cc^{(4)}$ is a medium-averaged trace of Wilson lines. In our approach, all adjoint Wilson lines are turned into fundamental ones using the identity $U_A^{ab}=2 \tr \left[t^a V_F t^b V_F^\dagger\right]$.\footnote{For problems involving only gluon lines, alternatively one can work entirely in the adjoint representation \cite{Arnold:2019qqc}.} If $\Cc^{(4)}$ contains $n$ pairs of Wilson lines in the fundamental representation there will be $n!$ different ways to connect the color of these Wilson lines. The correlator we are interested in calculating is only one of these states $\Cc^{(4)}=\Cc^{(4)}_i$, and can be found through a system of differential equations that involves all of the states
\begin{equation}
    \frac{\rmd}{\rmd t}\Cc_i(\u,\ub) = \M_{ij}(\u,\ub)\Cc_j(\u,\ub)\,.
\end{equation}
The sum runs over all of the $n!$ color states. In \cite{Isaksen:2020npj} we derived a general method of calculating the evolution matrix $\M$, which we refer the readers to for more details. In the large-$N_c$ limit the matrix $\M$ simplifies greatly and the system becomes analytically solvable. We will show a concrete example of this in Sec. \ref{sec:pair_prod_calculation}. 

The quadrupole $\Qc_i$ is a double path integral over the correlator $\Cc_i$.
In appendix \ref{appendix:schr-eq} it is shown that the path integral can be given equivalently in terms of a system of Schrödinger equations, namely
\begin{align}\label{eq:sch-eq-S4}
    \left[i\delta_{ij}\frac{\partial}{\partial t}+\delta_{ij}\frac{\partial^2_\u-\partial^2_{\bar \u}}{2\omega} -i \M_{ij}(\u,\ub)\right]
     \Qc_j(\U,\U_2)=i \bm1_i\delta(t-t_2)\delta^2(\u-\u_2)\delta^2({\bar \u}-{\bar \u}_2) \,,
\end{align}
where $\bm 1_i = [1,1,\dots,1]$ and we have defined $\U=(t,\u,\ub)$.

From this it is possible to derive Schrödinger equations for objects that contain the four-point function. Defining
\begin{align}\label{eq:object-to-solve-numerically}
    \Fc(\u,\ub|L) &= \int_0^L \rmd t_2 \int_0^{t_2}\rmd t_1 \int_{\u_2} \, \nn
    &\times(\partial_{\u_1}\cdot\partial_{\ub_2}) \Qc(\u,\ub,\u_2,\bar \u_2|L,t_2)\Kc(\u_2,\u_1|t_2,t_1)\big |_{\u_1=\bar\u_2=0}\,.
\end{align}
and acting on this object with the derivative operator $\mathbb{\hat D}[\cdot ] = i\frac{\partial}{\partial t}+\frac{\partial^2_\u-\partial^2_{\bar \u}}{2\omega} -i \M(\u,\ub)$ it turns into a non-homogeneous differential equation
\begin{align}\label{eq:schr-eq-with-k3}
    \mathbb{\hat D}\Big[\Fc(\u,\ub|L)\Big] &= -i \,\partial_{\ub}\delta^2(\ub)\cdot\partial_{\u_1}\int_0^L \rmd t_1 \Kc(\u,\u_1|L,t_1)\big |_{\u_1=0}\,,
\end{align}
where we have omitted the explicit form of the initial condition.
The derivative of the Dirac delta might look odd, but it can be dealt with numerically by simply choosing some representation of the delta function.
Using this object the in-in spectrum becomes
\begin{align}\label{eq:spectrum-with-F}
    (2\pi)^2\frac{\rmd I^{\rm{in-in}}}{\rmd z \rmd^2 \p} &= \frac{\alpha_s \, P_{ba}(z) }{ \omega^2}\rmR \int_{\u\ub}\rme^{-i (\u-\ub)\cdot\p}\, \Fc(\u,\ub|L)\,.
\end{align}
The recipe for calculating the spectrum is then the following: Calculate $\Fc(\u,\ub|L)$ numerically through the Schrödinger equation \eqref{eq:schr-eq-with-k3}, then do a numerical Fourier transform and insert the result into Eq.\eqref{eq:spectrum-with-F}. In the next section we will do this for a specific splitting process.

\subsection{The energy spectrum}
\label{sec:energy-spectrum}
In many cases we are not interested in the transverse momentum of the splitting, but only in the energy fraction carried by each parton. That is, we want to calculate $\frac{\rmd I}{\rmd z }=\int_\p(2\pi)^2 \frac{\rmd I}{\rmd z \rmd^2 \p}$. For now, we take the integral without any restriction on the phase space. Starting with the expression for the fully differential spectrum in momentum space \eqref{eq:p2-simplified-momentum-space}, we write
\begin{align}
    \frac{\rmd I}{\rmd z } &= \frac{\alpha_s}{ \omega^2}P_{ba}(z) \rmR \int_0^\infty\rmd t_1 \int_{t_1}^\infty \rmd t_2 \int_{\p} \int_{\l_1\l_2\bar\l_2} \l_1\cdot\bar\l_2 \, \Qc(\p,\l_2,\bar \l_2|t_\infty,t_2)\Kc(\l_2,\l_1|t_2,t_1)\,.
\end{align}
The relevant part to study is the part involving the quadrupole. Making use of the definitions of $\Qc(t,t_2)$, we find that
\begin{align}
    \int_{\p} \Qc(\p,\l_2,\bar \l_2|t,t_2) &= \int_{\P,\p} \, \Sc^{(4)}(\p,\l_2,\bar \l_2,\bar \p_2 - \P) \,,\nn
    &= \int_{\P,\p,\q_2} \, (\k,\q;\k,\q|S^{(4)}|\k_2,\q_2;\bar \k_2,\bar \p_2- \bar \k_2) \,,
\end{align}
where in the first line we have reinstated the original combination of momenta $\bar \p_2 - \P$. Using the definition of $S^{(4)}(t,t_2)$ in terms of the dressed in-medium propagators, see Eq.~\eqref{eq:S4-definition-1}, and using that $\int_\q (\bar \q_2 |\Gc^\dagger |\q) (\q|\Gc|\q_2) = (2\pi)^2 \delta(\bar \q_2 - \q_2)$ and is diagonal in color space, we can finally show that
\begin{equation}
    \int_\p \, \Qc(\p,\l_2,\bar \l_2|t,t_2) = (2\pi)^2 \delta(\l_2 - \bar \l_2) \,.
\end{equation}
The integral over the quadrupole has been reduced to a Dirac delta, and only the splitting kernel $\Kc(t_2,t_1)$ actually contributes to the energy spectrum, which now reads
\begin{align}
    \frac{\rmd I}{\rmd z } &= \frac{\alpha_s}{\omega^2}P_{ba}(z) \rmR \int_0^\infty\rmd t_1 \int_{t_1}^\infty \rmd t_2 \int_{\l_1\l_2}  \l_1\cdot \l_2\,\Kc(\l_2,\l_1|t_2,t_1)\,,\\
    &= \frac{\alpha_s}{\omega^2}P_{ba}(z) \rmR \int_0^\infty\rmd t_1 \int_{t_1}^\infty \rmd t_2 \, {\bs \partial}_\x \cdot {\bs \partial}_\y\,\Kc(\x,\y|t_2,t_1)_{\x=\y=0} \,,
\end{align}
in momentum and transverse-coordinate representation, respectively.

\section{Photon splitting in the harmonic oscillator approximation}\label{sec:concrete calculation}
So far our results have been completely general, valid for any potential and splitting process. In this section, we will employ our formalism to calculate a specific process in the medium and plot the results. 
For the potential we will use the harmonic oscillator (HO) approximation \cite{Baier:1996kr,Zakharov:1996fv}. 
Then we will narrow our calculation further and study the specific case of a photon splitting to a quark-antiquark pair. 

\subsection{Harmonic oscillator approximation}

The elastic scattering potential $\sigma(\r)$ can be expanded at short distances $|\r| \to 0$ as
\begin{align}
\label{eq:Def_v_potential}
    n(t) C_R\sigma(\r) &= g^2 n(t) C_R\int_\q\, \frac{\rmd\sigma_{\rm el}}{\rmd^2\q} \left(1 - \rme^{i \q \cdot \r} \right) \approx \frac14 \r^2 \hat q_R \log \frac{1}{\mu_\ast^2 \r^2} + \mathcal{O}(\r^4 \mu_\ast^2) \,,
\end{align}
where the form of the first term is universal for the scattering potentials used in the literature, while the scale $\mu_\ast$ is model-specific and depends on how the IR divergence of the scattering is screened by medium effects \cite{Barata:2020sav}. Assuming the logarithm is a slowly varying function, we drop it to arrive at
\begin{equation}
    n(t) C_R \sigma(\r) \simeq \frac14 \qhat_R \r^2\,,
\end{equation}
which is the so-called harmonic oscillator (HO) approximation \cite{Baier:1996kr,Zakharov:1996fv}, see \cite{Mehtar-Tani:2019tvy,Mehtar-Tani:2019ygg,Barata:2020sav,Barata:2020rdn,Barata:2021wuf,Isaksen:2022pkj} for a consistent treatment of the logarithmic corrections. The HO approximation is appropriate for accounting for multiple, soft interactions with the medium. This is usually what mainly governs the dynamics of soft splittings. Besides, it is a very useful approximation since it allows for analytical solutions, in particular for the three-point function $\Kc(t_2,t_1)$ in \ governing the splitting process.

Focusing solely on the three-point function for now, it follows that the potential of the path integral, given in Eq.~\eqref{eq:potential-finite-z}, becomes
\begin{equation}\label{eq:harmonic-pot}
    v_{ba}(\r,t)= \frac14 \qhat_{ba} \r^2\,,
\end{equation}
where
\begin{equation}
    \qhat_{ba} = \left[\frac{c_{cba}}{2C_R}+\frac{c_{acb}}{2C_R} z^2+\frac{c_{bac}}{2C_R}(1-z)^2\right]\qhat_R\,.
\end{equation}
For the process that we will consider in this paper, see Sec.~\ref{sec:pair_prod_calculation}, $\gamma \to q \bar q$ we have $\qhat_{q\gamma} = \qhat_F$.
 
Finally, the parameter $\qhat$ is in general a function of time. In this paper we will simplify this and use brick medium $\qhat(t) = \Theta(L-t)\qhat$, where we treat $\qhat \equiv \hat q_F$ as constant.

In this approximation the splitting kernel becomes $\Kc(t_2,t_1) = \Kc_{\text{\tiny HO}}(t_2,t_1)$, where
\begin{equation}\label{eq:3-point-HO}
    \Kc_{\text{\tiny{HO}}}(\u_2,\u_1|t_2,t_1) = \frac{\omega \Omega}{2 \pi i \,\sin (\Omega \Delta t)} 
    \rme^{\frac{i \omega \Omega}{2 \sin(\Omega\Delta t) }\left[\cos(\Omega\Delta t) \, (\u_1^2+\u_2^2)-2 \u_1 \cdot \u_2\right]} \,,
\end{equation}
where $\Delta t = t_2-t_1$ and $\Omega = \frac{1-i}{2}\sqrt{\qhat_{ba}/\omega}$. 

Hence, in the HO approximation one can obtain analytic expressions for the in-out spectrum and part of the in-in spectrum. The in-out spectrum, given in Eq.~\eqref{eq:in-out-general}, can be simplified by using the relation
\begin{equation}\label{eq:der-int-K3}
    \int_0^{t} \rmd t_1 \, \bdel_{\u_1} \Kc_{\text{\tiny{HO}}}(\u,\u_1|t,t_1)\vert_{\u_1 = 0} = \frac{\omega}{i\pi} \frac{\u}{\u^2} \rme^{i \frac{\omega \Omega}{2}\cot \Omega t \, \u^2} \,.
\end{equation}
After doing the integral over $\u$ the in-out contribution in the HO approximation becomes
\begin{align}\label{eq:in-out-HO}
    (2\pi)^2\frac{\rmd I^{\rm{in-out}}}{\rmd z \rmd^2 \p} 
    &=  -\frac{2\alpha_s }{\omega}P_{ba}(z)\,\rmR \,i \int_0^L \rmd t \,\frac{1}{\cos^2(\Omega t)}\rme^{-i \frac{\tan(\Omega t)}{2 \omega \Omega}\p^2}\nn
    &= -\frac{4 \alpha_s }{\p^2}P_{ba}(z)\,\rmR  \,\left[1-\rme^{-i \frac{\tan(\Omega L)}{2 \omega \Omega}\p^2} \right]\,.
\end{align}
Similarly, the in-in spectrum can also be simplified. Again the first time integral can be done in the HO approximation, and Eq. \eqref{eq:in-in-general} becomes
\begin{align}
    (2\pi)^2\frac{\rmd I^{\rm{in-in}}}{\rmd z \rmd^2 \p}&= \frac{\alpha_s}{\pi \omega}P_{ba}(z) \rmI \int_{0}^L\rmd t_2 \int_{\u_2\u_{\text{\tiny{L}}}\ub_{\text{\tiny{L}}}}\,
    \rme^{-i (\u_{\text{\tiny{L}}}-\ub_{\text{\tiny{L}}})\cdot\p} \,\rme^{i \frac{\omega \Omega}{2}\cot \Omega t_2 \, \u_2^2} \nn
    &\times\frac{\u_2}{\u_2^2}\cdot \partial_{\bar \u_2}\,
     \Qc(\u_{\text{\tiny{L}}},\ub_{\text{\tiny{L}}},\u_2,\bar \u_2|L,t_2)\big |_{\bar\u_2=0}\,.
\end{align}

In our numerical calculations we will compute the object $\Fc(\u,\ub|L)$ given in Eq. \eqref{eq:object-to-solve-numerically}. In the harmonic oscillator picture this is given by
\begin{align}\label{eq:object-to-solve-numerically-ho}
    \Fc(\u,\ub|L)
    = -i \frac{\omega}{\pi} \int_0^{t}\rmd t_2 \int_{\u_2}\, \,\rme^{i \frac{\omega \Omega}{2}\cot{(\Omega t_2)}\u_2^2} 
    \left(\frac{\u_2}{\u_2^2}\cdot\partial_{\ub_2}\right) \Qc(\u,\ub,\u_2,\bar \u_2|t,t_2) \big |_{\bar\u_2=0}
    \,.
\end{align}
This object can be computed through the Schrödinger equation
\begin{align}\label{eq:schr-eq-with-k3-ho}
    \left[i\frac{\partial}{\partial t}+\frac{\partial^2_\u-\partial^2_{\bar \u}}{2\omega} -i \M(\u,\ub)\right]\,\Fc(\u,\ub|L) =
    -\frac{\omega}{\pi} \frac{\u}{\u^2}\cdot \partial_{\ub}\delta^2(\ub) \,\,\rme^{i \frac{\omega \Omega}{2}\cot(\Omega L) \u^2}\,.
\end{align}
We finally stress that the HO approximation is not a necessary ingredient for our numerical procedure. When dealing with the full expression in \eqref{eq:Def_v_potential}, we would need to solve the three-point function $\Kc(t_2,t_1)$ with advanced resummation techniques or numerically, and provide numerical data as the non-homogeneous contribution to the evolution equation \eqref{eq:schr-eq-with-k3-ho}.

\subsection{Pair production}\label{sec:pair_prod_calculation}

Turning finally to the concrete goal of our numerical calculation, let us examine the process of a photon splitting into a quark-antiquark pair, $\gamma \to q \bar q$. The full derivation of this cross section is done in App. \ref{app:pair-prod}. 

The reason for choosing this process to study is mainly a practical one. We have to solve a system of differential equations numerically. For $\gamma \to q \bar q$ the system is $2\times 2$.
Naively, the size of the problem for $q \to q g$ splitting is $6\times 6$ and for $g \to gg$ it is $24\times 24$. But by a clever choice of basis \cite{Arnold:2019qqc,Lappi:2020srm}, these problems could be reduced to $3 \times 3$ and $8 \times 8$, respectively.
Since the numerical complexity increases, with limited computing resources, we chose the least complicated one for the current work.

Another reason is that $\gamma \to q \bar q$ is a pretty good proxy for studying both $q\to q g$ and $g \to gg$. The difference between the three cases is the Wilson line correlator $\Cc^{(4)}$ in Eq. \eqref{eq:4-point-pos-space}. For $\gamma \to q\bar q$ this is
\begin{equation}\label{eq:c4-pair-prod}
    \cC_{q \gamma}^{(4)}(L,t_2) = \frac{1}{N_c} \langle \tr [V_{1} V_{2}^{\dagger} V_{\bar{2}} V_{\bar{1}}^{\dagger}]\rangle\,,
\end{equation}
while for $q\to q g$ and $g\to gg$ it is \cite{Blaizot:2012fh,Apolinario:2014csa}
\begin{align}
    \Cc_{gq}^{(4)}(L,t_2) &= \frac{1}{N_c^2-1} \langle \tr[V_{\bar 1}^\dagger V_1V_2^\dagger V_{\bar 2}] \tr[V_{\bar 2}^\dagger V_2]-\frac{1}{N_c}\tr[V_{\bar 1}^\dagger V_1]\rangle   \\
    \cC_{gg}^{(4)}(L,t_2) &= \frac{1}{N_c(N_c^2-1)} \left\langle 
    \tr[V_1^\dagger V_{\bar 1}]  \tr[ V_2^\dagger V_{\bar 2} V_{\bar 1}^\dagger V_1] \tr[V_{\bar 2}^\dagger V_2]  
    -\tr[V_1^\dagger V_{\bar 1}V_2^\dagger V_{\bar 2} V_{\bar 1}^\dagger V_1 V_{\bar 2}^\dagger V_2] \right\rangle \,.
\end{align}
There is obviously a big difference in color complexity between these three systems. However, the large-$N_c$ limit can be used to simplify $\cC_{gq}^{(4)}$ and $\cC_{gg}^{(4)}$. In both cases the first term goes as $N_c^0$, while the second goes as $N_c^{-2}$, meaning that the latter is subleading in $N_c$. Additionally, in the large-$N_c$ limit, the first term becomes a product of several factors, namely
\begin{align}
    \Cc_{gq}^{(4)}(L,t_2)&\simeq \frac{1}{N_c^2} \langle \tr[ V_1V_2^\dagger V_{\bar 2}V_{\bar 1}^\dagger]\rangle \langle\tr[V_{\bar 2}^\dagger V_2]\rangle   \\
    \cC_{gg}^{(4)}(L,t_2) &\simeq \frac{1}{N_c^3} 
    \langle \tr[V_1^\dagger V_{\bar 1}] \rangle \langle\tr[V_1 V_2^\dagger V_{\bar 2} V_{\bar 1}^\dagger]\rangle \langle\tr[V_{\bar 2}^\dagger V_2]\rangle\,.
\end{align}
These consist of products of dipoles and a quadrupole. The dipoles are easily calculable
\begin{equation}
    \Pc_{1\bar1}(t,t_2)\equiv\frac{1}{N_c}\langle \tr[V_1^\dagger V_{\bar 1}] \rangle=\rme^{-C_F \int_{t_2}^L \dd s \, n(s) \sigma_{1\bar1}}\,,
\end{equation}
where we have introduced the notation $\sigma_{1\bar 1} = \sigma(\r_1 - \r_{\bar 1})$.
However, the quadrupole is not trivial. Notice that the quadrupole that appears in the $q \to q g$ and $g\to gg$ splittings is the same as the one for the $\gamma \to q \bar q$ process. One can therefore write
\begin{align}
    \cC_{gq}^{(4)}(L,t_2) &\simeq 
    \Pc_{2\bar2}(L,t_2) \cC_{q \gamma}^{(4)}(L,t_2)\\
    \cC_{gg}^{(4)}(L,t_2) &\simeq 
    \Pc_{1\bar1}(L,t_2) \Pc_{2\bar2}(L,t_2) \cC_{q \gamma}^{(4)}(L,t_2)\,.
\end{align}
Hence, by calculating the $\gamma \to q \bar q$ process we also calculate the non-trivial part of the $q \to q g$ and $g\to gg$ processes.

In the pair production case the potential in the path integral for the four-point function in \eqref{eq:4-point-pos-space} is a correlator of four Wilson lines, given in \eqref{eq:c4-pair-prod}. There are two ways of connecting the color of these four Wilson lines, leading to a system of Schrödinger equations \eqref{eq:sch-eq-S4} with two states. 
The two states are defined as $\Cc^{(4)}_1 = 1/N_c^2 \langle\tr[V_1 V_2^\dagger][V_{\bar 2} V_{\bar 1}^\dagger]\rangle$ and $\Cc^{(4)}_2 = 1/N_c \langle\tr[V_1 V_2^\dagger V_{\bar 2} V_{\bar 1}^\dagger]\rangle$, where the latter represents the physical state that we are interested in. 

To solve the Schrödinger equation given in Eq. \eqref{eq:schr-eq-with-k3-ho} we need to know the explicit form of the potential matrix $\mathbb{M}$.
In the $\gamma\to q\bar q$ case the indices are $i=1,2$ and the states $ \Qc_i$ are defined through Eq. \eqref{eq:4-point-pos-space} with the potentials $\Cc^{(4)}_i$. The second solution $\Qc_2$ is the physical state that is part of the emission spectrum, but it is coupled with the state $\Qc_1$ through the potential matrix in the Schrödinger equation. The potential matrix was derived in \cite{Isaksen:2020npj}, and for an arbitrary potential it is
\begin{equation}\label{eq:evolution-matrix}
    \mathbb{M} = -\frac12 n(t)
    \begin{bmatrix}
    2 C_F(\sigma_{12} + \sigma_{\bar{2}\,\bar{1}})+\frac{1}{N_c}\Sigma_1
    & -\frac{1}{N_c}\Sigma_1 \\
    -N_c \Sigma_2
    & 2C_F(\sigma_{1\bar1} + \sigma_{\bar{2}2})+\frac{1}{N_c}\Sigma_2
    \end{bmatrix} \,.
\end{equation}
Here we have introduced
\begin{align}
    \Sigma_1 &\equiv \sigma_{1\bar{2}} + \sigma_{2\bar{1}} - \sigma_{1\bar{1}} - \sigma_{2\bar{2}}  \nn
     \Sigma_2 &\equiv \sigma_{1\bar{2}} + \sigma_{\bar{1}2}- \sigma_{12} - \sigma_{\bar1\bar{2}}\,.
\end{align}
In the HO approximation, the potential matrix is given by
\begin{equation}\label{eq:potential-matrix-pair}
    \mathbb{M} =
    -\frac{\qhat}{4C_F} 
    \begin{bmatrix}
    C_F[\u^2+{\bar \u}^2]+\frac{1}{N_c}\u\cdot{\bar \u}
    & -\frac{1}{N_c}\u\cdot{\bar \u} \\
    N_c z(1-z)(\u-\ub)^2
    & [C_F-N_c z(1-z)](\u-\ub)^2
    \end{bmatrix}\,.
\end{equation}
The parameter $\qhat$ that appears here is in the fundamental representation $\qhat_F = \qhat_{q \gamma}$. 
In the potential matrix we have used the coordinate transformation \eqref{eq:coordinate-change-S4} to go to the $(\u,\ub)$ coordinates.
To get the full solution of the system we plug this matrix into Eq. \eqref{eq:schr-eq-with-k3-ho} and solve the differential equation numerically. This finite-$N_c$ result can then be compared to the large-$N_c$ calculation, which we will discuss now. 

\subsection{The large-$N_c$ limit}
In the large-$N_c$ limit the potential matrix simplifies to
\begin{equation}\label{eq:potential-matrix-pair-largeNc}
    \mathbb{M} =
    -\frac{\qhat}{4} 
    \begin{bmatrix}
    \u^2+{\bar \u}^2
    & 0 \\
    2 z(1-z)(\u-\ub)^2
    & [z^2+(1-z)^2](\u-\ub)^2
    \end{bmatrix}\,.
\end{equation}
This simplification actually makes it possible to reach analytical solutions for both of the states.

\paragraph{Calculating $\Qc_1$.} 
In the large-$N_c$ limit the equation for $ \Qc_1$ decouples from $ \Qc_2$, and in the path integral formulation it becomes a product of two independent path integrals
\begin{align}\label{eq:s41-large-Nc}
     \Qc_1(\U_{\text{\tiny{L}}},\U_2) &= \int_{\u_2}^{\u_{\text{\tiny{L}}}} \cD \u \int_{\ub_2}^{\ub_{\text{\tiny{L}}}} \cD \ub \rme^{i \int_{t_2}^{L} \rmd s\, \left[ \frac{\omega}{2} (\dot\u^2 - \dot {\bar \u}^2) +i\frac{\hat q}{4} (\u^2 + {\bar \u}^2) \right]} \nn
    &= \Kc_{\text{\tiny{HO}}}(\u_{\text{\tiny{L}}},\u_2|L,t_2)\Kc_{\text{\tiny{HO}}}^\ast(\ub_{\text{\tiny{L}}},\ub_2|L,t_2)\,,
\end{align}
with $\Kc_{\text{\tiny{HO}}}$ given in \eqref{eq:3-point-HO}, and $\Kc_{\text{\tiny{HO}}}^\ast$ is the complex conjugate with $\Omega^\ast \equiv \frac{1+i}{2}\sqrt{\qhat/\omega}$.

\paragraph{Calculating $\Qc_2$.} 
The second quadrupole $\Qc_2$ is the one that is present in the emission spectrum, and its calculation is therefore of some importance. It is given by the non-homogeneous Schrödinger equation
\begin{align}
    &\left[i\frac{\partial}{\partial t}+\frac{\partial^2_\u-\partial^2_{\bar \u}}{2\omega} +i \frac{\qhat}{4}[z^2+(1-z)^2](\u-\ub)^2\right]
     \Qc_2(\U,\U_2) \nn
     &=-i \frac{\qhat}{2}z(1-z)(\u-\ub)^2  \Qc_1(\U,\U_2) \,,
\end{align}
where $ \Qc_1(\U,\U_2)$ is given in \eqref{eq:s41-large-Nc}. The solution to this 
\begin{equation}\label{eq:Q2-hom-and-nonhom}
      \Qc_2(\U,\U_2) =   \Qc^{\rm{fac}}_2(\U,\U_2) + \int_{t_2}^{t} \rmd t_3 \int_{\v,{\bar \v}} \,  \Qc^{\rm{fac}}_2(\U,\V)  T(\V)  \Qc_1(\V,\U_2) \,,
\end{equation}
where the transition function is $T(\V) = -z(1-z)\qhat (\v - {\bar \v})^2/2$. These two terms were coined the factorizable and non-factorizable terms in \cite{Blaizot:2012fh}. Notice that the non-factorizable term is proportional to $z(1-z)$, so in the soft limit $z\to0$ or $z\to 1$ it becomes negligible. 

The factorizable problem is solved by a homogeneous Schrodinger equation
\begin{align}
    &\left[i\frac{\partial}{\partial t}+\frac{\partial^2_\u-\partial^2_{\bar \u}}{2\omega} +i \frac{\qhat}{4}[z^2+(1-z)^2](\u-\ub)^2\right]
     \Qc^{\rm{fac}}_2(\U,\U_2)\nn
     &=i \delta(t-t_2)\delta^2(\u-\u_2)\delta^2({\bar \u}-{\bar \u}_2) \,.
\end{align}
This is easier solved in the path integral form
\begin{align}
     \Qc^{\rm{fac}}_2(\U_{\text{\tiny{L}}},\U_2) = \int_{\u_2}^{\u_{\text{\tiny{L}}}} \cD \u \int_{\ub_2}^{\ub_{\text{\tiny{L}}}} \cD \ub \,
    \rme^{i \int_{t_2}^{t} \rmd s\, \left[ \frac{\omega}{2} (\dot\u^2 - \dot {\bar \u}^2) +i\frac{\hat q}{4} [z^2+(1-z)^2](\u-\ub)^2 \right]} \,.
\end{align}
After changing variables to $\y = \u-\ub$, $\x = 1/2(\u+\ub)$ the potential becomes independent of $\x$, and the path integrals can be performed. The result is
\begin{align}\label{eq:s4-2hom_xy}
     \Qc^{\rm{fac}}_2(\x,\y,\x_2,\y_2|L,t_2)&=
     \left(\frac{\omega}{2\pi\Delta t}\right)^2
    \rme^{i \frac{\omega}{\Delta t}\left[(\x-\x_2)\cdot(\y-\y_2)\right]}\rme^{-\frac{\qhat}{12}[z^2+(1-z)^2]\Delta t \left[\y^2+\y\cdot\y_2 + \y_2^2\right]}\,,
\end{align}
where $\Delta t=L-t_2$.

It is worth pausing at his point, and consider the form of this function in Fourier space, given by
\begin{equation}
    \Qc^{\rm fac}(\p,\l_2,\bar\l_2) =  \int_{\x,\y,\x_2,\y_2} \, \rme^{i (\x_2 + \frac12 \y_2)\cdot \l_2 - i(\x_2 - \frac12 \y_2)\cdot \bar \l_2 - i\y\cdot \p}\Qc^{\rm{fac}}_2(\x,\y;\x_2,\y_2) \,.
\end{equation}
We see that we can take the $\x$ integral directly on the four-point function. After further simplifications, we find
\begin{align}\label{eq:Q-fac-Peff}
    \Qc^{\rm fac}_2(\p,\l_2,\bar\l_2) &= (2\pi)^2 \delta(\l_2 -\bar \l_2)\, \frac{4\pi}{\hat q_{\rm eff} (L-t_2)}\rme^{-\frac{(\p-\l_2)^2}{\hat q_{\rm eff}(L-t_2)}} \nn
    &\equiv (2\pi)^2 \delta(\l_2 -\bar \l_2) \Pc_{\rm eff}(\p - \l_2|L,t_2)\,,
\end{align}
where the effective $\hat q$ parameter for the photon decay is $\hat q_{\rm eff} = \big[z^2 +(1-z)^2 \big] \hat q$. 

The non-factorizable piece, 
\begin{equation}
\label{eq:Q-nonfac-def}
    \Qc_2^{\rm non-fac}(\U,\U_2) = \int_{t_2}^{t} \rmd t_3 \int_{\v,{\bar \v}} \,  \Qc^{\rm{fac}}_2(\U,\V)  T(\V)  \Qc_1(\V,\U_2)\,,
\end{equation}
is much more complicated since it depends on both of the color-states of the four-point function. In the large-$N_c$ limit, we have an explicit solution for $\Qc_1$ and can therefore solve Eq.~\eqref{eq:Q-nonfac-def}. One can show that in momentum space it is given by
\begin{align}
    \Qc_{\rm non-fac}(\p,\l_2,\bar \l_2|L,t_2) 
    &= \int_{t_2}^L \rmd t_3 \int_{\l_3} \int_\u
     \rme^{-i(\p-\l_3)\cdot \u} \tilde \Pc(z \u|L,t_3) \tilde \Pc((1-z) \u|L,t_3)\nn
    &\times  T(\u|t_3)\Qc_1(\l_3,\l_2,\bar\l_2|t_3,t_2) \,,
\end{align}
where we have written the two-point broadening functions in coordinate space, i.e. $\tilde \Pc(\u) = \int_\p \rme^{-i \p \cdot \u} \Pc(\p)$. Since $T(\u) =-z(1-z)\qhat \u^2/2$, we can trade it for a double derivative and obtain
\begin{align}\label{eq:Q-non-fac-momentum}
    \Qc_{\rm non-fac}(\p,\l_2,\bar \l_2|L,t_2)&= \frac12 z(1-z)\qhat  \int_{t_2}^L \rmd t_3 \int_{\l_3} \nn
    &\times (\partial_\p\cdot\partial_\p) \Pc_{\rm eff}(\p-\l_3|L,t_3) \Qc_1(\l_3,\l_2,\bar\l_2|t_3,t_2) \,.
\end{align}
The quadrupole $\Qc_1$ was given in \eqref{eq:s41-large-Nc}, and in momentum space it is
\begin{equation}
    \Qc_1(\l_3,\l_2,\bar\l_2|t_3,t_2) = 
    \Kc_{\text{\tiny{HO}}}(\l_3,\l_2|t_3,t_2)\Kc_{\text{\tiny{HO}}}^\ast(\l_3,\bar\l_2|t_3,t_2)\,,
\end{equation}
where the splitting kernel in momentum space is
\begin{equation}
    \Kc_{\text{\tiny{HO}}}(\l_3,\l_2|t_3,t_2) = 
    \frac{2 \pi }{i \omega \Omega\,\sin (\Omega \Delta t)} 
    \rme^{\frac{i}{2\omega \Omega \sin(\Omega\Delta t) }\left[\cos(\Omega\Delta t) \, (\l_3^2+\l_2^2)-2 \l_2 \cdot \l_3\right]}\,.
\end{equation}
After taking the double derivative and doing the last Gaussian integral you arrive at the somewhat complicated expression
\begin{align}
    \Qc_{\rm non-fac}(\p,\l_2,\bar \l_2|L,t_2) &= -\frac12 z(1-z)\qhat  \int_{t_2}^L \rmd t_3  \frac{c^2}{A}\nn
    &\times \Qc_1(\p,\l_2,\bar\l_2|t_3,t_2) 
    \left[1-\frac{2 c}{A}+\frac{c}{A^2}\q^2\right]\rme^{-\frac{\q^2}{2 A}}\,,
\end{align}
where we have defined the time-dependent factors
\begin{align}
    c &= \frac{2}{\qhat_{\rm eff}(L-t_3)} \,,\\
    A &= \frac{2}{\qhat_{\rm eff}(L-t_3)}-i \omega[\Omega \cot \Omega (t_3-t_2)-\Omega^\ast \cot \Omega^\ast (t_3-t_2)] \,,\\
    \q &= \omega\left[\frac{\Omega}{\tan\Omega (t_3-t_2)}\left(\p -\frac{\l_2}{\sin\Omega (t_3-t_2)}\right)
    -\frac{\Omega^\ast}{\tan\Omega^\ast (t_3-t_2)}\left(\p -\frac{\bar\l_2}{\sin\Omega^\ast (t_3-t_2)}\right)\right]\,.
\end{align}

\subsection{Factorizable and non-factorizable contributions to the spectrum}

Hence, in this case, the factorizable part of the spectrum reads
\begin{align}
    (2\pi)^2\frac{\rmd I_{\rm{fac}}^{{\rm in}-{\rm in}}}{\rmd z \rmd^2\p} &= \frac{\alpha_s}{\omega^2}P_{ba} (z) \, \rmR \int_0^L \rmd t_2\int_{0}^{t_2}\rmd t_1\int_{\l_2,\l_1} \, \l_1\cdot \l_2 \mathcal{P}_{\rm eff}(\p - \l_2|L,t_2) \Kc(\l_2,\l_1|t_2,t_1) \,\\
    &= \frac{\alpha_s}{\omega^2}P_{ba} (z) \, \rmR \int_0^L \rmd t_2\int_{0}^{t_2}\rmd t_1 \, \rme^{-i \p\cdot \x} \mathcal{P}_{\rm eff}(\x|L,t_2) \,\bdel_\x\cdot \bdel_\y \Kc(\x,\y|t_2,t_1)_{\y=0} \,.
\end{align}
This describes a splitting of one parton into two at times $t_1 < t< t_2$ and the subsequent incoherent broadening of the two-body system, described by the effective jet coefficient $\hat q_{\rm eff}$, at times $t_2 < t< L$. 

After doing the remaining momentum integrals the factorizable part of the spectrum becomes
\begin{align}
\label{eq:in-in-hom-analytic}    
    (2\pi)^2\frac{\rmd I_{\rm{fac}}^{\rm{in-in}}}{\rmd z \rmd^2 \p}&= 
    \frac{2 \alpha_s}{\omega} P_{ba}(z) \rmR \,i
    \int_0^L \rmd t  \frac{2\omega\Omega \cot(\Omega t)}{2 \omega \Omega \cot(\Omega t)+i\qhat_{\rm eff}\Delta t}
    \rme^{-i \frac{\p^2}{2 \omega \Omega \cot(\Omega t)+i\qhat_{\rm eff}\Delta t}}\,.
\end{align}
Similarly, the non-factorizable part of the spectrum is
\begin{align}
\label{eq:in-in-nonfac-analytic}    
     (2\pi)^2 \frac{\rmd I_{\rm{non-fac}}^{\rm{in-in}}}{\rmd z \rmd^2 \p} 
     &= \frac{\alpha_s z(1-z)\qhat\, P_{ba}(z)}{2\omega^2} \rmR \int_0^L \rmd t_2\int_{0}^{t_2}\rmd t_1\int_{t_2}^L \rmd t_3 
     \int_{\l_1\l_2\bar \l_2\l_3}  \l_1\cdot\bar\l_2 \,  \nn
    &\times (\partial_\p\cdot\partial_\p) \Pc_{\rm eff}(\p-\l_3|L,t_3) \Qc_1(\l_3,\l_2,\bar\l_2|t_3,t_2)\Kc(\l_2,\l_1|t_2,t_1)\,.
\end{align}
The two expressions~\eqref{eq:in-in-hom-analytic} and \eqref{eq:in-in-nonfac-analytic} constitute the main analytical results in the large-$N_c$ limit. However, it is worth to stress that the full finite-$N_c$ result can similarly be interpreted as a factorizable contribution, corresponding to the diagonal elements of the interaction matrix $\mathbb{M}$, see \eqref{eq:potential-matrix-pair}, and the remaining diagonal and non-diagonal, or non-factorizable, contributions.

It is interesting to study the time scales of the different parts of the spectrum. The splitting kernel $\Kc(t_2,t_1)$ is governed by the time scale $t_{\rm br}=\sqrt{2\omega/\qhat}$, called the branching time. This which is evident from \eqref{eq:3-point-HO}, which contains the combination $|\Omega \tau| \sim \tau/t_{\rm br}$, where the time difference is $\tau = t_2 - t_1$. This constrains the extent of the time difference to be of the order of $t_{\rm br}$. The scaling of the spectrum with the medium length can be clarified by shifting the time integrals as
\begin{equation}
\int_{0}^L \rmd t_2 \int_0^{t_2} \rmd t_1 \big(\ldots\big) \to \int_0^L\rmd t \int_0^{L-t} \rmd \tau \big(\ldots\big)\,.
\end{equation}
For a large medium or for soft emissions, such that $L \gg t_{\rm br}$, the integration  over $\tau$ yields a factor $\sim t_{\rm br}$, while the remaining integration over $t$ scales with $L$. Hence, the factorizable piece is expected to scale linearly with $L$.

The time dependence of the non-factorizable piece is not so straightforward, as it is a convolution of two different quadrupoles, see Eq. \eqref{eq:Q-non-fac-momentum}. Between $t_2$ and $t_3$ the time dependence is set by $\Qc_1(t_3,t_2)$, which exhibits the same characteristic time scale as $\Kc(t_3,t_2)$, namely at the order of the branching time $t_{\rm br}$. 
One therefore expects the integration over the intermediate time $t_3$ to yield a factor $t_{\rm br}$. It can be shown numerically that the non-factorizable term vanishes in the $z \to 0$ limit, see Figs.~\ref{fig:Fmed-diff-diag-Nc} and \ref{fig:Lund-10}, since $t_{\rm br} \propto \sqrt{z}$. This has to be compensated by another length scale, which should emerge from the remaining dynamics between $t_3 \sim t_2 + t_{\rm br}$ and $L$. In \cite{Blaizot:2012fh} the authors argued that the non-factorizable part contributes only as $t_{\rm br}/L$, suggesting that the non-factorizable contribution should become less relevant for large media. In Sec.~\ref{sec:numerics} we will critically examine this proposed behavior.

\section{Numerical results}
\label{sec:numerics}

In this section we will present the results of our numerical calculations for the $\gamma \to q \bar q$ splitting. The system of coupled Schrödinger equations is solved on a dense grid using the fourth-order Runge-Kutta method. The coupled set of equations is defined in Eq.~\eqref{eq:schr-eq-with-k3-ho} and the potential matrix $\mathbb{M}$ entering these equations is given in \eqref{eq:potential-matrix-pair} at finite-$N_c$ and \eqref{eq:potential-matrix-pair-largeNc} at large-$N_c$. The simulation error was estimated by comparing the simulated value of the factorizable problem with the analytical solution given in \eqref{eq:in-in-hom-analytic}. The code we used to do these simulations can be found here: \url{https://github.com/johannesgutn/schrodinger}.

As mentioned above, we have chosen to focus on the $\gamma \to q \bar q$ process due to the small dimension of possible intermediate color states and, thus, the number of coupled equations; the $q\bar q$ pair can only be in two (singlet and octet) configurations. Our approach can be straightforwardly generalized to more complicated systems like $q\to q g$ and $g \to gg$, however, it is not a trivial task. The main problem is that in these cases the potential matrices $\mathbb{M}$ have a much higher dimension than in the photon case, and the number of coupled evolution equations grows rapidly. 
This also comes with a much higher demand for computing power and time, which is one of the reasons why we decided to focus on the photon case. On the other hand, as discovered for the calculations in the eikonal approximation \cite{Isaksen:2020npj}, one would expect more pronounced deviations from the large-$N_c$ limit in splitting processes involving more color.

The object we end up with after doing our numerical calculation is
\begin{equation}
    \Fc(\p|L) = \int_{\u\ub}\rme^{-i (\u-\bar \u)\cdot\p}\Fc(\u,\ub|L)\,,
\end{equation}
where $\Fc(\u,\ub|L)$ is defined in Eq.~\eqref{eq:object-to-solve-numerically-ho}. The code then simulates the time evolution of the object $\Fc(\p|L)$, as a function of the splitting fraction $z$ and the transverse momentum $\p$, or equivalently splitting angle $\theta$, as $|\p| \simeq \omega \theta$.

We focus on the medium modification factor $F_{\rm med}$, introduced in \eqref{eq:F-med-definintion}. It measures how much the medium-induced emission spectrum differs from the vacuum spectrum, and is defined as
\begin{equation}
\label{eq:fmed-2}
    F_{\rm med} = \left(\frac{\rmd I^{\rm{in-in}}}{\rmd z \rmd^2 \p} +\frac{\rmd I^{\rm{in-out}}}{\rmd z \rmd^2 \p}\right)\left.\middle/\frac{\rmd I^{\rm{vac}}}{\rmd z \rmd^2 \p}\right.\,.
\end{equation}
This ratio cancels directly out the soft and collinear divergences contained in the vacuum spectrum. In particular, since the Altarelli-Parisi splitting function is canceled out in \eqref{eq:fmed-2}, we expect that $F_{\rm med}$ should give a measure of the medium modifications that is, at least {\it qualitatively}, universal among the medium-induced splitting processes (up to the magnitude of the effects).

In the previous sections we have computed the vacuum and in-out contributions analytically in the HO approximation, see Eqs. \eqref{eq:vacuum-spectrum} and \eqref{eq:in-out-HO}. For the in-in contribution we have as mentioned calculated the object $\Fc(\p|L)$ numerically. Inserting this into the above equation gives
\begin{align}
    F_{\rm med} = \rmR \left(\frac{\p^2}{2 \omega^2}  \Fc(\p|L) - 2\left[1-\rme^{-i \frac{\tan(\Omega L)}{2 \omega \Omega}\p^2}\right]\right)\,.
\end{align}
We will now discuss some of the main observations from the results of our simulations.
\subsection{The in-in contribution}
\begin{figure}
    \centering
    \includegraphics[width=0.49\textwidth]{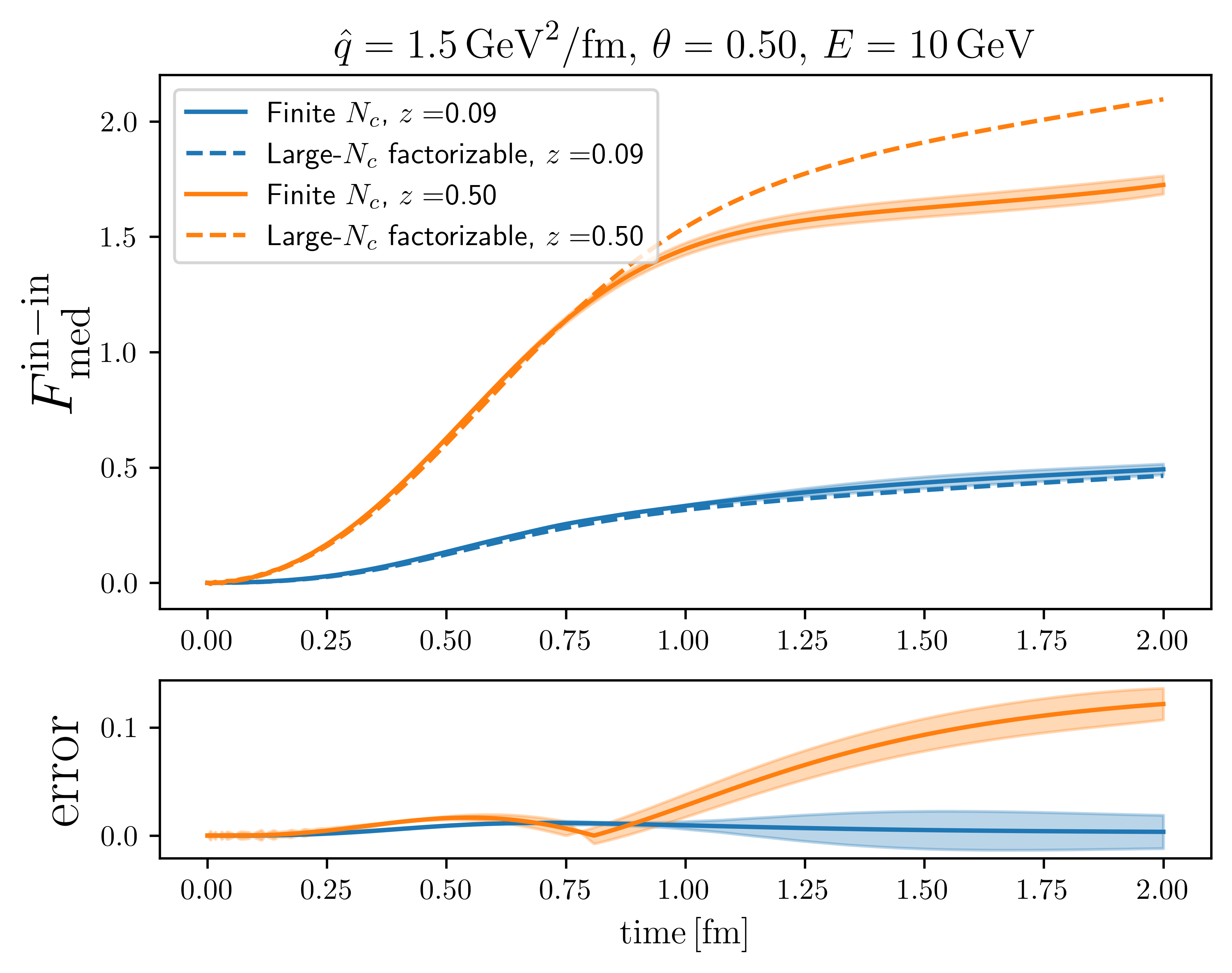}
    \includegraphics[width=0.49\textwidth]{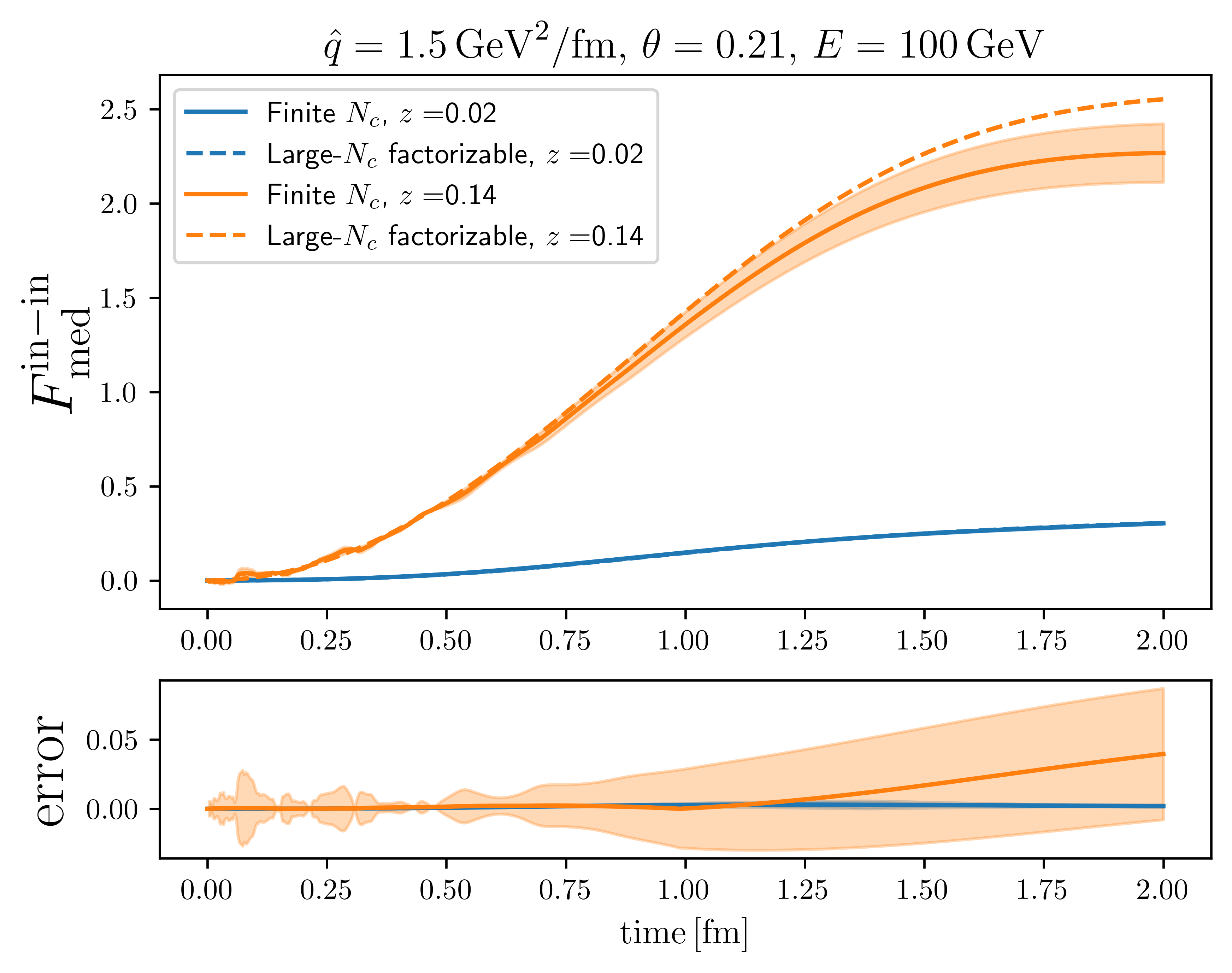}
    \caption{$\Fmed^{\rm in-in}$ as a function of time for different values of $\omega$.}
    \label{fig:fmed-inin-time}
\end{figure}

\begin{figure}
    \centering
    \includegraphics[width=0.9\textwidth]{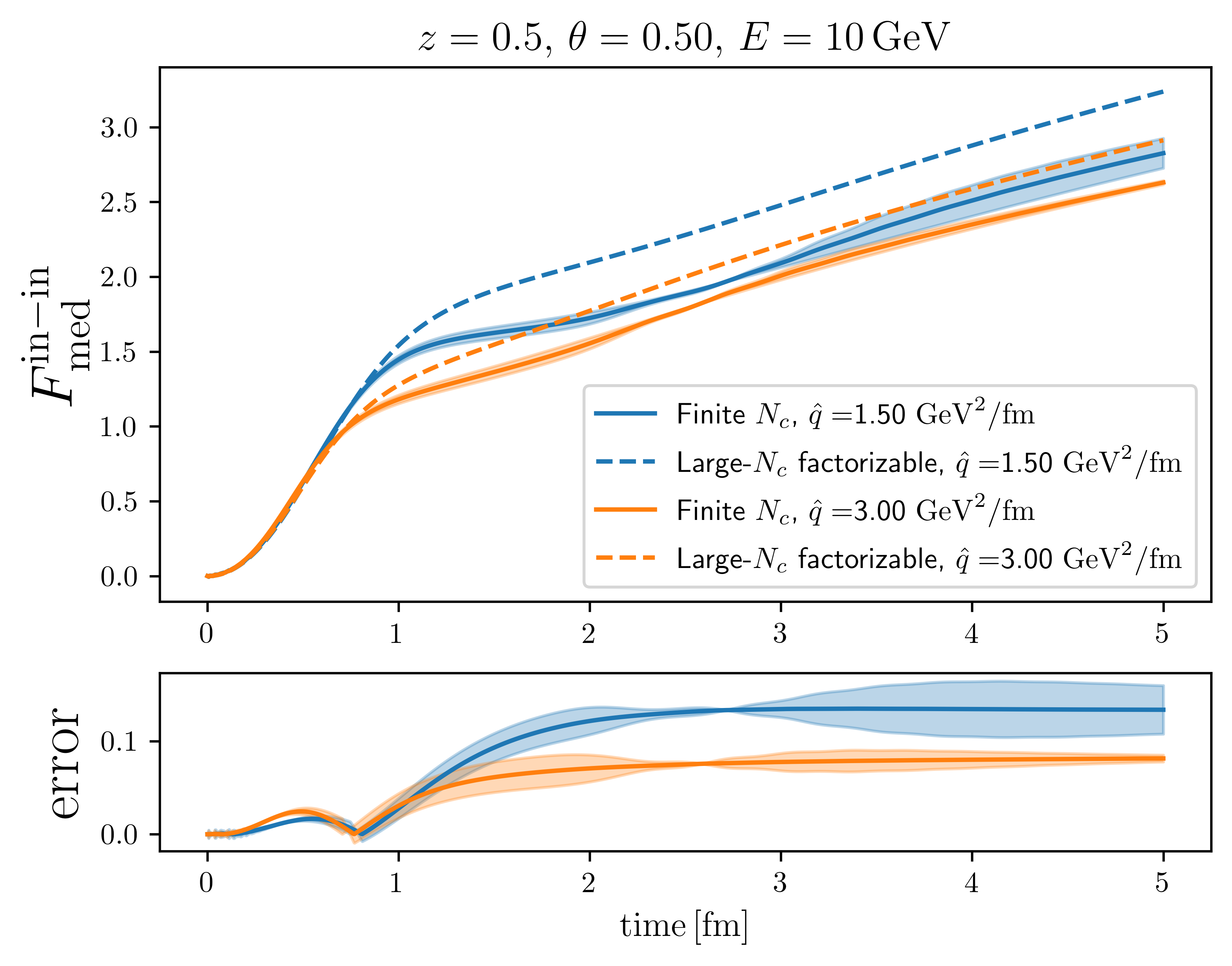}
    \caption{$\Fmed^{\rm in-in}$ as a function of time for two different values of $\qhat$.}
    \label{fig:fmed-inin-time-qhat}
\end{figure}

\begin{figure}
    \centering
    \includegraphics[width=0.49\textwidth]{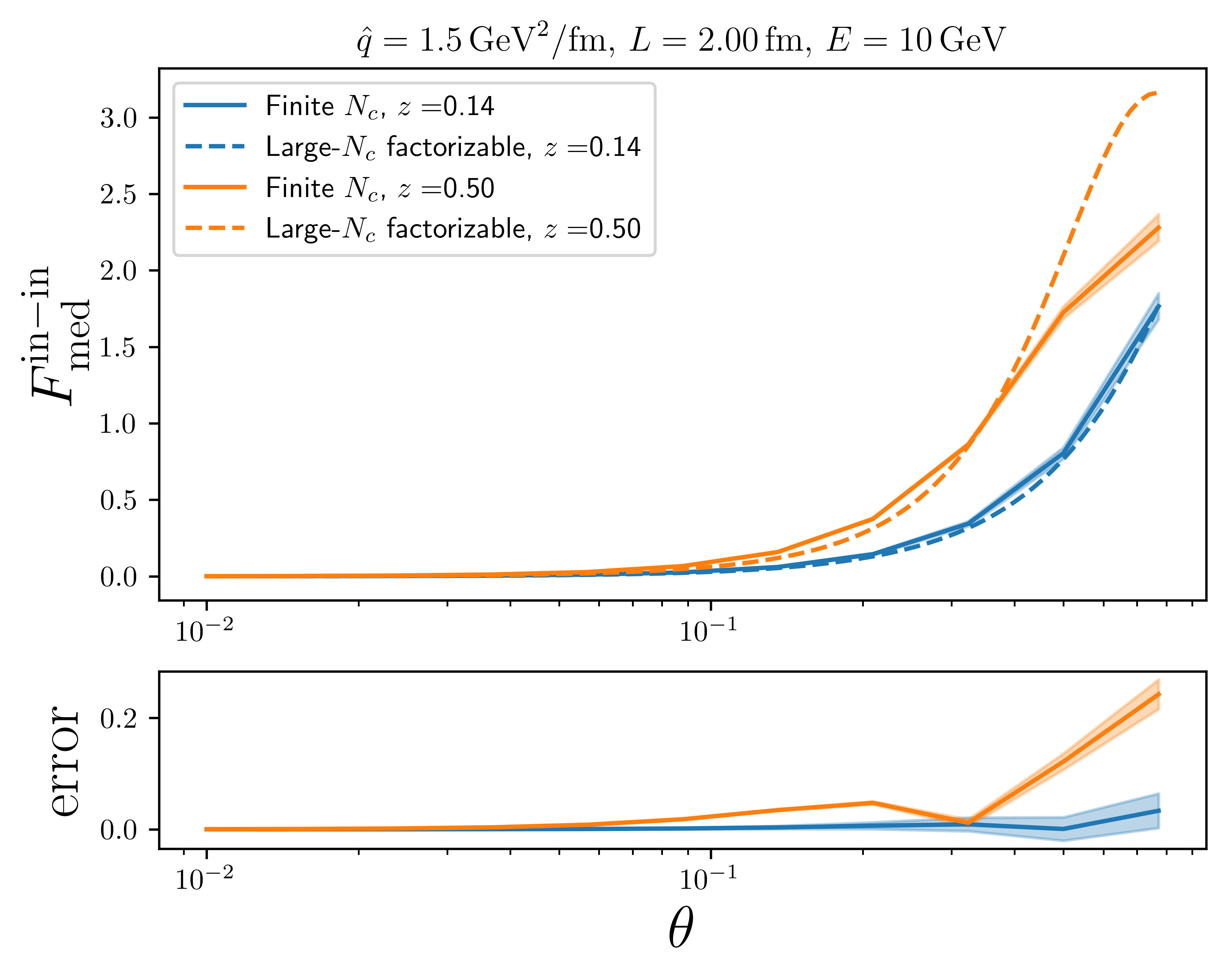}
    \includegraphics[width=0.49\textwidth]{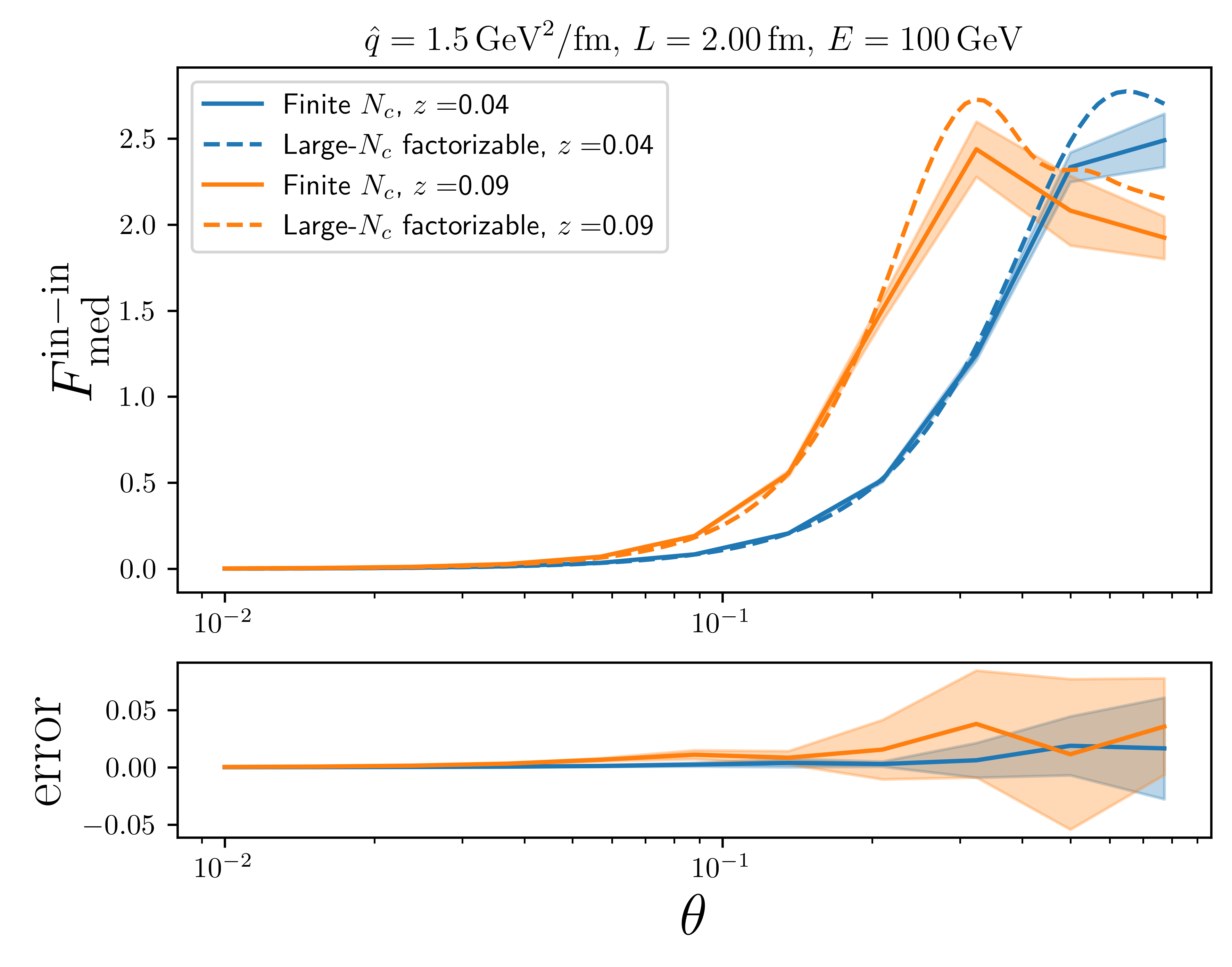}
    \caption{$\Fmed^{\rm in-in}$ as a function of $\theta$ for different values of $\omega$.}
    \label{fig:fmed-inin-theta}
\end{figure}

To start, it is interesting to look at just the in-in contribution, as this is what was studied in \cite{Blaizot:2012fh}. To do this we define $\Fmed^{\rm in-in} = \rmR \,\frac{\p^2}{2 \omega^2}  \Fc(\p|L)$, and plot this as a function of time and $\theta$ in Figs. \ref{fig:fmed-inin-time}-\ref{fig:fmed-inin-theta}. In \cite{Blaizot:2012fh} the authors argue that the non-factorizable part should be negligible as long as $L \gg t_{\rm br} = \sqrt{2 \omega/\qhat}$. The medium parameters were chosen to put focus on where the effects are most sizable. We can now use our numerical results to check this statement. 

Looking at the left plot in Fig. \ref{fig:fmed-inin-time} we have $t_{\rm br}\simeq 0.5$ fm and $t_{\rm br}\simeq 0.8$ fm for $z=0.09$ and $z=0.5$, respectively. 
We would therefore expect the finite-$N_c$ and the factorizable results to match closely at late times. However, this is not what we see. For $z=0.5$ the two solutions actually move away from each other at around 1 fm. The same is also true for the 100 GeV plot on the right, but since error bands are bigger it is harder to draw concrete conclusions.

To study the behavior at late times in more detail we have in Fig. \ref{fig:fmed-inin-time-qhat} plotted $\Fmed^{\rm in-in}$ up to 5 fm. Numerically this was only possible at 10 GeV, as the error blows up at late times for the 100 GeV case. In this figure we have used two different values of $\qhat$, which is a way of varying $t_{\rm br}$ while holding $\omega$ constant. From this figure it is clear that the difference between the finite-$N_c$ and factorizable solutions grows at around 1 fm, and stabilizes to a constant value. This contradicts the notion from \cite{Blaizot:2012fh} that the non-factorizable should become less important at late times. However, it is consistent with \cite{Isaksen:2020npj}, where we also found that the difference between the finite $N_c$ and factorizable versions of the quadrupole amounted to a constant shift after some time.

On the other hand, it is interesting to note that the error in the case where $\qhat=1.5$ GeV$^2/$fm is bigger than the error when $\qhat=3$ GeV$^2/$fm. This does seem to support the claim that the non-factorizable piece is less important when $\qhat$ is big, or equivalently $t_{\rm br}$ is small. Together, these observations seem to indicate that the size of the non-factorizable term as a function of the different parameters is more complicated than previously thought.  

In Fig. \ref{fig:fmed-inin-theta} we have plotted the in-in contribution as a function of the splitting angle $\theta$. This is also interesting to study, as the estimate that the non-factorizable piece should be small when $L\gg t_{\rm br}$ contains no information about the angle. We see that the factorizable piece matches the full solution one when $\Fmed$ itself is small, while the two solutions show differences around the peaks in the distributions.  Of the four curves in Fig. \ref{fig:fmed-inin-theta} only the one with $z=0.5$ and $E=10$ GeV ($t_{\rm br}\simeq 0.8$) exhibits a significant difference between the factorizable and full solution. However, in the four plots the branching time $t_{\rm br}$ ranges from 0.8 fm to 1.4 fm, which is smaller than $L$, but not very much smaller, so it is hard to draw conclusions.

It is clear in both Figs. \ref{fig:fmed-inin-time} and \ref{fig:fmed-inin-theta} that the non-factorizable part is small when $z$ is small, which is what we expect from the analytic calculation \eqref{eq:Q2-hom-and-nonhom}.

\subsection{Validation of the numerical results and comparison to approximate solutions}
Armed with our numerical simulation we have the opportunity to study several approximations. From now we plot the full $\Fmed$, given in \eqref{eq:fmed-2}, which also includes the in-out term. As a first test case, we study going from the full finite-$N_c$ result to the large-$N_c$ solution. This amounts to simplifying the potential matrix in the Schrödinger equation from \eqref{eq:potential-matrix-pair} to \eqref{eq:potential-matrix-pair-largeNc}. 

The next approximation we examine is comparing the magnitude of non-factorizable effects in the splitting. The factorizable piece corresponds to keeping only the first term of the quadrupole, see Eq. \eqref{eq:Q2-hom-and-nonhom}. This is what is usually done in most practical applications, and is therefore a very important test. For instance, for the special case of $g \to q\bar q$ (even with massive quarks), this is the {\it only} contribution at large-$N_c$ \cite{Attems:2022ubu}. 

Lastly, we will also compare our non-eikonal result to the fully eikonal approximation, where all the partons are put on classical paths. This is the approximation we used in \cite{Isaksen:2020npj}, where we performed a similar study. Here, we put this approximation to the test.

\begin{figure}[t!]
    \centering
    \includegraphics[width=0.47\textwidth]{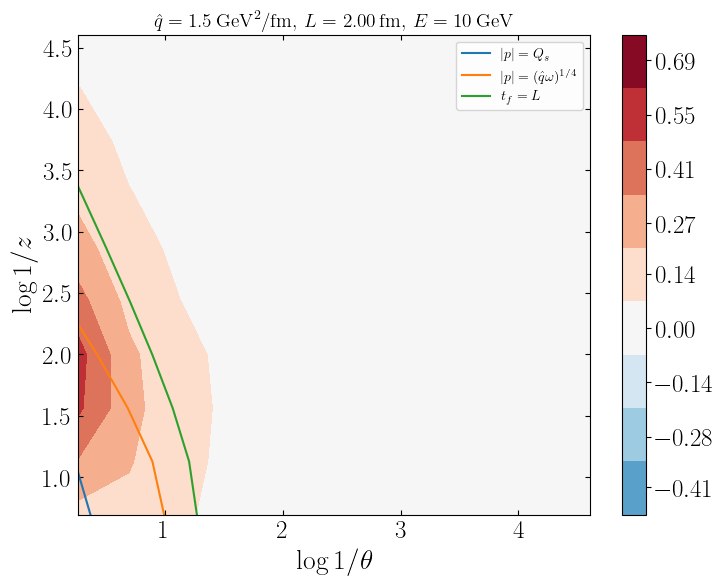}
    \includegraphics[width=0.47\textwidth]{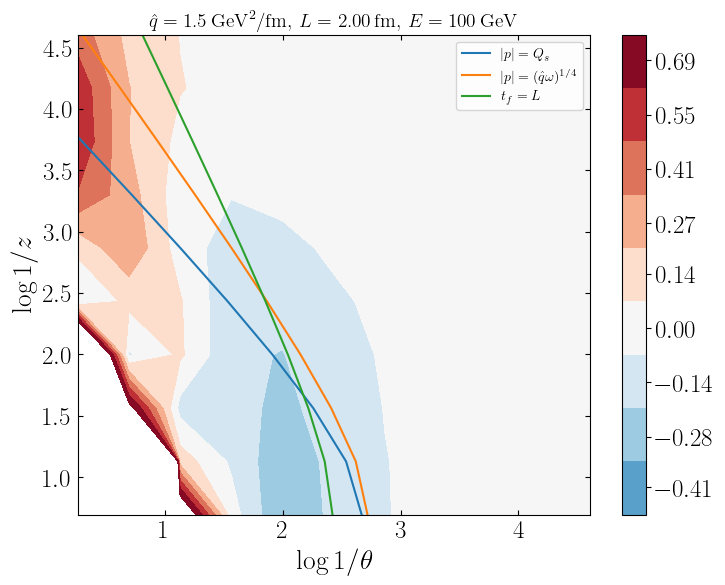}
    \caption{$F_{\rm med}$, given in Eq.~\eqref{eq:fmed-2}, simulated at finite-$N_c$ for two different energies, i.e. for $E=10$ GeV and $E = 100$ GeV. See text for further details. The two medium scales $|\p| = Q_s$ and $|\p| = (\omega \hat q)^{1/4}$ are also plotted, as well as $t_{\rm f} = L$.}
    \label{fig:Fmed-finiteNc}
\end{figure}
We evaluate the numerics for the medium parameters $\hat q = 1.5$ GeV$^2/$fm and $L = 2$ fm, and choose two, widely separated initial energies $E=10$ GeV and $E= 100$ GeV as representative test cases. The simulated $\Fmed$ at finite $N_c$ is shown in Fig.~\ref{fig:Fmed-finiteNc}, where we have plotted it in the two-dimensional $(\log \frac1z, \log\frac1\theta)$ plane (or Lund plane) for a wide range of splitting fractions and angles.\footnote{Following the high-energy notations, we are using natural logarithms throughout.} In this representation the vacuum spectrum alone, in the soft limit, would correspond to a constant $\propto \alpha_s C_R$. Hence, this provides a compact representation of where the medium effects are most pronounced.

In Fig.~\ref{fig:Fmed-finiteNc} we have also depicted the medium scales related to the transverse momentum scales from splitting, i.e. $|\p|=p_t= \omega \theta \sim (\omega \hat q)^{1/4}$ (thick, red line), and broadening, i.e. $p_t=\omega \theta \sim Q_s=\sqrt{\hat qL}$ (thick, blue line). Finally, we also delineate where the formation time $t_{\rm f} = 2 \omega/p_t^2$ becomes equal to the medium length $L$ (thick, green line).  Clearly, as is most visible for the higher parton energy (Fig~\ref{fig:Fmed-finiteNc}, right), medium effects are occurring at large angles and scale nicely with $Q_s$. However, whenever $Q_s \approx (\omega \hat q)^{1/4}$, i.e. for long branching times, we note a net negative effect of medium interactions. In this regime, both of these scales also become comparable to the condition on the formation time $t_{\rm f} <L$, or $p_t > \sqrt{2\omega/L}$. Note, however, that the full spectrum, i.e. vacuum plus medium, is positive.

Finally, we note that the simulation errors become quite large for large angles and large $z$ for $E= 100$ GeV, and our results are therefore not shown in the lower, left corner of Fig~\ref{fig:Fmed-finiteNc} (right). We now turn to discuss exactly the simulation errors.

\paragraph{Estimating the simulation error.}
\begin{figure}[b!]
    \centering
    \includegraphics[width=0.4\textwidth]{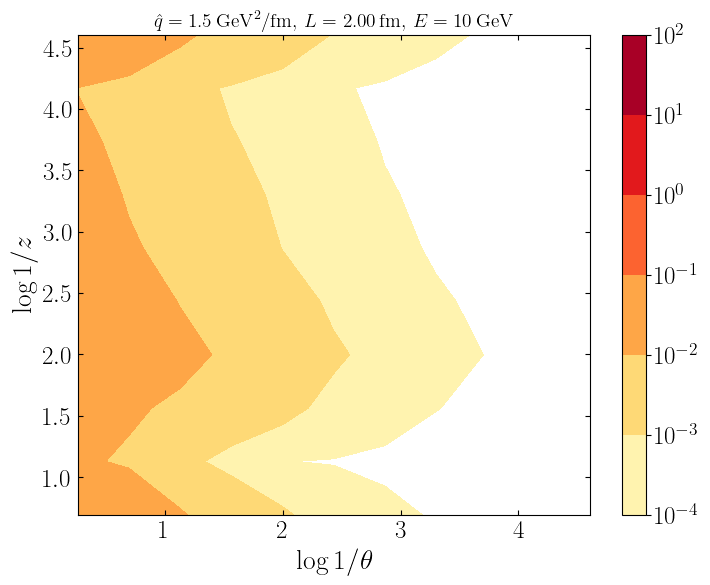}
    \includegraphics[width=0.4\textwidth]{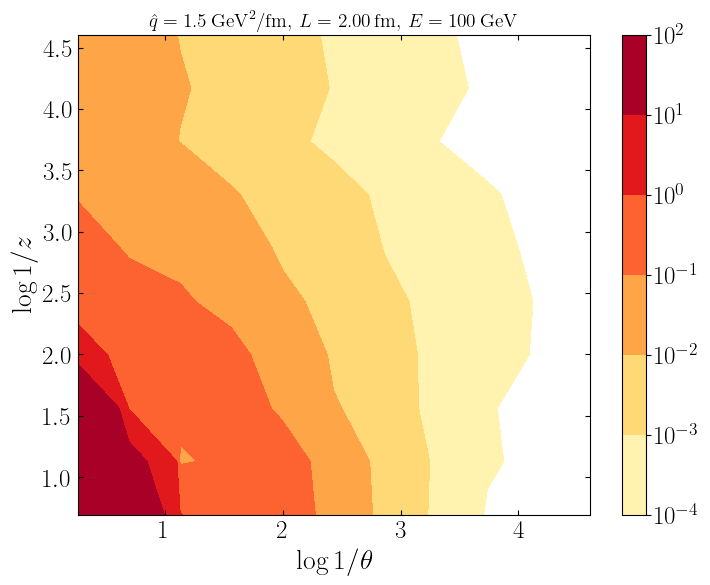}
    \caption{The simulation error $\Delta_{\rm num}$, defined in Eq.~\eqref{eq:delta-num}, for two different energies.}
    \label{fig:simulation-error}
\end{figure}
Before we do any of these comparisons we will discuss the error in our simulation. In this work, we have done three numerical simulations, corresponding to \texttt{i)} the finite-$N_c$ result, \texttt{ii)} the  large-$N_c$ result, and \texttt{iii)} the factorizable part of the large-$N_c$ result. The third point corresponds to a special case where we solve the coupled Schrödinger equations with the potential matrix $\mathbb{M}$ at large-$N_c$, Eq.~\eqref{eq:potential-matrix-pair-largeNc}, while also setting the non-diagonal contributions to zero. In this case we also have an analytic formula, Eq. \eqref{eq:in-in-hom-analytic}. This gives us a way to compare our numerical simulations, in this special case, to a well-defined answer and the difference between these two will then give us an estimate of the simulation error. 

Hence, as a proxy for the full numerical error, we define the error of the simulation as $1+F_{\rm med}$ calculated analytically and numerically, according to prescription \texttt{iii)} above. Hence, we define the accuracy as,
\begin{equation}
\label{eq:delta-num}
    \Delta_{\rm num} \equiv \frac{|F_{\rm {med,analytic}}-F_{\rm {med,simulated}}|}{1+|F_{\rm {med,analytic}}|} \,.
\end{equation}
This quantity is plotted on the Lund plane in Fig.~\ref{fig:simulation-error}.

As one can see from Fig.~\ref{fig:simulation-error} the simulation error is generally small for the 10 GeV case. For the 100 GeV case the error grows large for big $\theta$ and $z\sim 0.5$. We therefore do not expect to get accurate results in this part of the phase space. This comes from the fact that the Schrödinger equation \eqref{eq:schr-eq-with-k3-ho} contains a non-homogeneous term with a complex phase which is proportional to $\sqrt{\omega}$. This phase will oscillate rapidly when $\omega$ is big, meaning we need an increasingly detailed grid to capture these oscillations. Further work is needed to address this corner of the phase space.

However, except for this corner of the phase space for the 100 GeV case, the simulation error is sufficiently small to be able to draw meaningful conclusions from the numerical results. 

\paragraph{Comparing finite-$N_c$ and large-$N_c$.}
\begin{figure}[t!]
    \centering
    \includegraphics[width=0.4\textwidth]{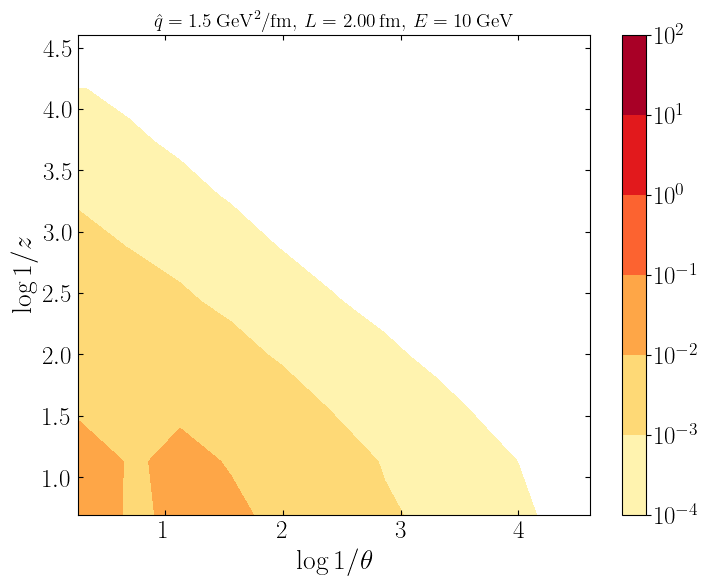}
    \includegraphics[width=0.4\textwidth]{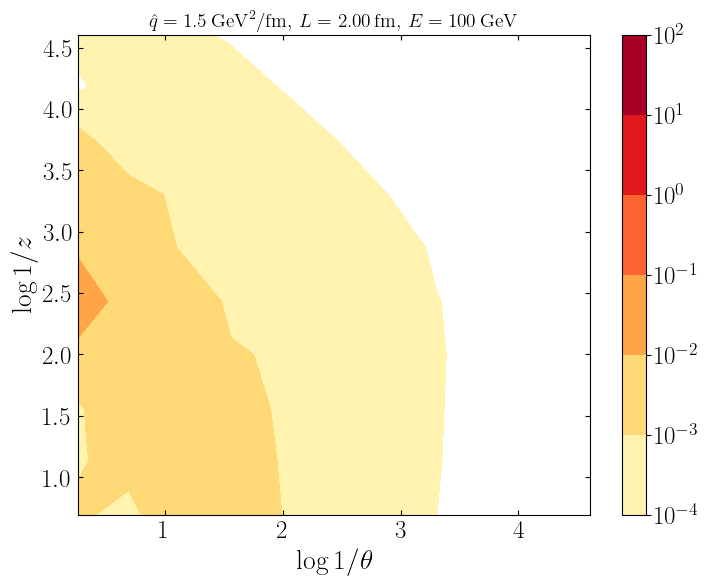}
    \caption{The difference $\Delta_{N_c}$ between $1+F_{\rm med}$ simulated at finite $N_c$ and large-$N_c$, defined in Eq.~\eqref{eq:delta-nc}.}
    \label{fig:Fmed-diff-Nc-full}
\end{figure}
In \cite{Isaksen:2020npj} we found that there is only a very small difference between the spectrum for finite $N_c$ and large-$N_c$ in the $\gamma \to q\bar q$ case. In that paper we used the eikonal approximation for all the partons, so it is interesting to see if anything changes when we here include the possibility to accumulate transverse momentum along the partonic lines. Again we define the difference between the two schemes on the level of $1+F_{\rm med}$, and plot
\begin{equation}
\label{eq:delta-nc}
    \Delta_{N_c} \equiv \frac{|F_{\rm {med,finite-}N_c}-F_{\rm {med,large-}N_c}|}{1+|F_{\rm {med,finite-}N_c}|} \,,
\end{equation}
in Fig.~\ref{fig:Fmed-diff-Nc-full}. As we can see from  the difference between the finite-$N_c$ and large-$N_c$ results are very small in the whole $(\theta,z)$ plane, mostly under 1 \%.  This is consistent with our earlier findings in \cite{Isaksen:2020npj}. However, in that work we found substantial differences in the cases of $q \to q g$ and $g \to gg$, so we would expect the same in this case.

Since the difference between the results at finite $N_c$ and large-$N_c$ is so small we will subsequently only plot the result at finite $N_c$.

\paragraph{Comparing large-$N_c$ and the factorizable contribution.}
\begin{figure}[t!]
    \centering
    \includegraphics[width=0.4\textwidth]{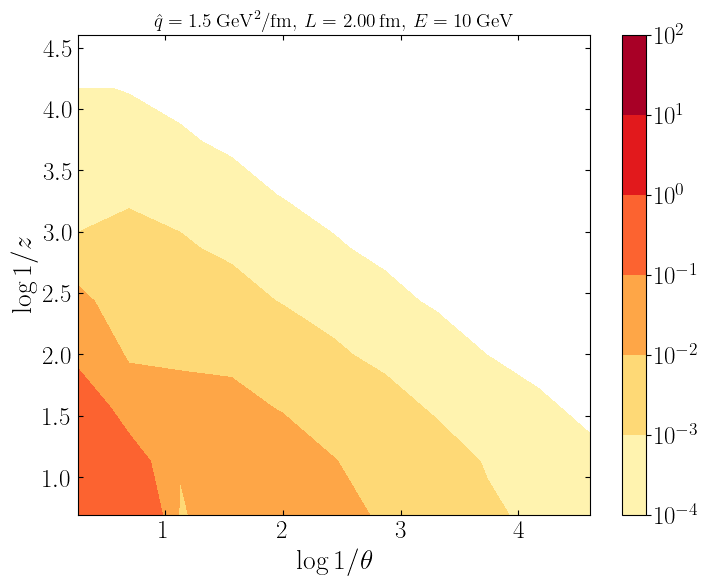}
    \includegraphics[width=0.4\textwidth]{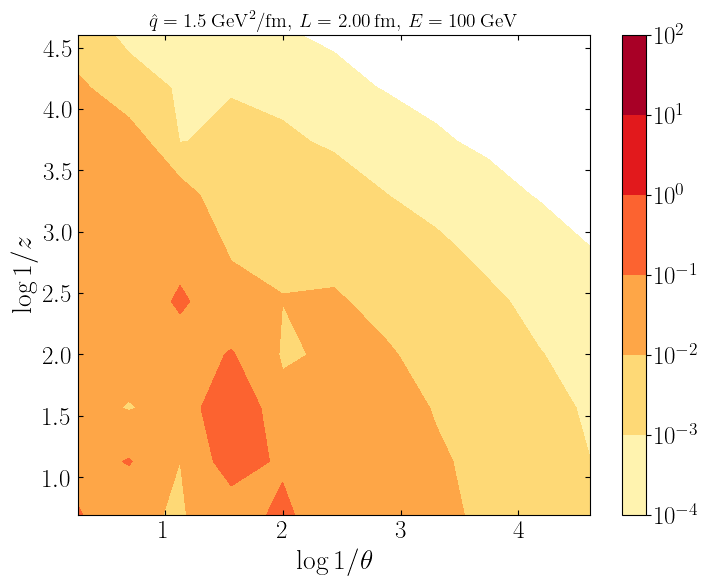}
    \caption{The difference $\Delta_{\rm non-fac}$ between $1+F_{\rm med}$ calculated at large-$N_c$ and keeping only the factorizable term of the large-$N_c$ solution, defined in Eq.~\eqref{eq:delta-non-fac}.}
    \label{fig:Fmed-diff-diag-Nc}
\end{figure}
The quadrupole at large-$N_c$ is given in Eq.~\eqref{eq:Q2-hom-and-nonhom}. In most analytic calculations of the emission spectrum the second non-factorizable term is dropped, and only the first term is considered. We directly see from the definition that the non-factorizable term vanishes for soft emissions $z \to 0$. For soft or unbalanced splittings, $z \ll 1$, we expect the non-factorizable contributions to be suppressed by an inverse length, i.e. $t_{\rm br}/L$ (at least this was argued in the large-$N_c$ limit) \cite{Blaizot:2012fh}.  However, there could be important differences at finite $z$ and large angles.

In order to gauge the importance of the full range of non-factorizable contributions at finite-$N_c$, we define
\begin{equation}
\label{eq:delta-non-fac}
    \Delta_{\rm non-fac} \equiv \frac{|F_{\rm {med,large-}N_c}-F_{\rm {med,factorizable}}|}{1+|F_{\rm {med,large-}N_c}|} \,.
\end{equation}
which is plotted in Fig.~\ref{fig:Fmed-diff-diag-Nc}. We see from comparing Figs.~\ref{fig:Fmed-diff-Nc-full} and \ref{fig:Fmed-diff-diag-Nc} that dropping the non-factorizable part of the large-$N_c$ solution introduces a considerably bigger error than what is introduced by going from finite-$N_c$ to large-$N_c$. 

It is also clear that this approximation works well in the soft limit, as expected. This indicates that it is safe to only use the factorizable piece when calculating soft emissions. However, for finite $z$ one should be more careful, as there might be significant contributions from the non-factorizable term, leading to significant deviations.

\paragraph{Comparing the eikonal approximation and non-eikonal corrections.}
In \cite{Isaksen:2020npj} we also studied many of the same effects we have presented here. However, in that paper we used the eikonal approximation, where all the partons travel on straight lines through the medium. In App. \ref{app:eikonal} we show how the calculation simplifies in the eikonal limit. It is interesting to examine how well the eikonal approximation actually works. 
We expect that the eikonal approximation works best when $\omega$ is big, or when $z\sim 0.5$.  
\begin{figure}[t!]
    \centering
    \includegraphics[width=0.4\textwidth]{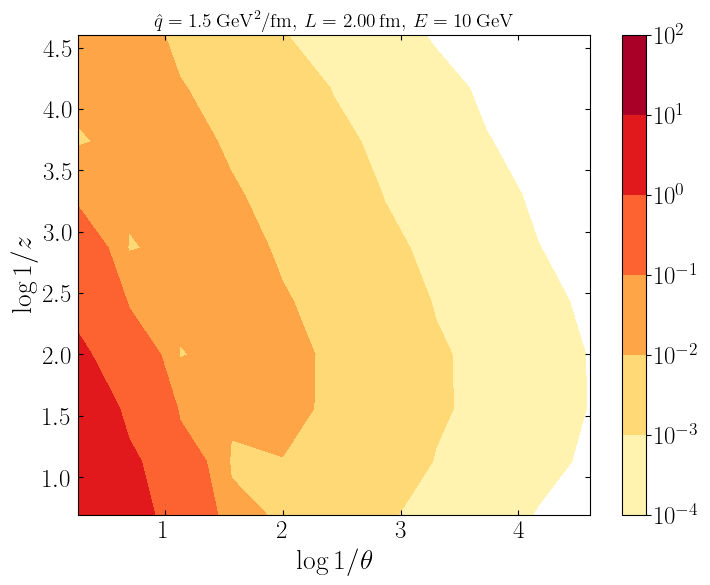}
    \includegraphics[width=0.4\textwidth]{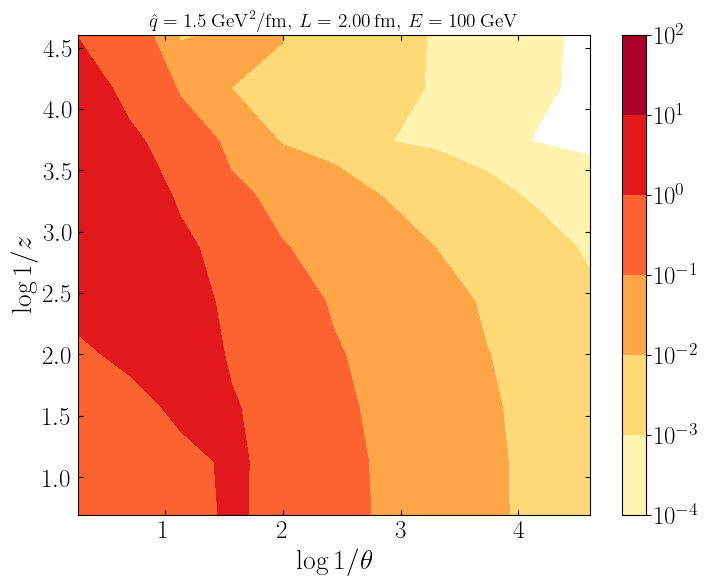}
    \caption{The difference $\Delta_{\rm eik}$ between eikonal and non-eikonal $1+F_{\rm med}$ of the large-$N_c$ solution, defined in Eq.~\eqref{eq:delta-eik}. The non-eikonal result is obtained from numerical simulations, while the eikonal results is calculated using the analytical formulas for the factorizable and non-factorizable terms \eqref{eq:spectrum-fac-eikonal} and \eqref{eq:spectrum-nonfac-eikonal}.}
    \label{fig:Fmed-eikonal}
\end{figure}
In Fig. \ref{fig:Fmed-eikonal} we have plotted the error introduced by using the eikonal approximation, compared to the non-eikonal version, defined as 
\begin{equation}
\label{eq:delta-eik}
    \Delta_{\rm eik} \equiv \frac{|F_{\rm {med,non-eikonal}}-F_{\rm {med,eikonal}}|}{1+|F_{\rm {med,non-eikonal}}|} \,.
\end{equation}
As one can see the eikonal approximation overestimates the contribution by a big margin, especially along the line $\omega \theta \sim Q_s$, i.e. around the peak of the spectrum. We can therefore conclude that using the eikonal approximation does not work well for the values of the parameters we have chosen in this paper.

\begin{figure}[t!]
    \centering
    \includegraphics[width=0.5\textwidth]{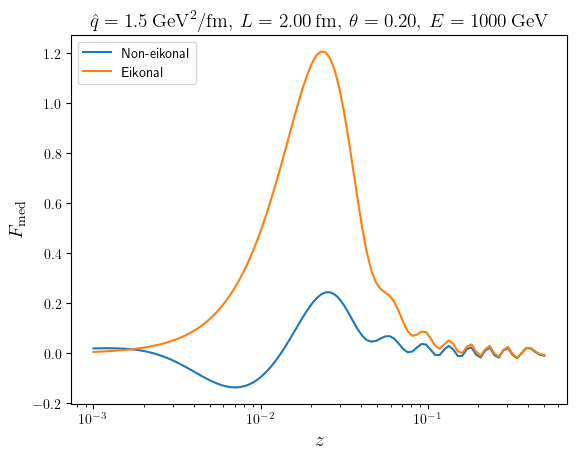}
    \caption{Comparison of the non-eikonal and fully eikonal $F_{\rm med}$, calculated using the factorizable part of the large-$N_c$ solution, as a function of $z$, for a much higher energy $E=1000$ GeV. Only the factorizable part was used here, because the non-factorizable term is expected to be small at large energies.}
    \label{fig:Fmed-eikonal-z}
\end{figure}
In Fig. \ref{fig:Fmed-eikonal-z} we have plotted $\Fmed$ as a function of $z$ at a higher energy of $E=1000$ GeV, to study whether the eikonal approximation is accurate at this energy scale. From the figure it is clear that it does indeed work well for $z$ close to $0.5$. However, it still fails to capture the main contribution of $\Fmed$, which is present at lower $z$.

We would also like to point out that we have here used the eikonal approximation on all of the partons. It is common to study the case where one of the emitted partons is soft, while the other is hard. In that case the eikonal approximation could be used on the hard partons, while the full transverse dependence of the soft parton should be kept. We expect this to give a more accurate result than the fully eikonal case, but we have decided not to pursue this scheme further here.

\subsection{Precision calculation of splitting function in the Lund plane}
\label{sec:precision-calculation}

Finally, we turn to the state-of-the-art calculations of the splitting function, evaluated in the HO approximation without any other further approximations. In this section we discuss our results, and plot slices of the Lund plane as a function of different parameters.

\begin{figure}[t!]
    \centering
    \includegraphics[width=0.49\textwidth]{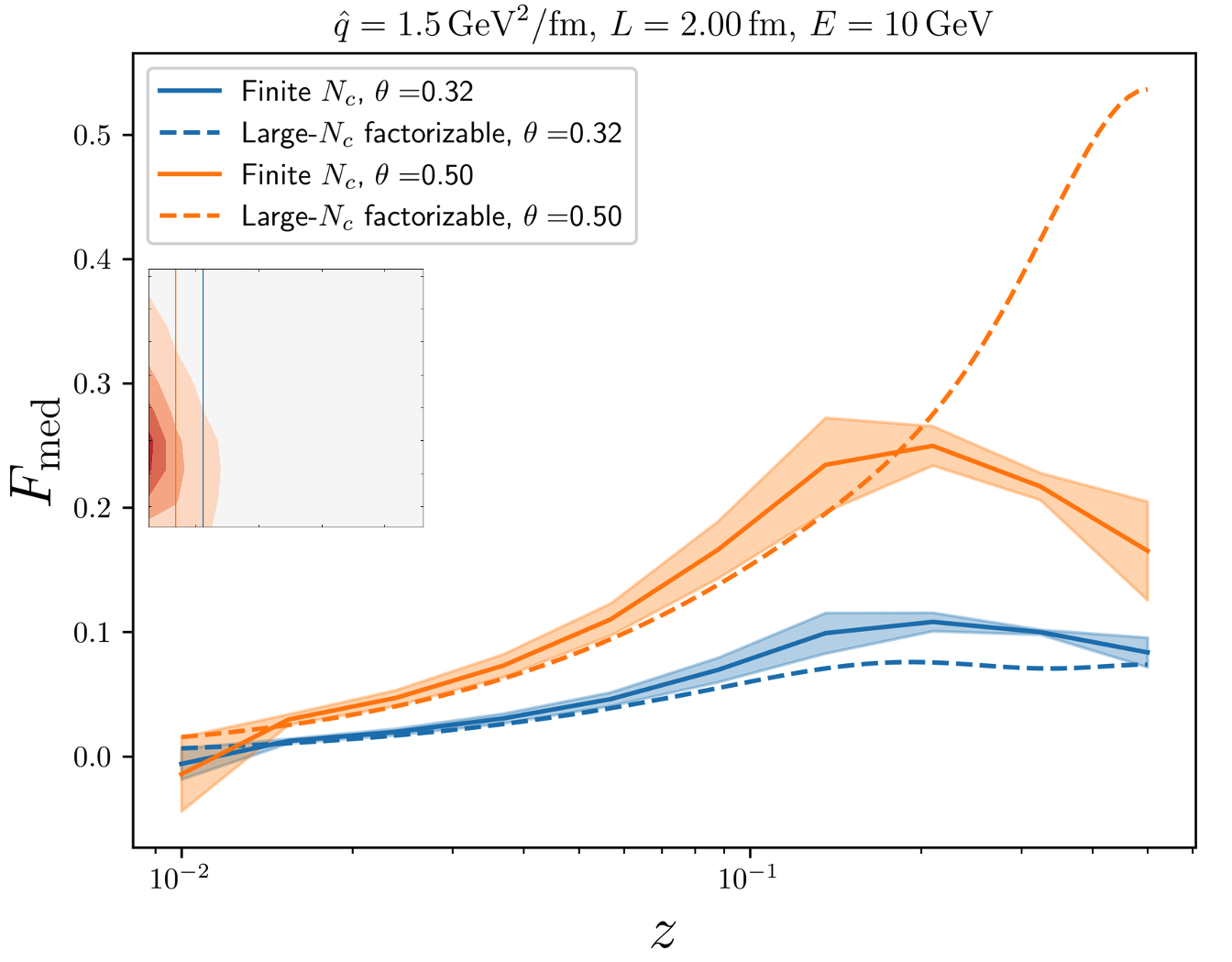}
    \includegraphics[width=0.49\textwidth]{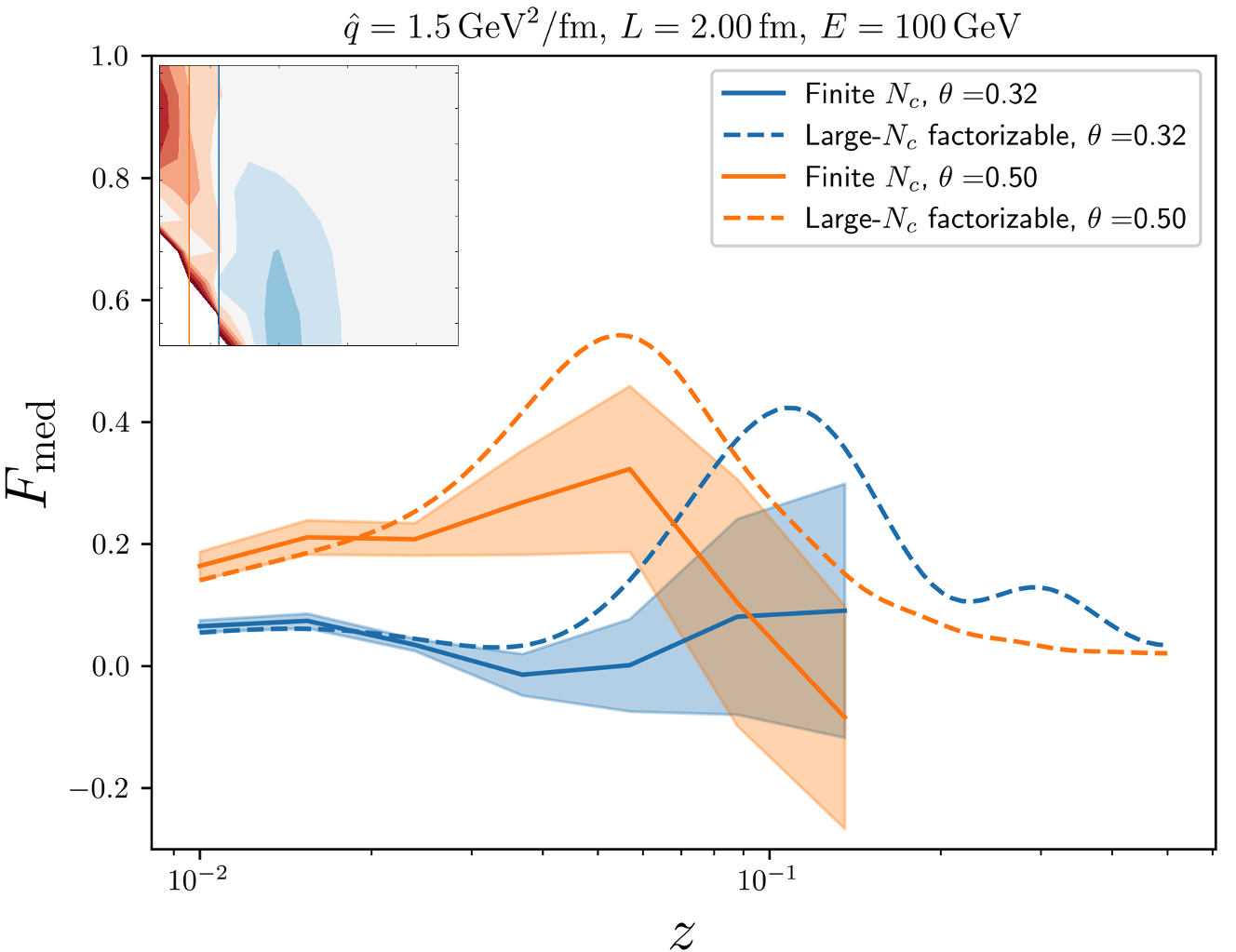}
    \caption{$F_{\rm med}$ calculated at finite $N_c$ and the factorizable part of the large-$N_c$ solution. On the Lund plane we have indicated constant $\theta$ slices that we plot as a function of $z$.}
    \label{fig:Lund-10}
\end{figure}
In Fig. \ref{fig:Lund-10} we have plotted $\Fmed$ simulated at finite $N_c$. We have plotted a slice of this at constant $\theta$, as a function of $z$. We have plotted both the simulated finite-$N_c$ result and the analytical formula for the factorizable piece of the large-$N_c$ solution. It is interesting to note that the two different solutions are within error bands of each other at low-$z$, i.e. $z<0.1$. 

However, as $z$ approaches 0.5 the two solutions diverge. The factorizable part overestimates the finite-$N_c$ value by a significant margin. In addition, the two solutions peak at different $z$-values. This again shows that the factorizable part is not as accurate at finite $z > 0.1$.

\begin{figure}[t!]
    \centering
    \includegraphics[width=0.49\textwidth]{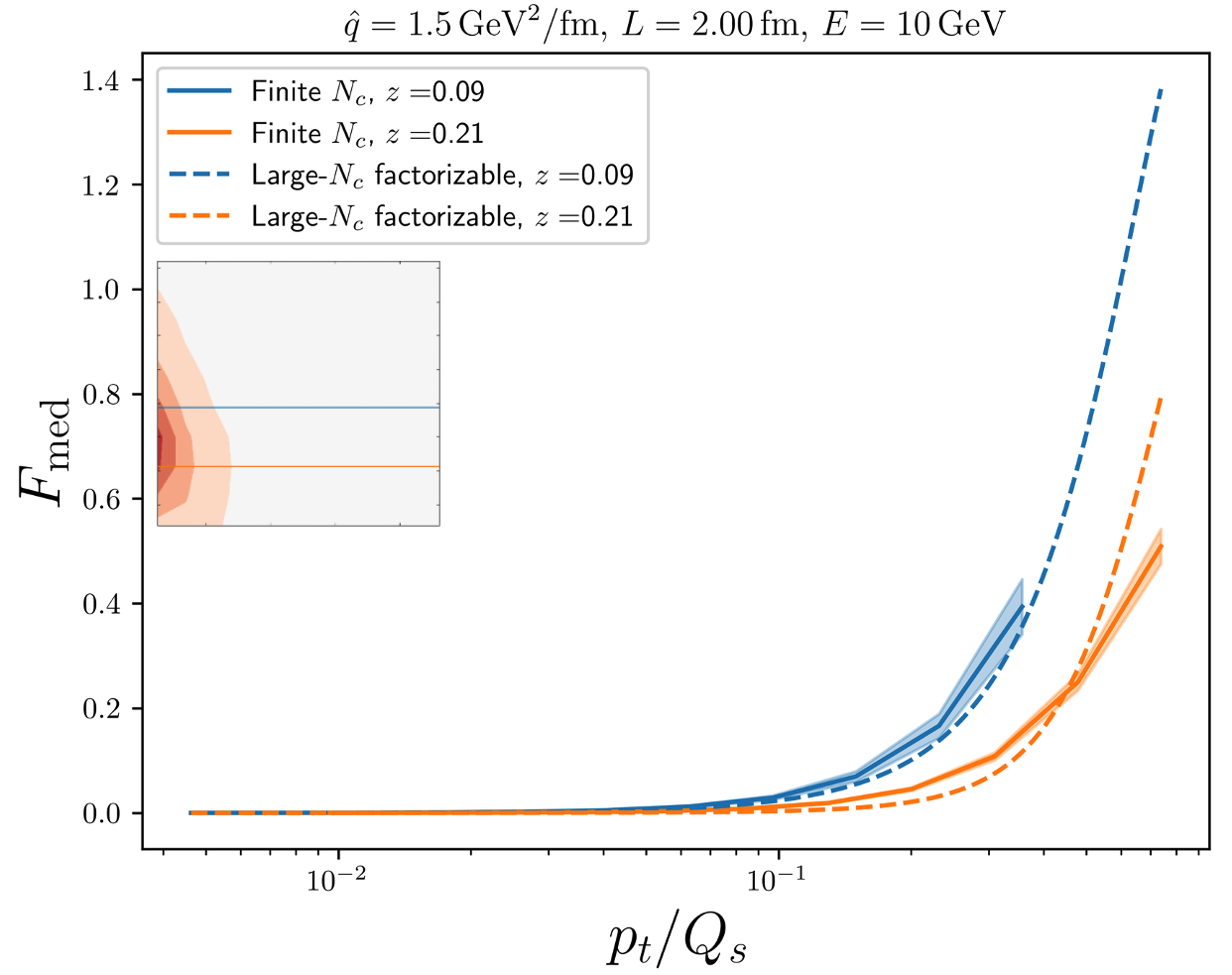}
    \includegraphics[width=0.49\textwidth]{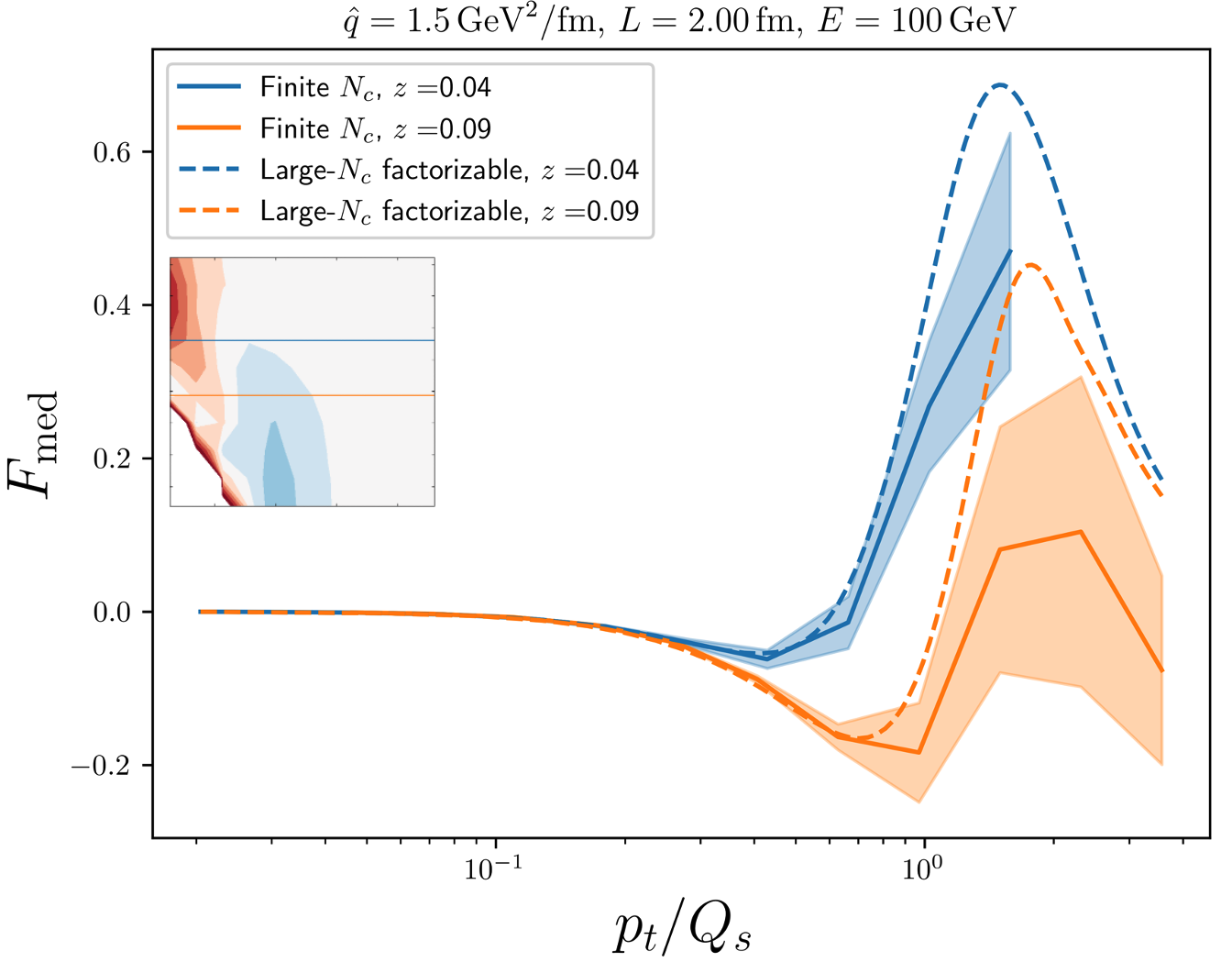}
    \caption{$F_{\rm med}$ calculated at finite $N_c$ and the factorizable part of the large-$N_c$ solution. On the Lund plane we have indicated constant $z$ slices that we plot as a function of $p_t/Q_s$. }
    \label{fig:Lund-100}
\end{figure}
Next, in Fig.~\ref{fig:Lund-100}, we plot the medium modification $F_{\rm med}$ for two slices of constant $z$, as indicated on the Lund plane in the inset. We plot this as a function of the dimensionless ratio $p_t/Q_s$, where $p_t \equiv |\prel| = \omega \theta$. In the soft limit, $z \ll 1$, one expects that the distribution should peak around $p_t/Q_s \approx 1$ \cite{Barata:2020rdn}. This is indeed the case for both the factorizable part (now at finite-$z$) and for the full, finite-$N_c$ simulation, although the peak is less distinguishable due to the numerical errors. However, the factorizable solution again overshoots the true value, especially at large $p_t$ or angles.

\begin{figure}[t!]
    \centering
    \includegraphics[width=0.47\textwidth]{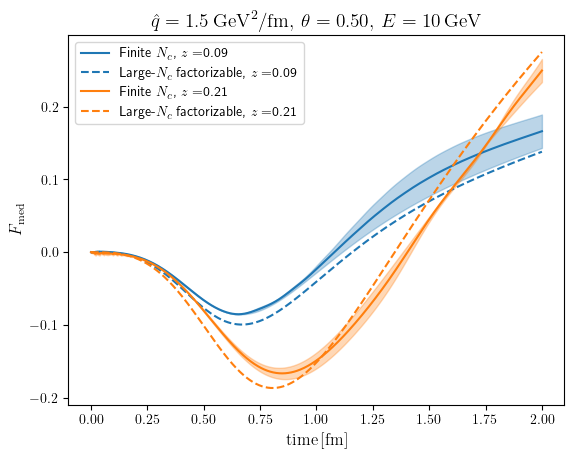}
    \includegraphics[width=0.47\textwidth]{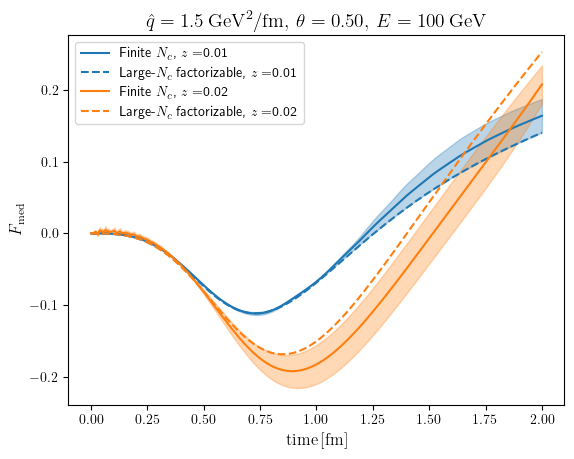}
    \caption{$F_{\rm med}$ as a function of time for two different energies.}
    \label{fig:Fmed-time}
\end{figure}

As our last numerical results, we show the time evolution of $\Fmed$ in Fig.~\ref{fig:Fmed-time}. Here we have chosen $z$ and $\theta$ to be close to the respective peaks in the distributions at $L=2$ fm. From these plots it is clear that the peak of the distribution moves with time. As we sit in a constant point in the $(\theta,z)$ plane the distribution is first negative, and then becomes positive as time progresses. 

We note also that the large-$N_c$/factorizable result agrees with the full calculation at early times, to diverge from it at later times. This is related to the physics of multiple scattering. At early times in the medium, there were simply not enough scatterings to cause strong color rotations so as to probe the many possible intermediate states of the two-body final particle system. This, however, changes when time progresses and the full color dynamics become apparent. 

As is well known \cite{Barata:2021wuf}, hard medium-induced emissions occur mostly at early times in the medium. It can be explicitly checked that the mixing of different color states only occurs at higher orders ($N>1$) in the opacity expansion. In order to induce the color dynamics responsible for factorization-breaking effects one needs frequent scatterings to occur; some to induce the splitting and others in charge of de-correlating and broadening the final splitting products.

\section{Conclusion and outlook}

In this paper we have derived the spectrum for a medium-induced emission without invoking the large-$N_c$ or eikonal approximations. This was done in Sec. \ref{sec:splitting}. The spectrum can be divided into a vacuum component (called out-out) and a medium component, consisting of an in-out piece and in-in piece, which for a general potential are given in Eqs. \eqref{eq:vacuum-spectrum}, \eqref{eq:in-out-general} and \eqref{eq:in-in-general}, respectively. The in-in contribution contains the quadrupole correlation function $\Qc(L,t)$, which currently cannot be solved analytically without using the large-$N_c$ approximation. We then showed how the in-in contribution can be solved through a Schrödinger equation in Eq.~\eqref{eq:schr-eq-with-k3}. 

In Sec. \ref{sec:concrete calculation}, we focused our efforts by applying the general results we derived in the previous sections to a state-of-the-art calculation of a concrete splitting process. We chose to study the process $\gamma \to q \bar q$, in the harmonic oscillator approximation. We here found that the large-$N_c$ approximation works extremely well for the photon case, which also echoes the results we found in \cite{Isaksen:2020npj}. Our main result of this calculation was shown in Fig. \ref{fig:Fmed-diff-Nc-full}.

We also examined how well the full spectrum can be approximated using only the first of the two terms in the large-$N_c$ solution of the quadrupole, given in Eq. \eqref{eq:Q2-hom-and-nonhom}. This is important, as usually only the first term, referred to as the factorizable term, is kept in calculations of the emission spectrum, as the quadrupole then simplifies substantially. The factorizable and full solutions were shown in Fig. \ref{fig:Fmed-diff-diag-Nc} and also Figs. \ref{fig:Lund-10}-\ref{fig:Fmed-time}. As expected from the analytical formula \eqref{eq:Q2-hom-and-nonhom} we found that the factorizable term provides an excellent approximation of the full solution in the soft limit when $z\sim 0$. However, we do see a small deviation at finite $z$, at the order of 10 \%, and even higher at large angles. It could therefore be important to include both terms in studies that focus on corrections where $z$ is not small. 

Our results, also summarized in Fig.~\ref{fig:fmed-inin-time-qhat}, indicate that the non-factorizable terms are sizable and persistent even for large media. Furthermore, Fig.~\ref{fig:fmed-inin-theta} demonstrates that these contributions decrease when decreasing the branching time. This indicates that the color-dynamics taking place from moment of the branching to the end of the medium is non-trivial. These effects are expected to be bigger for partons with a higher color charge, such as $g\to gg$ branching, and deserves further studies.

We also compared our result to the fully eikonal result, where all the partons are put on straight lines. This is what was done in \cite{Isaksen:2020npj}, so keeping the full transverse dynamics is the main improvement over that paper. We found in Fig. \ref{fig:Fmed-eikonal} that the fully eikonal approximation is not generally very effective at capturing the spectrum, especially at low $\omega$. However, it should be noted that in many calculations the soft limit is used, where you keep the non-eikonal behavior of the soft line, while putting the rest on straight lines. We expect this to work better for small $z$, and is something that could be studied in future work.

Having established a working methodology, this work therefore paves the way for precise calculations of all QCD and QED/EW splitting functions in the medium. As a next important step, we will attempt to solve the necessary evolution equations for $g \to gg$ and $q \to qg$ which both contain soft divergences and should therefore occur frequently in the QGP. On the level of spectra differential only in the momentum-sharing fractions, this would also be relevant for studying overlapping formation times in $g \to ggg$ splitting beyond the soft and/or large-$N_c$ limits \cite{Arnold:2019qqc}. A further perspective worth considering is to simulate more partons as many-body and multi-level (in the sense of color) quantum-mechanical systems in a background field. This effort seems however daunting due to the exponentially growing size of the system of coupled evolution equations, describing the available color representations of the involved partons. 

Going beyond the HO approximation is also, in principle, straightforward. In order to achieve a full resummation one should use the complete elastic medium scattering potential, and use the most general interaction matrix $\mathbb{M}$ as given in Eq.~\eqref{eq:evolution-matrix}. In this case, one also needs to evaluate the three-point function $\Kc(\u,\u_2|t_2,t_1)$ numerically with a completely analogous numerical Schrödinger equation. Hence, the non-homogeneous contribution to $\Fc$, as in Eq.~\eqref{eq:schr-eq-with-k3}, would be provided as numerical data.
In principle, one could also use scattering potentials extracted from lattice simulations of high-temperature QCD \cite{Schlichting:2021idr}.

\acknowledgments
We would like to thank Fabio Dominguez for helpful discussions and private communication on this subject.
We would also like to thank Adam Takacs and Alexandre Falc\~ao for useful discussions and feedback. Finally, we acknowledge financial support from the Trond Mohn Foundation (BFS2018REK01).

\appendix

\section{Simplifying n-point functions}\label{appendix:n-point}
Here we will show how the n-point functions present in Eq. \eqref{eq:emission-spectrum-not-simplified} can be simplified. It is adapted from the procedure shown in \cite{Blaizot:2012fh}. 

\subsection{Four-point function}\label{appendix:4-point}
The four-point function up to some final time $t_f$ is given by
\begin{align}
    (\k,\q;\k,\q|S^{(4)}(t_f,t_2)|\k_2,\q_2;\bar\k_2,\bar\p_2-\bar\k_2)=d_{bc} 
    \left\langle(\k|\cG_b|\k_2)(\q|\cG_c|\q_2)(\bar\p_2-\bar\k_2|\cG_c^\dagger|\q)(\bar\k_2|\cG_b^\dagger|\k)\right\rangle\,,
\end{align}
where the color factor $d_{bc}$ is process dependent, and we have neglected all color indices. The propagators are usually formulated in position space, see Eq.~\eqref{eq:prop-G}, so after a Fourier transform it becomes
\begin{align}\label{eq:4-point-Fourier}
    &(\k,\q;\k,\q|S^{(4)}(t_f,t_2)|\k_2,\q_2;\bar\k_2,\bar\p_2-\bar\k_2) = \nn
    &\int_{\x_2\y_2\bar \y_2 \bar \x_2\x\y\bar \y\bar \x} 
    \rme^{i \k_2 \cdot \x_2-i \bar \k_2 \cdot \bar \x_2+i \q_2 \cdot \y_2-i (\bar\p_2-\bar\k_2) \cdot \bar \y_2 - i \k \cdot (\x-\bar \x)-i \q\cdot (\y-\bar \y)}\nn
    &\times(\x,\y;\bar\x,\bar\y|S^{(4)}(t_f,t_2)|\x_2,\y_2;\bar\x_2,\bar\y_2)\,.
\end{align}
The four-point function in position space is then given by
\begin{align}
    &(\x,\y;\bar\x,\bar\y|S^{(4)}(t_f,t_2)|\x_2,\y_2;\bar\x_2,\bar\y_2) = d_{bc} 
    \left\langle(\x|\cG_b|\x_2)(\y|\cG_c|\y_2)(\bar \y_2|\cG_c^\dagger|\bar\y)(\bar\x_2|\cG_b^\dagger|\bar\x)\right\rangle \nn
    &= \int^{\x}_{\x_2} \cD\r_1 \int^{\y}_{\y_2} \cD\r_2 \int^{\bar \y}_{\bar \y_2} \cD\bar \r_{2} \int^{\bar \x}_{\bar \x_2} \cD\bar\r_{1}\,
    \rme^{i\frac{E}{2}\int_{t_2}^{t_f} \dd s\, \left[z(\dot\r^2_1-\dot {\bar \r}^2_{1})+(1-z)(\dot \r^2_2-\dot {\bar \r}^2_2)\right]}\Cc^{(4)}(\r_1,\r_2,\bar \r_2, \bar \r_1)\,,
\end{align}
where the potential term $\Cc^{(4)}$ is a correlator of Wilson lines
\begin{equation}
    \Cc^{(4)}(\r_1,\r_2,\bar \r_2, \bar \r_1) = d_{bc} \left\langle V_b(\r_1) V_c(\r_2) V_c^\dagger(\bar\r_2) V_b^\dagger(\bar \r_1) \right\rangle\,.
\end{equation}
The Wilson lines $V(\r)$ can be in the fundamental or adjoint representations depending on the process. 

In \cite{Isaksen:2020npj} it was shown that all Wilson line correlators can be written as a system of differential equations
\begin{equation}
    \frac{\rmd}{\rmd t} \Cc^{(4)}_i = \M_{ij} \Cc^{(4)}_j\,.
\end{equation}
Here $\Cc_j$ indicates some other color configuration of the same Wilson lines. It was also shown in \cite{Isaksen:2020npj} that the evolution matrix only depends on the differences of the coordinates
\begin{equation}
    \M_{ij} = \M_{ij}(\sigma_{12},\sigma_{\bar1\bar2},\sigma_{1\bar1},\sigma_{2\bar2},\sigma_{1\bar2},\sigma_{\bar1 2})\,.
\end{equation}
Here we have used the notation $\sigma_{12} \equiv \sigma(\r_1 - \r_2)$.  
This implies that $\Cc^{(4)}$ also only depends on the differences of the coordinates. It is therefore natural to change to the following coordinates, with unit Jacobian
\begin{align}\label{eq:coordinate-change-S4}
  \u &= \r_1-\r_2 \nn
  \ub &= \bar \r_1-\bar \r_2\nn
  \v &= z(\r_1-\bar\r_1) +(1-z)(\r_2-\bar \r_2)\nn
  \w &= \frac12[z(\r_1+\bar\r_1) +(1-z)(\r_2+\bar \r_2)]\,.
\end{align}
Now the correlator $\Cc^{(4)}$ only depends on the coordinates $\u,\ub$ and $\v$, and there is no dependence on $\w$. The four-point function then becomes
\begin{align} 
    &(\x,\y;\bar\x,\bar\y|S^{(4)}(t_f,t_2)|\x_2,\y_2;\bar\x_2,\bar\y_2) = \nn
    &\int \cD\w \int \cD\v \int \cD\bm u \int \cD\ub\,
    \rme^{i\frac{E}{2} \int_{t_2}^{t_f} \dd s\, \left[2\dot\v \cdot \dot\w +z(1-z)(\dot \u^2-\dot {\bar \u}^2)\right]}\Cc^{(4)}(\u,\ub,\v) \,.
\end{align}
Since the potential does not depend on $\w$ the path integral over $\w$ can be done, which has the effect of forcing $\v$ to be on the classical path $\v \to \v_{\rm cl}$. The result is
\begin{align}
    &(\x,\y;\bar\x,\bar\y|S^{(4)}(t_f,t_2)|\x_2,\y_2;\bar\x_2,\bar\y_2) \nn
    &= \left(\frac{E}{2 \pi (t_f-t_2)}\right)^2
    \rme^{i \frac{E}{(t_f-t_2)}\Delta \v\cdot \Delta \w}
    \int \cD\u \int \cD{\bar \u}
    \,\rme^{i\frac{\omega}{2}\int_{t_2}^{t_f} \dd s\, (\dot \u^2-\dot {\bar \u}^2)} \Cc^{(4)}(\u,{\bar \u,\v_{\rm cl}})\,,
\end{align}
where $\Delta \w = \w(t_f)-\w(t_2)$ and we have defined $\omega=z(1-z)E$. Now we can return to momentum space. After performing the same coordinate change on the Fourier components in Eq.~\eqref{eq:4-point-Fourier} the integrals involving the $\w$ components are
\begin{align}
    &\int_{\w_L \w_2} \rme^{i \w_2\cdot(\q_2+\k_2-\bar\p_2)}\rme^{i \frac{E}{(t_f-t_2)}\Delta \v\cdot(\w_L-\w_2)}\nn 
    &= \left(\frac{2\pi(t_f-t_2)}{E}\right)^2(2\pi)^2\delta^2(\Delta \v)(2\pi)^2\delta^2(\q_2+\k_2-\bar\p_2)\,.
\end{align}
Inserting this into the expression for the four-point function in momentum space Eq.~\eqref{eq:4-point-Fourier} and doing one $\v$ integral leads to Eq. \eqref{eq:s4-simplified}
\begin{align}
\label{eq:S4-structure}
    &(\k,\q;\k,\q|S^{(4)}(t_f,t_2)|\k_2,\q_2;\bar\k_2,\bar\p_2-\bar\k_2)=(2\pi)^2\delta^2(\q_2+\k_2-\bar\p_2)\nn \, 
    &\times \int_{\u_2 \u_f \ub_2 \ub_f \v}
    \rme^{i \v \cdot(\bar\p_2-\q-\k)+i \u_2\cdot(\k_2-z \bar \p_2)
    -i \ub_2\cdot(\bar\k_2-z \bar\p_2)-i (\u_f-\bar \u_f)\cdot((1-z)\k-z\q)}\nn
    &\times \int_{\u_2}^{\u_f} \cD\u \int_{\ub_2}^{\ub_f} \cD{\bar \u}
    \,\rme^{i\frac{\omega}{2}\int_{t_2}^{t_f} \dd s\, (\dot \u^2-\dot {\bar \u}^2)} \Cc^{(4)}(\u,{\bar \u},\v)\nn
    &\equiv (2\pi)^2\delta^2(\q_2+\k_2-\bar\p_2) 
    \Sc^{(4)}((1-z)\k-z\q,\k_2-z\bar\p_2,\bar\k_2-z\bar\p_2,\bar\p_2-\k-\q|t_f,t_2) \,.
\end{align}
Here it is clear that it is more convenient to use the momentum variables $\l_2 = \k_2-z\bar\p_2$, $\bar\l_2 = \bar\k_2-z\bar\p_2$, $\p = (1-z)\k-z\q$ and $\P=\q+\k$, where the four-point function is $\Sc^{(4)}(\p,\l_2,\bar \l_2,\bar\p_2-\P|t_f,t_2)$.

\subsection{Three-point function}
The three-point function is given by
\begin{align}
    (\k_2,\q_2;\bar \p_2 |S^{(3)}(t_2,t_1)|\k_1, \p_1-\k_1;\bar \p_1)=d_{abc} 
    \left\langle(\k_2|\Gc_b|\k_1)(\q_2|\Gc_c|\p_1-\k_1)(\bar\p_1|\Gc_a^{\dagger}|\bar \p_2)\right\rangle\,,
\end{align}
where the color factor $d_{abc}$ is process dependent, and we have neglected all color indices. The propagators are usually formulated in position space, see Eq.~\eqref{eq:prop-G}, so after a Fourier transform it becomes
\begin{align}\label{eq:3-point-Fourier}
    &(\k_2,\q_2;\bar \p_2 |S^{(3)}(t_2,t_1)|\k_1, \p_1-\k_1;\bar \p_1) = \nn
    &\int_{\x_1\x_2\y_1\y_2\bar \z_1\bar \z_2} 
    \rme^{i \k_1 \cdot \x_1-i \k_2 \cdot \x_2+i (\p_1-\k_1) \cdot \y_1-i \q_2 \cdot \y_2-i \bar \p_1 \cdot \bar\z_1+i \bar\p_2 \cdot \bar\z_2}\nn
    &\times(\x_2,\y_2;\bar \z_2 |S^{(3)}(t_2,t_1)|\x_1, \y_1;\bar \z_1)\,.
\end{align}
The three-point function in position space is then given by
\begin{align}
    &(\x_2,\y_2;\bar \z_2 |S^{(3)}(t_2,t_1)|\x_1, \y_1;\bar \z_1) = d_{abc} 
    \left\langle(\x_2|\Gc_b|\x_1)(\y_2|\Gc_c|\y_1)(\bar\z_1|\Gc_a^{\dagger}|\bar \z_2)\right\rangle \nn
    &= \int^{\x_2}_{\x_1} \cD\r_1 \int^{\y_2}_{\y_1} \cD\r_2 \int^{\bar \z_2}_{\bar \z_1} \cD\bar \r_{0}\,
    \rme^{i\frac{E}{2}\int_{t_1}^{t_2} \dd s\, \left[z\dot\r^2_1+(1-z)\dot \r^2_2-\dot {\bar \r}^2_0\right]}\Cc^{(3)}(\r_1,\r_2,\bar \r_0)\,,
\end{align}
where the correlator $\Cc^{(3)}$ is given by
\begin{equation}
    \Cc^{(3)}(\r_1,\r_2,\bar \r_0) = d_{abc} \left\langle V_b(\r_1) V_c(\r_2) V_a^\dagger(\bar\r_0)\right\rangle\,.
\end{equation}
The Wilson lines $V(\r)$ can be in the trivial, fundamental, or adjoint representations depending on the process. 

In \cite{Isaksen:2020npj} it was shown that all Wilson line correlators can be written as a system of differential equations
\begin{equation}
    \frac{\rmd}{\rmd t} \Cc^{(3)}_i(\r_1,\r_2,\bar \r_0) = \M_{ij}(\r_1-\bar\r_0,\r_2-\bar\r_0,\r_2-\r_1) \Cc^{(3)}_j(\r_1,\r_2,\bar \r_0)\,.
\end{equation}
Here $\Cc^{(3)}_j$ indicates some other color configuration of the same Wilson lines. The fact that the evolution matrix only depends on the differences of the coordinates implies that the same is true for the correlator: $\Cc(\r_1,\r_2,\bar \r_0) = \Cc(\r_1-\bar\r_0,\r_2-\bar\r_0,\r_2-\r_1)$.
This leads naturally to a variable change, with unit Jacobian
\begin{align}
  \u &= \r_1-\r_2 \nn
  \v &= z\r_1 +(1-z)\r_2-\bar \r_0 \nn
\end{align}
Now the correlator $\Cc^{(3)}$ only depends on the coordinates $\u$ and $\v$, and there is no dependence on $\bar \r_0$. The three-point function then becomes
\begin{align} 
    &(\x_2,\y_2;\bar \z_2 |S^{(3)}(t_2,t_1)|\x_1, \y_1;\bar \z_1) = \nn
    &\int \cD\u \int \cD\v \int \cD\bar \r_0\,
    \rme^{i\frac{E}{2} \int_{t_1}^{t_2} \dd s\, \left[z(1-z)\dot\u^2 +\dot\v^2+2 \dot\v \cdot \dot {\bar \r}_0\right]}\Cc^{(3)}(\u,\v) \,.
\end{align}
Since the potential does not depend on $\bar \r_0$ the path integral over $\bar \r_0$ can be done, which has the effect of forcing $\v$ to be on the classical path $\v \to \v_{\rm cl}$. The result is
\begin{align}
    &(\x_2,\y_2;\bar \z_2 |S^{(3)}(t_2,t_1)|\x_1, \y_1;\bar \z_1) \nn
    &= \left(\frac{E}{2 \pi (t_2-t_1)}\right)^2
    \rme^{i \frac{E}{(t_2-t_1)}\Delta \v\cdot (\Delta \v+2 (\bar \z_2-\bar \z_1))}
    \int \cD\u 
    \,\rme^{i\frac{\omega}{2}\int_{t_1}^{t_2} \dd s\, \dot \u^2} \Cc^{(3)}(\u,\v_{\rm cl})\,,
\end{align}
where $\Delta \v = \v_2-\v_1$ and we have defined $\omega=z(1-z)E$. Now we can return to momentum space. After performing the same coordinate change on the Fourier components in Eq.~\eqref{eq:4-point-Fourier} the integrals involving the $\z$ components are
\begin{align}
    &\int_{\bar \z_1 \bar \z_2} \rme^{i \bar \z_1\cdot(\p_1-\bar \p_1)}\rme^{-i \bar \z_2\cdot(\k_2+\q_2-\bar \p_2)}\rme^{i \frac{E}{(t_2-t_1)}\Delta \v\cdot(\bar\z_2-\bar \z_1)}\nn 
    &= \left(\frac{2\pi(t_2-t_1)}{E}\right)^2(2\pi)^2\delta^2(\Delta \v)(2\pi)^2\delta^2(\p_1-\bar\p_1)\,.
\end{align}
Here we have used that $\bar \p_2=\k_2+\q_2$, see Eq. \eqref{eq:s4-simplified}.
Inserting this into the expression for the three-point function in momentum space Eq.~\eqref{eq:3-point-Fourier} and doing one $\v$ integral leads to Eq. \eqref{eq:k3-simplified}
\begin{align}\label{eq:3-point-simplified}
    &(\k_2,\q_2;\bar \p_2 |S^{(3)}(t_2,t_1)|\k_1, \p_1-\k_1;\bar \p_1)=(2\pi)^2\delta^2(\p_1-\bar\p_1)\nn \, 
    &\times \int_{\u_1 \u_2\v}\rme^{i \v \cdot(\p_1-\bar \p_2)+i \u_1\cdot(\k_1-z\p_1)
    -i \u_2\cdot(\k_2- z\bar\p_2)}\int \cD\u 
    \,\rme^{i\frac{\omega}{2}\int_{t_1}^{t_2} \dd s\, \dot \u^2} \Cc^{(3)}(\u,\v)\nn
    &\equiv(2\pi)^2\delta(\p_1-\bar\p_1)
    \Sc^{(3)}(\k_2-z\bar \p_2,\k_1-z\p_1,\p_1-\bar \p_2|t_2,t_1)\,.
\end{align}
After defining the momentum variables $\l_2 = \k_2-z\bar\p_2$ and $\l_1 = \k_1-z\p_1$ this becomes $\Sc^{(3)}(\l_2,\l_1,\p_1-\bar \p_2|t_2,t_1)$. 

\subsection{Two-point function}
The two-point function is
\begin{align}
    (\p_1;\p_1 |S^{(2)}(t_1,t_0)|\p_0;\bar \p_0) =
    d_a^{(2)} \left\langle(\p_1|\Gc_a(t_1,t_0)|\p_0)(\bar \p_0|\Gc_a^\dagger(t_0,t_1)|\bar\p_1)\right\rangle\,,
\end{align}
where we have used that $\bar \p_1 =\p_1$ from \eqref{eq:3-point-simplified}.
After Fourier transforming this becomes
\begin{equation}
    ( \p_1;\p_1 | S^2(t_1-t_0) | \p_0 ; \bar \p_0) = \int_{\x_1 \x_0 \bar \x_1 \bar \x_0} 
    \rme^{-i \p_1 \cdot (\x_1-\bar \x_1) +i \p_0 \cdot \x_0 -i \bar \p_0 \cdot \bar \x_0}
    ( \x_1;\bar\x_1 | S^2(t_1-t_0) | \x_0 ; \bar \x_0)\,.
\end{equation}
In position space it can be written in terms of path integrals
\begin{align}
    ( \x_1;\bar\x_1 | S^2(t_1-t_0) | \x_0 ; \bar \x_0) &= 
    d_a^{(2)}  \langle (\x_1 | \Gc_a(t_1,t_0) | \x_0) (\bar\x_0 | \Gc_b^\dagger(t_1,t_0) | \bar \x_1)\rangle\nn 
    &= \int_{\x_0}^{\x_1} \cD \r \int_{\bar\x_0}^{\bar\x_1} \cD \bar \r \,
    \rme^{i \frac{E}{2}\int_{t_0}^{t_1} \rmd s (\dot \r^2-\dot{\bar \r}^2)} \Cc^{(2)}(\r-\bar\r|t_1,t_0)\,,
\end{align}
where $\Cc^{(2)}$ is a Wilson line correlator that has a simple solution
\begin{align}
    \Cc^{(2)}(\r-\bar \r|t_1,t_0) &= d_a^{(2)} \left\langle V_a V_{\bar a}^\dagger\right\rangle = \rme^{-C_R \int_{t_0}^{t_1} \rmd s \, n(s) \sigma(\r-\bar \r)}\,.
\end{align}
Changing coordinates to $\u = \r-\bar \r$ and $\v = 1/2(\r+\bar\r)$ the potential does not depend on $\v$, and the path integral over $\v$ can be performed. This has the effect of forcing $\u$ to be on the classical path, and the full two-point function becomes
\begin{align}
    ( \p_1;\p_1 | S^2(t_1-t_0) | \p_0 ; \bar \p_0) &= \left(\frac{E}{2\pi\Delta t}\right)^2
    \int_{\u_1\u_0\v_1\v_0} \rme^{i \v_0\cdot(\p_0-\bar\p_0)-i\u_1\cdot\p_1+\frac12 i \u_0\cdot(\p_0+\bar\p_0)}\nn
    &\times\rme^{i \frac{E}{\Delta t}(\u_1-\u_0)\cdot(\v_1-\v_0)-C_R\int_{t_0}^{t_1} \rmd s \,n(s) \sigma(\u_{\rm cl})}\,.
\end{align}
All but one of the integrals can be done immediately, leading to a momentum-conserving delta function 
\begin{equation}
    (\p_1; \p_1|S^{(2)}(t_1-t_0)|\p_0;\bar\p_0) = (2\pi)^2\delta(\p_0-\bar\p_0)\Pc(\p_1-\p_0|t_1,t_0)\,.
\end{equation}
Here we have defined the broadening distribution $\Pc$ as
\begin{equation}\label{eq:p-broadening}
    \Pc(\p_1-\p_0|t_1,t_0) = \int_\u \rme^{-i \u \cdot(\p_1-\p_0)-C_R\int_{t_0}^{t_1} \rmd s \,n(s) \sigma(\u) }\,.
\end{equation}
\subsection{Summary of color structure} \label{app:color-summary}
In the main text we have written the n-point functions in a very general way, namely
\begin{align}
    (\p_1;\bar \p
    _1 |S^{(2)}(t_1,t_0)|\p_0;\bar \p_0) &= 
    d_a^{(2)} \langle(\p_1|\Gc_a(t_1,t_0)|\p_0)(\bar \p_0|\Gc_a^\dagger(t_0,t_1)|\bar\p_1)\rangle\nn
    (\k_2,\q_2;\bar \p_2 |S^{(3)}(t_2,t_1)|\k_1, \p_1-\k_1;\bar \p_1)
    &=d_{abc}^{(3)} \langle(\k_2|\Gc_b|\k_1)(\q_2|\Gc_c|\p_1-\k_1)(\bar\p_1|\Gc_a^{\dagger}|\bar \p_2)\rangle\nn
    (\k,\q;\k,\q|S^{(4)}(t_\infty,t_2)|\k_2,\q_2;\bar\k_2,\bar\p_2-\bar\k_2)
    &=d_{bc}^{(4)} \langle(\k|\cG_b|\k_2)(\q|\cG_c|\q_2)(\bar\p_2-\bar\k_2|\cG_c^\dagger|\q)(\bar\k_2|\cG_b^\dagger|\k)\rangle\,.
\end{align}
Here we have neglected all the color indices. Here we present the n-point correlators for the two processes $\gamma \to q\bar q$ and $g\to g g$ with full color dependence. The process dependent color factors $d_a^{(2)}$, $d_{abc}^{(3)}$ and $d_{bc}^{(4)}$ then be inferred from this.

In the case of $\gamma \to q \bar q$ the two-, three-, and four-point functions are given as
\begin{align}
    S^{(2)}(t_1,t_0) &= 
     (\p_1|\Gc_0|\p_0)(\bar \p_0|\Gc_0^{\dagger}|\bar\p_1)\nn
    S^{(3)}(t_2,t_1)
    &=\frac{1}{N_c} \langle\tr[(\k_2|\Gc_F|\k_1)(\p_0-\k_1|\bar \Gc_F|\q_2)]\rangle (\bar\p_0|\Gc_0^{\dagger}(t_1,t_2)|\bar \p_2)]\rangle \nn
   S^{(4)}(t_\infty,t_2)
    &=\frac{1}{N_c}\langle\tr[(\k|\Gc_F|\k_2)(\q_2|\bar \Gc_F|\q)(\q|\Gc_F^\dagger|\bar\p_2-\bar \k_2)(\bar\k_2|\bar \Gc_F^\dagger|\k)]\rangle\,.
\end{align}
Finally, for the $g \to gg$ process we have 
\begin{align}
    S^{(2)}(t_1,t_0) &= 
     \frac{1}{N_c^2-1}\langle\tr[(\p_1|\Gc_A|\p_0)(\bar \p_0|\Gc_A|\bar\p_1) ]\rangle\nn
    S^{(3)}(t_2,t_1)
    &=\frac{1}{N_c(N_c^2-1)} f^{\bar a_2 b_2 c_2}f^{a_1 b_1 c_1} 
    \langle (\k_2|\Gc_A^{b_2b_1}|\k_1)(\q_2|\Gc_A^{c_2c_1}|\p_1-\k_1)(\bar\p_1|\Gc_A^{\dagger a_1\bar a_2}|\bar \p_2)\rangle \nn
     S^{(4)}(t_\infty,t_2)
    &=\frac{1}{N_c(N_c^2-1)} f^{\bar a_2 b_2 c_2}f^{\bar a_2 \bar b_2 \bar c_2} \nn
    &\times\langle(\k|\Gc_A^{b b_2}|\k_2)(\q|\Gc_A^{c c_2}|\q_2)(\bar \k_2|\Gc_A^{\dagger \bar b_2 b}|\k)(\bar \p_2-\bar\k_2|\Gc_A^{\dagger \bar c_2 c}|\q)\rangle\,.
\end{align}
Here we have neglected the momentum dependence on the left-hand side to make the color structure more clear. 

\section{Calculation of different processes}\label{app:calc-of-processes}
Here we will show how to derive the emission spectrum for two different physical processes.
\subsection{Pair production}\label{app:pair-prod}
The pair production matrix element is 
\begin{align}
    \Mc_{s_i,s_j}^{i j }(\k,\q) &=  \int_0^{\infty} \dd t_1 \int_{\p_0\p_1\k_1\q_1}\, 
    \rme^{i \frac{\k^2}{2zE}t_\infty}(\k|\Gc_F^{i i_1}(t_\infty,t_1)|\k_1)
    \rme^{i \frac{\q^2}{2(1-z)E}t_\infty}(\q_1|\bar \Gc_F^{j_1 j}(t_1,t_\infty)|\q)\nn
    &\times (\q_1,\k_1|V_{\lambda s_i s_j}^{i_1 j_1}|\p_1)
    \frac{1}{2 E} (\p_1|\Gc_0(t_1,t_0)|\p_0) \Mc_{0\lambda}(\p_0)\,.
\end{align}
Here the photon propagator is simply the free propagator, given in \eqref{eq:G0-mom}.
The vertex is given by
\begin{equation}
    (\q_1,\k_1|V_{\lambda s_i s_j}^{i_1 j_1}|\p_1) = \delta^{i_1 j_1} (2\pi)^2\delta(\p_1-\k_1-\q_1) e \,\Gamma_{\lambda \s_i s_j}(\k_1-z\p_1)\,,
\end{equation}
where
\begin{equation}
    \Gamma_{\lambda s_i s_j}(\Q) = \delta_{-s_j s_i} (z \delta_{\lambda s_i}-(1-z)\delta_{\lambda s_j})\frac{2 i}{\sqrt{z(1-z)}} \Q\cdot \epsilon_\lambda\,.
\end{equation}
The cross section is, after summing over flavor
\begin{align}
    \frac{\rmd^2 \sigma}{\rmd \Omega_\k \rmd \Omega_\q} = n_f \sum \langle |\Mc(\k,\q)|^2 \rangle\,.
\end{align}
After squaring the amplitude the cross section is 
\begin{align}
    &\frac{\rmd^2 \sigma}{\rmd \Omega_\k \rmd \Omega_\q} = \frac{e^2 n_f}{(2E)^2} \rmR \int_0^\infty \rmd t_2 \int_0^{t_2} \rmd t_1
    \int_{\p_0\p_1\k_1\q_2\k_2\bar\p_0\bar\p_2\bar\k_2\bar\p_1} 
    \langle\Mc_{0\lambda}(\p_0)\Mc_{0\bar \lambda}^*(\bar \p_0)\rangle\nn
    &\times\Gamma_{\lambda s_i s_j}(\k_1-z\p_1)\Gamma_{\bar \lambda s_i s_j}(\bar \k_2-z\bar \p_2)
    \langle(\p_1|\Gc_0(t_1,t_0)|\p_0)(\bar \p_0|\Gc_0^{\dagger}(t_0,t_1)|\bar\p_1)\rangle\nn
    &\times \langle(\k_2|\Gc_F^{i_2i_1}(t_2,t_1)|\k_1)(\p_1-\k_1|\bar \Gc_F^{i_1j_2}(t_1,t_2)|\q_2)(\bar\p_1|\Gc_0^{\dagger}(t_1,t_2)|\bar \p_2)\rangle\nn
    &\times \langle(\k|\Gc_F^{i i_2}(t_\infty,t_2)|\k_2)(\q_2|\bar \Gc_F^{j_2 j}(t_\infty,t_2)|\q)(\q|\Gc_F^{\dagger j \bar i_2}(t_2,t_\infty)|\bar \p_2-\bar \k_2)(\bar\k_2|\bar \Gc_F^{\dagger \bar i_2 i}(t_2,t_\infty)|\k)\rangle\,.
\end{align}
This equation contains a lot of information and should be understood in the following way: The two-point function between $t_0$ and $t_1$ describes the propagation of the initial photon. Then the splitting happens in the amplitude at $t_1$, and in the complex conjugate amplitude at $t_2$, which is described by the three-point function. Finally, the quark-antiquark system broadens until the end of the medium and propagates until $t_\infty$.

The two-point function is simply
\begin{equation}
    \langle(\p_1|\Gc_0(t_1,t_0)|\p_0)(\bar \p_0|\Gc_0^{\dagger}(t_0,t_1)|\bar\p_1)\rangle = (2\pi)^4\delta(\p_1-\p_0)\delta(\bar \p_1-\bar\p_0) 
    \rme^{-i \frac{\p_0^2-\bar \p_0^2}{2 E}(t_1-t_0)}\,.
\end{equation}
The three-point function is
\begin{align}
    &\langle(\k_2|\Gc_F^{i_2i_1}(t_2,t_1)|\k_1)(\p_0-\k_1|\bar \Gc_F^{i_1j_2}(t_1,t_2)|\q_2)(\bar\p_0|\Gc_0^{\dagger}(t_1,t_2)|\bar \p_2)\rangle\nn
    &= \frac{\delta^{i_2j_2}}{N_c} \langle\tr[(\k_2|\Gc_F(t_2,t_1)|\k_1)(\p_0-\k_1|\bar \Gc_F(t_1,t_2)|\q_2)]
    (\bar\p_0|\Gc_0^{\dagger}(t_1,t_2)|\bar \p_2)\rangle\nn
    &=\delta^{i_2j_2}(\k_2,\q_2;\bar \p_2|S^{(3)}(t_2,t_1)|\k_1,\p_0-\k_1;\bar \p_0)\nn
    &=\delta^{i_2j_2}(2\pi)^2\delta (\bar \p_0-\p_0)
    \Sc^{(3)}(\k_2-z\bar \p_2,\k_1-z\p_0,\p_0-\bar \p_2|t_2,t_1)\,.
\end{align}
Finally, the four-point function becomes
\begin{align}
    &\langle\tr[(\k|\Gc_F(t_\infty,t_2)|\k_2)(\q_2|\bar \Gc_F(t_\infty,t_2)|\q)(\q|\Gc_F^\dagger(t_2,t_\infty)|\bar\p_2-\bar \k_2)(\bar\k_2|\bar \Gc_F^\dagger(t_2,t_\infty)|\k)]\rangle\nn
    &=N_c(\k,\q;\k,\q | S^{(4)}(t_\infty,t_2)|\k_2,\q_2;\bar\k_2,\bar\p_2-\bar\k_2)\nn
    &=N_c(2\pi)^2\delta(\bar\p_2-\k_2-\q_2)
    \Sc^{(4)}((1-z)\k-z\q,\k_2-z\bar\p_2,\bar\k_2-z\bar\p_2,\bar\p_2-\k-\q|t_\infty,t_2)\,.
\end{align}
Plugging this into the cross section we get
\begin{align}
    \frac{\rmd^2 \sigma}{\rmd \Omega_\k \rmd \Omega_\q} &= \frac{n_f e^2}{(2E)^2} N_c \rmR \int_0^\infty \rmd t_2 \int_0^{t_2} \rmd t_1
    \int_{\p_0\k_1\k_2\bar\k_2\bar\p_2}  
    \langle\Mc_{0\lambda}(\p_0)\Mc_{0\bar \lambda}^*(\p_0)\rangle\nn
    &\times\Gamma_{\lambda s_i s_j}(\k_1-z\p_0)\Gamma_{\bar \lambda s_i s_j}(\bar \k_2-z\bar\p_2)\nn
    &\times \Sc^{(3)}(\k_2-z\bar \p_2,\k_1-z\p_0,\p_0-\bar \p_2|t_2,t_1)\nn
    &\times \Sc^{(4)}((1-z)\k-z\q,\k_2-z\bar\p_2,\bar\k_2-z\bar\p_2,\bar\p_2-\k-\q|t_\infty,t_2)\,.
\end{align}
After using that the vertices combine as
\begin{equation}
    \Gamma_{\lambda s_i s_j}(\Q_1)\Gamma_{\bar \lambda s_i s_j}(\Q_2) = \delta^{\lambda \bar \lambda}(z^2+(1-z)^2) \frac{4}{z(1-z)}\Q_1\cdot \Q_2\,,
\end{equation}
we end up with
\begin{align}
    P_2(\k,\q;\p_0) &= \frac{e^2}{z(1-z)E^2} P_{q \gamma}(z) \rmR \int_0^\infty \rmd t_2 \int_0^{t_2} \rmd t_1
    \int_{\k_1\k_2\bar\k_2\bar\p_2}  (\k_1-z\p_0)\cdot(\bar \k_2-z\bar\p_2)\nn
   &\times \Sc^{(3)}(\k_2-z\bar \p_2,\k_1-z\p_0,\p_0-\bar \p_2|t_2,t_1)\nn
    &\times \Sc^{(4)}((1-z)\k-z\q,\k_2-z\bar\p_2,\bar\k_2-z\bar\p_2,\bar\p_2-\k-\q|t_\infty,t_2)\,,
\end{align}
where we have introduced the Altarelli-Parisi splitting function
\begin{equation}
    P_{q \gamma}(z)= n_f N_c[z^2+(1-z)^2]\,.
\end{equation}

\subsection{Gluon-gluon splitting}
This is calculated in detail in \cite{Blaizot:2012fh}, but we will repeat the most relevant parts here. The matrix element is
\begin{align}
    &\Mc_{\lambda_b,\lambda_c}^{b c}(\k,\q) =  \int_0^{\infty} \dd t_1 \int_{\p_0\p_1\k_1\q_1}\, 
    \epsilon_{\lambda_b}^{* j} \rme^{i \frac{\k^2}{2zE}t_\infty}(\k|\Gc^{b b_1}(t_\infty,t_1)|\k_1)
    \nn
    &\times \epsilon_{\lambda_c}^{* l} \rme^{i \frac{\q^2}{2(1-z)E}t_\infty}(\q|\Gc^{c c_1}(t_\infty,t_1)|\q_1)(\q_1,\k_1|V_{a_1 b_1 c_1}^{ijl}|\p_1)
    \frac{1}{2 E} (\p_1|\Gc^{a_1 a_0}(t_1,t_0)|\p_0) \Mc_0^{i a_0}(\p_0)\,,
\end{align}
where all the propagators are gluon propagators. The vertex is given by
\begin{equation}
    (\q_1,\k_1|V_{a_1 b_1 c_1}^{ijl}|\p_1) = (2\pi)^2\delta(\p_1-\k_1-\q_1)g f^{a_1 b_1 c_1} \Gamma^{ijk}(\k_1-z\p_1)\,,
\end{equation}
and
\begin{equation}
    \Gamma^{ijk}(\bm Q) = 2\left(\frac{1}{z} \bm Q^j \delta^{il}+\frac{1}{1-z}\bm Q^l \delta^{ij} -\bm Q^i \delta^{jl}\right)\,.
\end{equation}
The cross section is simply achieved by squaring the amplitude and averaging over the initial and summing over the final quantum numbers
\begin{align}
    \frac{\rmd^2 \sigma}{\rmd \Omega_\k \rmd \Omega_\q} = \sum \langle |\Mc(\k,\q)|^2 \rangle\,.
\end{align}
After squaring the amplitude and use polarization sums $\sum_\lambda \epsilon_\lambda^i \epsilon_\lambda ^{* j} = \delta^{ij}$ this is
\begin{align}
    &\frac{\rmd^2 \sigma}{\rmd \Omega_\k \rmd \Omega_\q} = \frac{g^2}{(2E)^2} \rmR \int_0^\infty \rmd t_2 \int_0^{t_2} \rmd t_1
    \int_{\p_0\p_1\k_1\q_2\k_2\bar\p_0\bar\p_2\bar\k_2\bar\p_1}  
    \Gamma^{ijk}(\k_1-z\p_1)\Gamma^{\bar ijk}(\bar \k_2-z\bar \p_2) \nn
    &\times f^{a_1 b_1 c_1} f^{\bar a_2 \bar b_2 \bar c_2} \Mc_0^{i a_0}(\p_0)\Mc_0^{* \bar i \bar a_0}(\bar \p_0)
    (\p_1|\Gc^{a_1a_0}(t_1,t_0)|\p_0)(\bar \p_0|\Gc^{\dagger \bar a_0 \bar a_1}(t_0,t_1)|\bar\p_1)\nn
    &\times (\k_2|\Gc^{b_2b_1}(t_2,t_1)|\k_1)(\q_2|\Gc^{c_2c_1}(t_2,t_1)|\p_1-\k_1)(\bar\p_1|\Gc^{\dagger \bar a_1\bar a_2}(t_1,t_2)|\bar \p_2)\nn
    &\times (\k|\Gc^{b b_2}(t_\infty,t_2)|\k_2)(\q|\Gc^{c c_2}(t_\infty,t_2)|\q_2)(\bar \k_2|\Gc^{\dagger \bar b_2 b}(t_2,t_\infty)|\k)(\bar \p_2-\bar\k_2|\Gc^{\dagger \bar c_2 c}(t_2,t_\infty)|\q)\,.
\end{align}
We have divided into three regions by using the property
\begin{equation}
    (\k|\Gc^{b b_1}(t_\infty,t_1)|\k_1) = \int_{\k_2} (\k|\Gc^{b b_2}(t_\infty,t_2)|\k_2)(\k_2|\Gc^{b_2b_1}(t_2,t_1)|\k_1)\,.
\end{equation}
The cross section again looks quite complicated, but it can be divided into three distinct physical processes.

The initial state can be simplified by using
\begin{equation}
    \Mc_0^{i a_0}(\p_0)\Mc_0^{* \bar i \bar a_0}(\bar \p_0) = \frac{\delta^{a_0 \bar a_0}}{N_c^2-1}\Mc_0^{i}(\p_0)\Mc_0^{* \bar i}(\bar \p_0)\,.
\end{equation}
This makes it possible to simplify the 2-point function
\begin{align}
    &\delta^{a_0 \bar a_0} (\p_1|\Gc^{a_1a_0}(t_1,t_0)|\p_0)(\bar \p_0|\Gc^{\dagger \bar a_0 \bar a_1}(t_0,t_1)|\bar\p_1) 
    = \delta^{a_1 \bar a_1} (\p_1;\bar \p_1 |S^{(2)}(t_1,t_0)|\p_0;\bar \p_0)\nn
    &=\delta^{a_1 \bar a_1} (2\pi)^2\delta(\bar \p_0-\p_0)\Pc(\p_1-\p_0|t_1,t_0)\,.
\end{align}
With this, the 3-point function becomes
\begin{align}
    &f^{a_1 b_1 c_1} (\k_2|\Gc^{b_2b_1}(t_2,t_1)|\k_1)(\q_2|\Gc^{c_2c_1}(t_2,t_1)|\p_1-\k_1)(\bar\p_1|\Gc^{\dagger a_1\bar a_2}(t_1,t_2)|\bar \p_2)\nn
    &= f^{\bar a_2 b_2 c_2} (\k_2,\q_2;\bar \p_2 |S^{(3)}(t_2,t_1)|\k_1, \p_1-\k_1;\bar \p_1)\nn
    &= f^{\bar a_2 b_2 c_2} (2\pi)^2\delta(\bar \p_1-\p_1) \Sc^{(3)}(\k_2-z\bar \p_2,\k_1-z\p_1,\p_1-\bar \p_2|t_2,t_1)\,,
\end{align}
Finally, the 4-point function is 
\begin{align}
    &f^{\bar a_2 b_2 c_2} f^{\bar a_2 \bar b_2 \bar c_2} (\k|\Gc^{b b_2}(t_\infty,t_2)|\k_2)(\q|\Gc^{c c_2}(t_\infty,t_2)|\q_2)(\bar \k_2|\Gc^{\dagger \bar b_2 b}(t_2,t_\infty)|\k)(\bar \p_2-\bar\k_2|\Gc^{\dagger \bar c_2 c}(t_2,t_\infty)|\q)\nn
    &=N_c(N_c^2-1) (\k,\q;\k,\q | S^{(4)}(t_\infty,t_2)|\k_2,\q_2;\bar\k_2,\bar\p_2-\bar\k_2)\nn
    &=N_c(N_c^2-1) (2\pi)^2\delta(\bar\p_2-\k_2-\q_2)\Sc^{(4)}((1-z)\k-z\q,\k_2-z\bar\p_2,\bar\k_2-z\bar\p_2,\bar\p_2-\k-\q|t_\infty,t_2)\,.
\end{align}
After using these relations the cross section becomes
\begin{align}
    \frac{\rmd^2 \sigma}{\rmd \Omega_\k \rmd \Omega_\q} &= \frac{g^2}{(2E)^2} N_c \rmR \int_0^\infty \rmd t_2 \int_0^{t_2} \rmd t_1
    \int_{\p_0\p_1\k_1\k_2\bar\p_2\bar\k_2}  
    \Gamma^{ijk}(\k_1-z\p_1)\Gamma^{\bar ijk}(\bar \k_2-z\bar \p_2) \nn
    &\times \Mc_0^{i}(\p_0)\Mc_0^{* \bar i}(\p_0)
    \Pc(\p_1-\p_0|t_1,t_0)\nn
    &\times \Sc^{(3)}(\k_2-z\bar \p_2,\k_1-z\p_1,\p_1-\bar \p_2|t_2,t_1)\nn
    &\times \Sc^{(4)}((1-z)\k-z\q,\k_2-z\bar\p_2,\bar\k_2-z\bar\p_2,\bar\p_2-\k-\q|t_\infty,t_2)\,.
\end{align}
\begin{align}
    N_c \Gamma^{ijk}(\Q_1)\Gamma^{\bar ijk}(\Q_2) &= 
    \left[\frac{1}{z^2}+\frac{1}{(1-z)^2}\right] \Q_1\cdot \Q_2 \delta^{i \bar i}+2 Q_1^i Q_2^{\bar i}  \nn
    &=\frac{4}{z(1-z)}P_{\textrm{gg}}(z)\delta^{i \bar i}\Q_1\cdot \Q_2  
\end{align}
The last step is true if we only consider inclusive cross sections and average over azimuthal angles.

We can now combine the initial hard processes in $\frac{\rmd \sigma_0}{\rmd \Omega_{\p_0}}=|\Mc_0(\p_0)|^2$, and extract this from the equation.
Using this together with Eqs. \eqref{eq:cross-section-P} we get the generalized splitting function
\begin{align}
    P_2(\k,\q;\p_0) &= \frac{g^2}{z(1-z)E^2} P_{gg}(z) \rmR \int_0^\infty \rmd t_2 \int_0^{t_2} \rmd t_1
    \int_{\p_1\k_1\k_2\bar\p_2\bar\k_2}  \nn
    &\times (\k_1-z\p_1)\cdot(\bar \k_2-z\bar \p_2)\Pc(\p_1-\p_0|t_1,t_0)\nn
    &\times \Sc^{(3)}(\k_2-z\bar \p_2,\k_1-z\p_1,\p_1-\bar \p_2|t_2,t_1)\nn
    &\times \Sc^{(4)}((1-z)\k-z\q,\k_2-z\bar\p_2,\bar\k_2-z\bar\p_2,\bar\p_2-\k-\q|t_\infty,t_2)\,.
\end{align}
Here we have introduced the Altarelli-Parisi splitting function
\begin{equation}
    P_{gg}(z)=N_c\frac{[1-z(1-z)]^2}{z(1-z)}\,.
\end{equation}


\section{Deriving the Schrödinger equation}\label{appendix:schr-eq}
Here we show how to derive the Schrödinger equation \eqref{eq:sch-eq-S4} starting with the path integral \eqref{eq:4-point-pos-space}. We will do this for a general system, which the specific system in question is only a special case of. Let the path integral go from some initial state at $(t_0,\u_0,{\bar \u}_0)$ to some final state at $(t+\epsilon,\u_f,{\bar \u}_f)$.
\begin{equation}
\Qc_i(\u_f,{\bar \u}_f,\u_0,{\bar \u}_0|t+\epsilon,t_0) \equiv \int^{\u_f}_{\u_0} \cD\bm u \int^{{\bar \u}_f}_{{\bar \u}_0} \cD \ub
\,\rme^{i\frac{\omega}{2}\int_{t_0}^{t+\epsilon} \dd s\, (\dot \u^2-\dot {\bar \u}^2)}\Cc_i(\u,\ub|t+\epsilon)\,.
\end{equation}
In this equation $\Cc_i$ indicates some Wilson line correlator. In \cite{Isaksen:2020npj} it was shown that all Wilson line correlators can be written as a system of differential equations.
\begin{equation}
    \frac{\rmd}{\rmd t}\Cc_i(t) = \M_{ij}\Cc_j(t)\,.
\end{equation}
Here $\Cc_j$ indicates some other color configuration of the same Wilson lines. Notice that this implies that $\Cc_i(t+\epsilon)=\Cc_i(t)+\epsilon \M_{ij}\Cc_j(t)$ when $\epsilon\to0$. Start by discretizing the path integral with $N$ time intervals with length $\epsilon$. Let the whole path integral go from $t_0$ to $t+\epsilon$. Then we separate the very last interval from the $N-1$ preceding ones. Then we have
\begin{align}
&\Qc_i(\u_f,{\bar \u}_f,\u_0,{\bar \u}_0|t+\epsilon,t_0) = \nn
&\frac{1}{A_\u A_{\bar \u}} \int \dd\u_{N-1}\int \dd{\bar \u}_{N-1} 
\exp\left\{i\frac{\omega}{2}\int_{t}^{t+\epsilon} \dd s(\dot \u_N^2-\dot {\bar \u}_N^2)\right\} \nn
&\times \int^{\u_{N-1}}_{\u_0}\cD\u \int^{{\bar \u}_{N-1}}_{{\bar \u}_0}\cD{\bar \u}  \,\exp\left\{i\frac{\omega}{2}\int_{t_0}^{t} \dd s(\dot \u^2-\dot {\bar \u}^2) \right\}\nn
&\times\left(\Cc_i(\u,\ub|t)+\epsilon \M_{ij}(\u_{N-1},\ub_{N-1})\Cc_j(\u,\ub|t)\right)\nn
&=\frac{1}{A_\u A_{\bar \u}} \int \dd\u_{N-1}\int \dd{\bar \u}_{N-1} 
\exp\left\{i\frac{\omega}{2}\epsilon \left[\left(\frac{\u_{N}-\u_{N-1}}{\epsilon}\right)^2-\left(\frac{{\bar \u}_{N}-{\bar \u}_{N-1}}{\epsilon}\right)^2\right]\right\} \nn
&\times \left(\Qc_i(\u_{N-1},\ub_{N-1}|t)+\epsilon \M_{ij}(\u_{N-1},\ub_{N-1}) \Qc_j(\u_{N-1},\ub_{N-1}|t)\right)\,.
\end{align}
We have used the more compact notation $\Qc_i(\u_{N-1},{\bar \u}_{N-1}|t)=\Qc_i(\u_{N-1},{\bar \u}_{N-1},\u_0,{\bar \u}_0|t,t_0)$. The normalization factors are
\begin{align} \label{eq:norm-factors}
A_\u&= -A_{\bar \u}=\frac{2 \pi i \epsilon}{\omega}
\end{align}
Using that $\u_N = \u_f$ we see that the integral over $\u_{N-1}$ is dominated by terms where $\u_f-\u_{N-1}$ is small (same for ${\bar \u}_{N-1}$ and ${\bar \u}_f$). We define new variables through $\u_{N-1}=\u_f + \fxi$ and ${\bar \u}_{N-1}={\bar \u}_f + \feta$ so the integration becomes
\begin{align}\label{eq:path-int-derivation}
&\Qc_i(\u_f,{\bar \u}_f,\u_0,{\bar \u}_0|t+\epsilon,t_0) \nn
&=\frac{1}{A_\u A_{\bar \u}} \int \dd\fxi\int \dd\feta \exp\left\{i \frac{\omega}{2\epsilon}\left(\fxi^2 - \feta^2 \right)\right\}
\Qc_i(\u_f + \fxi,{\bar \u}_f + \feta|t)\nn
&+\frac{\epsilon}{A_\u A_{\bar \u}} 
\int \dd\fxi\int \dd\feta \exp\left\{i \frac{\omega}{2\epsilon}\left(\fxi^2 - \feta^2 \right)\right\}\M_{ij}(\u_f + \fxi,{\bar \u}_f + \feta)\Qc_j(\u_f + \fxi,{\bar \u}_f + \feta|t)\,.
\end{align}
Now we Taylor expand $\Qc_i$ and $\M_{ij}$ in $\fxi$ and $\feta$ in the following way
\begin{align}
    f(\u_f + \fxi,{\bar \u}_f + \feta|t)&=
    \Biggl[1+\fxi\cdot\partial_\u +\feta\cdot\partial_{\bar \u}  \nn
    &+\left.\frac12\left(\xi_1^2 \frac{\partial^2}{\partial u_1^2}+\xi_2^2 \frac{\partial^2}{\partial u_2^2}+\eta_1^2 \frac{\partial^2}{\partial v_1^2}+\eta_2^2 \frac{\partial^2}{\partial v_2^2}\right)\right]f(\u_f,{\bar \u}_f|t)\,.
\end{align}
After using the normalization factors \eqref{eq:norm-factors} the Gaussian integrals become
\begin{align}
\frac{1}{A_\u A_{\bar \u}} &\int \dd\fxi \int \dd\feta \, \exp\left\{i \frac{\omega}{2\epsilon}\left(\fxi^2 - \feta^2 \right)\right\}= 1 \nn
\frac{1}{A_\u A_{\bar \u}} &\int \dd\fxi\int \dd\feta \, (\fxi\cdot\partial_\u +\feta\cdot\partial_{\bar \u} ) \exp\left\{i \frac{\omega}{2\epsilon}\left(\fxi^2 - \feta^2 \right)\right\}= 0 \nn
\frac{1}{A_\u A_{\bar \u}} &\int \dd\fxi\int \dd\feta \, \left(\xi_1^2 \frac{\partial^2}{\partial u_1^2}+\xi_2^2 \frac{\partial^2}{\partial u_2^2}+\eta_1^2 \frac{\partial^2}{\partial v_1^2}+\eta_2^2 \frac{\partial^2}{\partial v_2^2}\right) \exp\left\{i \frac{\omega}{2\epsilon}\left(\fxi^2 - \feta^2 \right)\right\} \nn
&= \frac{i \epsilon}{\omega}\left(\partial_\u^2-\partial_{\bar \u}^2\right)\,.
\end{align}
Going back to \eqref{eq:path-int-derivation} one can see that after expanding all the terms linear in $\fxi$ and $\feta$ are zero. The second term in \eqref{eq:path-int-derivation} already goes as $\epsilon$, so the integrals quadratic in $\fxi$ and $\feta$ give something going as $\epsilon^2$, and can be discarded. In the end we get
\begin{align}
&\Qc_i(\u,{\bar \u}, \u_0,{\bar \u}_0|t,t_0)+\epsilon\frac{\partial}{\partial t}\Qc_i(\u,{\bar \u}, \u_0,{\bar \u}_0|t,t_0) \nn
&=\left[1+\frac{i \epsilon}{2\omega}\left(\partial^2_\u-\partial^2_{\bar \u}\right)\right]\Qc_i(\u,{\bar \u}, \u_0,{\bar \u}_0|t,t_0)+\epsilon \M_{ij}\Qc_j(\u,{\bar \u}, \u_0,{\bar \u}_0|t,t_0)\,.
\end{align}
Gathering all the terms linear in $\epsilon$ we get the Schrödinger equation
\begin{align}
    &\left[i\frac{\partial}{\partial t}+\frac{\partial^2_\u-\partial^2_{\bar \u}}{2\omega} \right]\Qc_i(\u,{\bar \u},\u_0,{\bar \u}_0|t,t_0)
    -i \M_{ij}(\u,\ub)\Qc_j(\u,{\bar \u},\u_0,{\bar \u}_0|t,t_0)\nn
    &=i \Qc_i(\u,{\bar \u},\u_0,{\bar \u}_0|t_0,t_0)\,.
\end{align}
Notice that there is a minus sign in front of $\partial^2_{\bar \u}$. This is something that would only normally appear if the mass is negative. The initial condition is a delta function at the starting point in the transverse plane
\begin{equation}
    i \Qc_i(\u,{\bar \u},\u_0,{\bar \u}_0|t_0,t_0) = i \delta(t-t_0)\delta^2(\u-\u_0)\delta^2(\ub-\ub_0)\,.
\end{equation}
\section{The eikonal limit}\label{app:eikonal}
If the energy of both of the daughter partons is big one can use the eikonal approximation. The eikonal approximation assumes that the partons travel on straight lines through the medium, and neglects the effects of momentum broadening. The medium propagator Eq. \eqref{eq:prop-G} then reduces to the eikonal propagator Eq. \eqref{eq:eikonal}. The splitting process simplifies greatly in the eikonal approximation. Starting with Eq. \eqref{eq:emission-spectrum-not-simplified} all of the momentum integrals can be done, and you end up with
\begin{align}
    \Pc_2(\k,\q;\p_0) &= \frac{g^2}{ z(1-z)E^2}P_{ba}(z) (2\pi)^2 \delta^2(\p_0-\q-\k) 
    \rmR \int_0^\infty\rmd t_1 \int_{t_1}^\infty \rmd t_2 \nn
    &\times ((1-z)\k-z\q)^2 \rme^{-i \frac{1}{2 \omega}(z\q-(1-z)\k)^2(t_2-t_1)}
    \Cc^{(4)}(t_\infty,t_2)\Cc^{(3)}(t_2,t_1)\,.
\end{align}
The objects $\Cc^{(3)}$ and $\Cc^{(4)}$ represent Wilson line correlators, given in Eq. \eqref{eq:C-Wilson-correlators}.

After defining $\p = z\q-(1-z)\k$ and $\P=\q+\k$, and integrating out $\P$ we end up with the eikonal analogue of Eq. \eqref{eq:p2-simplified-momentum-space}
\begin{align}\label{eq:spectrum-eikonal}
    (2\pi)^2\frac{\rmd I}{\rmd z \rmd^2 \p} &= 
    \frac{\alpha_s \p^2  }{ \omega^2}P_{ba}(z)
    \rmR \int_0^\infty\rmd t_1 \int_{t_1}^\infty \rmd t_2 \nn
    &\times \rme^{-i \frac{\p^2}{2 \omega}(t_2-t_1)}
   \Cc^{(4)}(t_\infty,t_2) \Cc^{(3)}(t_2,t_1)\,.
\end{align}

This is perhaps more conveniently written in terms of the angle $\theta$ between the two daughter partons, using the relation $\p^2 \simeq (\theta \omega)^2$. This is accurate for small angles $\theta\ll 1$. Then the emission spectrum is
\begin{align}\label{eq:spectrum-eikonal-theta}
    &\frac{\rmd I}{\rmd z \rmd \theta} = 2 \pi \theta \omega^2 \frac{\rmd I}{\rmd z \rmd^2 \p} \nn 
    &= \frac{\alpha_s}{ 2\pi} \omega^2\theta^3 P_{ba}(z)
    \rmR \int_0^\infty\rmd t_1 \int_{t_1}^\infty \rmd t_2 
    \,\rme^{-i \frac{\omega \theta^2 }{2}(t_2-t_1)}
   \Cc^{(4)}(t_\infty,t_2) \Cc^{(3)}(t_2,t_1)\,.
\end{align}
This spectrum was calculated for three different processes in \cite{Isaksen:2020npj}. In this paper we will write the spectra in terms of the transverse momentum $\p$. 

Again, we can divide it into the out-out (vacuum), in-out, and in-in contributions. For the vacuum contribution we have $\Cc^{(4)} = \Cc^{(3)} = 1$, and we get the same as in the non-eikonal case
\begin{align}
   (2\pi)^2\frac{\rmd I^{\rm{out-out}}}{\rmd z \rmd^2 \p} = \frac{2 \alpha_s}{\p^2}P_{ba}(z)\,.
\end{align}
For the out-out spectrum we have $\Cc^{(4)} = 1$, and we get
\begin{align}
    (2\pi)^2\frac{\rmd I^{\rm{in-out}}}{\rmd z \rmd^2 \p} &= \frac{\alpha_s\p^2 }{ \omega^2}P_{ba}(z)
    \rmR \int_0^L\rmd t_1 \int_{t_1}^L \rmd t_2 \nn
    &\times \rme^{-i \frac{\p^2}{2 \omega}(t_2-t_1)}\Cc^{(3)}(L,t_1)\,.
\end{align}
Lastly, the in-in spectrum is
\begin{align}
    (2\pi)^2\frac{\rmd I^{\rm{in-in}}}{\rmd z \rmd^2 \p} &= \frac{\alpha_s \p^2 }{ \omega^2}P_{ba}(z)
    \rmR \int_0^L\rmd t_1 \int_{t_1}^L \rmd t_2 \nn
    &\times \rme^{-i \frac{\p^2}{2 \omega}(t_2-t_1)}
   \Cc^{(4)}(L,t_2) \Cc^{(3)}(t_2,t_1)\,.
\end{align}

Let us consider the $\gamma \to q \bar q$ process and use the harmonic approximation. The partons travel on straight lines, so $\u^2(t)=(t-t_1)^2\theta^2$ and $\ub^2(t) = (t-t_2)^2\theta^2$. The three-point function is
 \begin{equation}
    \Cc^{(3)}(t_2,t_1) = \rme^{-\frac{1}{12} \frac{\qhat \p^2}{\omega^2}  (t_2-t_1)^3 } \,.
\end{equation}
Hence, the in-out contribution is given by
\begin{align}
    (2\pi)^2\frac{\rmd I^{\rm{in-out}}}{\rmd z \rmd^2 \p} &= -\frac{2\alpha_s}{z(1-z)E}P_{ba}(z)
    \rmR \,i \int_0^L\rmd \tau \,\rme^{-i \frac{\p^2}{2 \omega}\tau}\rme^{-\frac{1}{12} \frac{\qhat \p^2}{\omega^2}  \tau^3 }\,.
\end{align}
The four-point function is not trivial to calculate in the eikonal approximation either. However, it can now be calculated through an ordinary differential equation, instead of a more complicated Schrödinger equation
\begin{equation}
    \frac{\rmd}{\rmd t}\Cc^{(4)}_i(t,t_2) = \M_{ij}(t)\Cc_j(t)\,.
\end{equation}
The matrix in Eq. \eqref{eq:potential-matrix-pair} now has a simple time dependence
\begin{equation}\label{eq:potential-matrix-pair-eikonal}
    \mathbb{M}(t) =
    -\frac{\qhat \theta^2}{4C_F} 
    \begin{bmatrix}
    C_F[(t-t_1)^2+(t-t_2)^2]+\frac{1}{N_c}(t-t_1)(t-t_2)
    & -\frac{1}{N_c}(t-t_1)(t-t_2) \\
    N_c z(1-z)(t_2-t_1)^2
    & [C_F-N_c z(1-z)](t_2-t_1)^2
    \end{bmatrix}\,.
\end{equation}
This can be solved numerically.

In the large-$N_c$ approximation the above matrix simplifies
\begin{equation}\label{eq:potential-matrix-pair-eikonal-Nc}
    \mathbb{M}(t) =
    -\frac{\qhat \theta^2}{4} 
    \begin{bmatrix}
    (t-t_1)^2+(t-t_2)^2
    & 0 \\
    2z(1-z)(t_2-t_1)^2
    & [1-2 z(1-z)](t_2-t_1)^2
    \end{bmatrix}\,.
\end{equation}
The zero in the upper right entry means that $\Cc^{(4)}_1$ can be solved analytically
\begin{equation}
    \Cc^{(4)}_1(L,t_2) = \rme^{-\frac{1}{12}\frac{\qhat \p^2}{\omega^2} \left[(L-t_2)^3+(L-t_1)^3-(t_2-t_1)^3\right]}\,.
\end{equation}
The physical solution that we need in the emission spectrum is $\Cc^{(4)}_2$. This can be solved using the solution of 
$\Cc^{(4)}_1$ through
\begin{equation}
    \Cc^{(4)}_2(L,t_2) = \Cc^{(4)}_{2,\rm{fac}}(L,t_2) -\frac12 \frac{\qhat \p^2}{\omega^2}z(1-z) (t_2-t_1)^2
    \int_{t_2}^L \rmd s \,\Cc^{(4)}_{2,\rm{fac}}(L,s) \Cc^{(4)}_1(s,t_2)\,,
\end{equation}
where the factorizable solution is
\begin{equation}
    \Cc^{(4)}_{2,\rm{fac}}(L,s) = \rme^{-\frac{1}{4}\frac{\qhat \p^2}{\omega^2} [1-2z(1-z)](L-s)(t_2-t_1)^2}\,,
\end{equation}
and the second term constitutes the non-factorizable solution. 

The factorizable contribution to the in-in spectrum is
\begin{align}\label{eq:spectrum-fac-eikonal}
   (2\pi)^2\frac{\rmd I_{\rm{fac}}^{\rm{in-in}}}{\rmd z \rmd^2 \p} &= 
    \frac{4\alpha_s }{\qhat[1-2z(1-z)]} P_{ba}(z)\nn
    &\times\rmR \, \int_0^L\rmd \tau \,\frac{1}{\tau^2}\rme^{-i \frac{\p^2}{2 \omega}\tau}\rme^{-\frac{1}{12} \frac{\qhat \p^2}{\omega^2}  \tau^3 }  \left(1-\rme^{-\frac{1}{4}\frac{\qhat \p^2}{\omega^2} [1-2z(1-z)](L-\tau)\tau^2}\right)\,.
\end{align}

Similarly, the non-factorizable contribution is given by the slightly more complicated expression
\begin{align}\label{eq:spectrum-nonfac-eikonal}
   &(2\pi)^2\frac{\rmd I_{\rm{non-fac}}^{\rm{in-in}}}{\rmd z \rmd^2 \p} = 
    -\frac{2\alpha_s \theta^2 z(1-z)}{[1-2z(1-z)]} P_{ba}(z)\nn
    &\times\rmR \, \int_0^L\rmd \tau  \int_\tau^L\rmd \sigma \,\rme^{-i \frac{\p^2}{2 \omega}\tau}\rme^{-\frac{1}{12} \frac{\qhat \p^2}{\omega^2}  [(\sigma-\tau)^3+\sigma^3] }  \left(1-\rme^{-\frac{1}{4}\frac{\qhat \p^2}{\omega^2} [1-2z(1-z)](L-\sigma)\tau^2}\right)\,.
\end{align}

\bibliography{references.bib} 
\bibliographystyle{jhep} 
\end{document}